\documentclass[11pt]{article}
\usepackage[letterpaper, left=1in, right=1in, top=1in,bottom=1in]{geometry}
\usepackage{amsmath}    %
\usepackage{amsfonts}   %
\usepackage{amssymb}    %
\usepackage{amsthm}     %
\usepackage{graphicx}   %
\usepackage{hyperref}   %
\usepackage{xcolor}
\usepackage{titletoc}
\usepackage{tikz}
\usepackage{float}
\usepackage{mathtools}
\usepackage{caption}
\usepackage{subcaption}
\usepackage{bbm}
\usepackage[capitalize]{cleveref}
\usetikzlibrary{angles,quotes} 
\usepackage{gensymb}
\usepackage{enumitem}
\usepackage{placeins}

\usepackage[round]{natbib}
\usepackage{setspace}
\usepackage[T1]{fontenc}
\usepackage[utf8]{inputenc}
\usepackage{auxiliary}

\newtheoremstyle{mytheoremstyle}
  {\topsep} %
  {\topsep} %
  {\itshape} %
  {} %
  {\bfseries} %
  {.} %
  {.5em} %
  {} %
\theoremstyle{mytheoremstyle}
\newtheorem{proposition}{Proposition}[section]
\newtheorem{definition}{Definition}[section]
\newtheorem{lemma}{Lemma}[section]
\newtheorem{theorem}{Theorem}[section]

\newtheorem{remark}{Remark}[section]

\crefname{figure}{Figure}{Figures}
\Crefname{figure}{Figure}{Figures}

\usepackage{color}              %

\DeclareMathOperator*{\argmax}{arg\,max}

\NewDocumentEnvironment{myproof}{o}
  {\IfNoValueTF{#1}{\emph{{{Proof.}}}} {\emph{{{#1}} }} }
  {\hfill$\qed$}

\title{
Misspecified Estimate-then-Optimize Leads to \\ Supra-Competitive Prices
}
\author{
Jackie Baek\thanks{Stern School of Business, New York University, \texttt{baek@stern.nyu.edu}} \and 
Vivek F. Farias\thanks{Massachusetts Institute of Technology, \texttt{vivekf@mit.edu}} \and Farrell Wu\thanks{Massachusetts Institute of Technology, \texttt{farrellw@mit.edu}}
}
\date{}

\begin{document}

\maketitle

\begin{abstract}
  We study whether simple algorithmic pricing systems can systematically produce collusive-like prices in multi-firm markets. 
We consider firms that price using a myopic estimate-then-optimize rule: each repeatedly fits a demand model to its own price and sales history and sets the price that maximizes estimated profit. This demand model is misspecified, omitting competitors' prices. We analyze the dynamics of this rule when it is initialized by an exploration phase of independent random prices. 
We characterize when this pipeline converges to supra-competitive prices above the Nash equilibrium, 
via a fluid-limit ordinary differential equation analysis.
We show that supra-competitive prices arise when firms initially explore within similar price ranges on the same side of the Nash price.
Moreover, prices can be substantially above the Nash price; we show that prices can reach monopoly levels under symmetric exploration.
Simulations calibrated to a real multifamily rental market confirm that supra-competitive outcomes arise robustly beyond our theoretical assumptions, including under finite horizons, heterogeneous products, and nonlinear logit demand.
\end{abstract}

\section{Introduction}
Businesses increasingly rely on algorithms to automate pricing decisions \citep{brown2023competition}.
This trend has raised a fundamental question: can algorithmic pricing produce collusive-like outcomes?
A growing literature suggests that the answer can be yes for certain learning algorithms.
For example, the seminal work of \cite{calvano2020artificial} demonstrates that Q-learning agents can reach and sustain supra-competitive prices through reward-and-punishment dynamics. These results are highly sensitive to the explicit learning dynamics assumed\footnote{For instance, in the case of \cite{calvano2020artificial}, the definition of agent state, discount factor and exploration probability must all be carefully calibrated to get sustained collusion.}. 
However, it remains unclear whether these findings apply to the pricing systems most commonly deployed in practice: classic
reinforcement learning approaches such as Q-learning are complex to implement, require long and costly training, and they are outperformed by simpler pricing alternatives \citep{den2024artificial}.

In practice, many firms use far simpler, model-based approaches to algorithmic pricing: estimate a demand curve from their historical data of past prices and realized sales, then apply a myopic ``estimate-then-optimize'' rule to update prices, often without explicitly modeling competitors' prices or strategic responses. This approach has been widely studied in the single-agent learning literature under various names such as \textit{certainty-equivalence} in adaptive control \citep{simon1956dynamic}, and \textit{greedy} or \textit{exploitation} policies in reinforcement learning \citep{sutton1998reinforcement}. It is also a performant baseline in the revenue management and inventory control literature \citep{lariviere1999stalking, aviv2002pricing, farias2010dynamic}.

It is unclear whether these more realistic dynamics systematically produce supra-competitive outcomes.  To this end, \cite{cooper2015learning}, building on earlier work by \cite{kirman1975learning, kirman1986mistaken, kirman1995learning}, investigate sellers that use monopoly (i.e., misspecified) demand models under estimate-then-optimize dynamics. 
They show that, depending on the initial conditions, the long-run price can be sub-competitive, Nash, or supra-competitive. This is a statement about what is \emph{possible}: they establish it by constructing, for each target price, a short initial history under which prices freeze at that target within a few periods. Their result thus characterizes which prices \emph{can} persist, but does not analyze how prices evolve from natural initial conditions, and so does not identify which outcome actually \emph{arises}. 
Therefore, whether realistic pricing dynamics systematically produce supra-competitive prices remains an open question.

We answer this question by analyzing the dynamics directly. 
We study firms that begin with an exploration phase, posting independent random prices, before settling into the estimate-then-optimize dynamic above. 
We view the exploration phase as a model of uncoordinated initial price variation that initializes the learning dynamics, rather than a coordinated algorithm firms deliberately adopt.
In a fluid scaling where the exploration and exploitation horizons grow large, we prove that \emph{when firms initially explore within a well-defined region of similar prices, the limiting prices of all firms are supra-competitive}. Crucially, the relevant region of exploration covers a nontrivial fraction of the exploration space and arises naturally when prices are clustered.
The mechanism for supra-competitive prices does not require punishment, communication, or explicit coordination; the misspecified pricing dynamic suffices.  Computational evidence on both synthetic and real-world-calibrated markets further shows that supra-competitive outcomes persist beyond the conditions of our analytical results and in realistic rental-market calibrations.

\subsection{Summary of model and results}

\paragraph{Model setup.}
We study a competitive market with $N$ symmetric firms over $K + T$ time periods under a linear demand model.
Each firm prices by a myopic estimate-then-optimize rule (``exploitation'') for $T$ periods, initialized by an exploration phase of $K$ periods.
In the exploration phase, firm $i$'s price is drawn independently over time from a distribution with mean $\mu_i$.
Then, throughout the exploitation phase, each firm estimates a monopoly-style demand curve in each period using only its own past prices and realized sales. This demand curve omits competitor prices
and is thus misspecified.
This misspecification reflects a common real-world constraint where firms have limited visibility into competitors' real-time pricing or choose a simpler model for statistical or computational efficiency.

We analyze a ``fluid'' scaling regime where both the exploration and exploitation horizons grow large.
Under this asymptotic scaling, we show that terminal prices converge to a deterministic limit described by a \textit{price-moments ordinary differential equation (ODE)}. 
This ODE keeps track of the running price means and covariances for all firms, and a careful analysis of this ODE yields the results we describe below.

\begin{figure}[!htbp]
  \centering
  \begin{tikzpicture}[x=3.1cm,y=3.1cm]
  \pgfmathsetmacro{\xMax}{4/3}
  \pgfmathsetmacro{\yMax}{4/3}
  \pgfmathsetmacro{\pNE}{2/3}
  \pgfmathsetmacro{\pM}{1}
  \pgfmathsetmacro{\xCut}{5/6}
  \pgfmathsetmacro{\xZero}{1/2}
  \pgfmathsetmacro{\yZero}{1/2}
  
  \definecolor{green}{RGB}{0,120,0}
  \definecolor{navy}{RGB}{0,0,180}
  
  \pgfmathsetmacro{\mLo}{sqrt(2)-1} %
  \pgfmathsetmacro{\mHi}{sqrt(2)+1} %
  
  \pgfmathsetmacro{\yLoLeft}{\pNE*(1-\mLo)}                 %
  \pgfmathsetmacro{\yLoRight}{\pNE + \mLo*(\xMax-\pNE)}     %
  \pgfmathsetmacro{\xHiBottom}{\pNE*(1-1/\mHi)}             %
  \pgfmathsetmacro{\xHiTop}{\pNE + (\yMax-\pNE)/\mHi}       %
  
  \fill[red!10] (\pNE,\pNE) -- (\xMax,\xCut) -- (\xMax,\yMax) -- (\xCut,\yMax) -- cycle; %
  \fill[red!10] (0,0) -- (\xZero,0) -- (\pNE,\pNE) -- (0,\yZero) -- cycle; %
  
  \draw[->] (0,0) -- (\xMax+0.08,0) node[right] {$\mu_1$};
  \draw[->] (0,0) -- (0,\yMax+0.08) node[above] {$\mu_{2}$};
  
  \draw[thick] (0,\yZero) -- (\xMax,\xCut);
  
  \draw[thick] (\xZero,0) -- (\xCut,\yMax);
  
  \draw[thick, dashed, gray!70] (0,\yLoLeft) -- (\xMax,\yLoRight);
  \draw[thick, dashed, gray!70] (\xHiBottom,0) -- (\xHiTop,\yMax);
  
  \fill (\pNE,\pNE) circle[radius=0.02];
  
  \coordinate (I) at (\pNE,\pNE);
  \coordinate (A) at (\xMax,\xCut);   %
  \coordinate (B) at (\xCut,\yMax);   %
  \pic[draw, line width=0.9pt, "$\theta$", angle radius=0.4cm, angle eccentricity=1.6]
    {angle = A--I--B};
  
  \coordinate (C45) at ({\pNE-0.55},{\pNE-\mLo*0.55});  %
  \coordinate (D45) at ({\pNE-0.25},{\pNE-\mHi*0.25});  %
  \pic[draw, dashed, gray!70, "$45^\circ$", angle radius=0.30cm, angle eccentricity=1.75]
    {angle = C45--I--D45};
  
  \draw[dashed] (0,\pNE) -- (\pNE,\pNE);
  \draw[dashed] (\pNE, 0) -- (\pNE,\pNE);
  \draw (\pNE,0.03) -- (\pNE,-0.03);
  \node[below] at (\pNE,-0.02) {Nash};
  \draw (0.03,\pNE) -- (-0.03,\pNE);
  \node[left] at (-0.02,\pNE) {Nash};
  
  \draw[gray!55] (\xMax,0) -- (\xMax,\yMax) -- (0,\yMax);

  \begin{scope}[xshift=7.0cm] %
    \draw[->] (0,0) -- (\xMax+0.08,0) node[right] {$s$};
    \draw[->] (0,0) -- (0,\yMax+0.08) node[above] {Final price};
  
    \draw[gray!55] (\xMax,0) -- (\xMax,\yMax) -- (0,\yMax);
  
    \draw[dashed] (0,\pNE) -- (\pNE,\pNE);
    \draw[dashed] (\pNE, 0) -- (\pNE,\pNE);
    \draw[dashed] (\pM, 0) -- (\pM,\pM);
  
    \draw (\pNE,0.03) -- (\pNE,-0.03);
    \node[below] at (\pNE-0.05,-0.02) {Nash};
  
    \draw (\pM,0.04) -- (\pM,-0.03);
    \node[below] at (\pM+0.12,-0.02) {Monopoly};
  
    \draw[thick, navy] (0,\pM) -- (\pNE,\pM);
  
    \draw[thick, navy] (\pNE,\pNE) -- (\pM,\pM);
  
    \draw[thick, navy] (\pM,\pM) -- (\xMax,\pM);
  
    \fill[navy] (\pNE,\pNE) circle[radius=0.02, navy];      %
    \draw[navy, fill=white] (\pNE,\pM) circle[radius=0.02]; %
  
    \fill[navy] (\pM,\pM) circle[radius=0.02];
  
    \draw (0.03,\pNE) -- (-0.03,\pNE);
    \node[left] at (-0.02,\pNE) {Nash};
  
    \draw (0.03,\pM) -- (-0.03,\pM);
    \node[left] at (-0.02,\pM) {Monopoly};
  
  \end{scope}
  
  \end{tikzpicture}
  \caption{
    \emph{Left:} the shaded regions depict the best-response cones in the $(\mu_1,\mu_2)$ plane, where $\mu_i$ is firm $i$'s average exploration price. We show that the terminal prices are supra-competitive whenever $(\mu_1,\mu_2)$ lies in the shaded region. The angle $\theta$ depends on the demand parameters but is always at least $45^\circ$, so the cones cover more than one quarter of the feasible exploration-mean space. \emph{Right:} Final price under symmetric exploration, in the regime of vanishing exploration variance and an infinite exploitation horizon, where $s$ is the average price during exploration.
  }
  \label{fig:br_cones_intro}
  \end{figure}
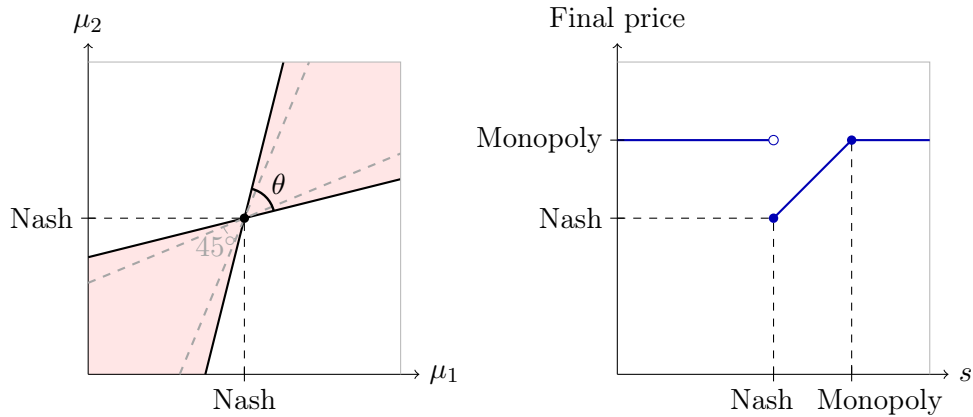

\paragraph{Main theoretical result.}
Our primary result (\cref{thm:supra_sufficient}) states that limiting prices are supra-competitive when firms' exploration means $\mu_i$ lie on the same side of the Nash price within what we call the \textit{best-response cones} (left plot of \cref{fig:br_cones_intro}). 
The upper cone yields supra-competitive prices immediately, while the lower cone yields supra-competitive prices after a finite waiting period.
Informally, the cone condition is that all firms explore with ``similar prices'' on the same side of the Nash price, with the allowable price dispersion across firms increasing as prices move further from Nash.
More formally, these cones are regions where every firm's best-response price, if it were computed against the exploration means $\mu$, would move all firms' prices in the same direction.

Crucially, under our dynamic, firms {\em do not actually best-respond, nor do they model competitor prices}. Rather, firms simply update their prices to maximize revenue as predicted by a misspecified demand model fit to their own historical prices and sales. The core innovation in our analysis lies precisely in analyzing such a dynamic when exploration occurs within a best-response cone.

In a duopoly, the best-response cones cover at least one quarter of the feasible price space for any demand parameters.
We also show that in a duopoly, the cone arises endogenously under our dynamics:
under \textit{any} initial conditions, the limit price falls in the closure of the best-response cones (\cref{prop:duopoly_limit_points_cones}). This has an iterative implication: even when initial exploration falls outside a cone, the resulting terminal prices land inside one; so if firms re-explore around those prices, supra-competitive outcomes are guaranteed in the next round.
For general $N$, we show that naturally clustered exploration profiles land in a best-response cone with nontrivial probability. 
In particular, under a random interval prior in which firms' exploration means are drawn from a common random price band, the probability of landing in a best-response cone is at least $1/4$, and this probability increases with $N$ (\cref{prop:random_interval_cone_probability}).

\paragraph{Quantifying the price gap.}
Beyond establishing supra-competitive prices, 
we quantify the size of the price gap above Nash under symmetric exploration.
We consider a setting where all firms have the same mean exploration price $s$, where we prove a sharp analytical characterization of the final price as exploration noise vanishes and the exploitation horizon grows (right plot of \Cref{fig:br_cones_intro}, \cref{thm:symmetric_explore_zero_noise}).
We show that prices converge to the monopoly price if $s$ is either below the Nash price or above the monopoly price, and when $s$ is between the Nash and monopoly prices, prices converge to $s$.
This reveals that limiting prices can be \textit{substantially} above the Nash price, potentially reaching the monopoly benchmark.
We confirm that the price gap above Nash is substantial beyond the theorem's assumptions, using numerical ODE evaluation (\cref{sec:duopoly_ode_sims}) and stochastic simulations (\cref{sec:stochastic_eval}).

  \paragraph{Key mechanism: persistent correlation.}
  The bias channel is omitted-variable bias: when a firm's prices are positively correlated with its competitors' prices, omitting competitors' prices from the misspecified OLS regression biases the fitted demand slope, leading the firm to price above its true best response. Exploration in a best-response cone initializes common-sign price movements relative to firms' running means, thereby inducing positive cross-firm price correlation. The central novelty of our analysis is to show that this correlation does not dissipate: the common-sign movement is forward invariant, so the correlation persists throughout the exploitation phase. This persistent upward bias ultimately yields supra-competitive limiting prices.

\paragraph{Simulations calibrated to a real rental market.}
In assessing the robustness of our theory there are several issues that merit consideration. In no particular order these include: (1) real-world time scales vs. the fluid regime; (2) nonlinear demand and firm asymmetry; (3) exploration outside of the best-response cones; and (4) a richer supply side with product-specific costs.
To address these considerations, we run empirically calibrated simulations to the multifamily rental market in the Greater Boston area (\cref{sec:sim_design}). Specifically, using demand estimates and market structure from \citet{calder2024algorithmic}, we simulate firms deploying the same explore--then--exploit pricing pipeline in a realistic environment with heterogeneous products, heterogeneous customers, nonlinear logit demand, and product-specific shadow costs calibrated so that observed rents form the Nash benchmark.

This analysis shows that the qualitative predictions of our theory persist: supra-competitive terminal prices arise across a broad range of exploration parameters, and the effect is strongest when firms experiment with similar prices on the same side of the Nash benchmark.
The finite-horizon simulations further show that these effects appear quickly, not only in the asymptotic fluid regime.
These results confirm that supra-competitive outcomes arise robustly across a range of problem parameters and model features that extend beyond the assumptions required for our analytical results, including parameters calibrated to a real-world rental market.

\subsection{Related Work}

There is a burgeoning literature studying how pricing algorithms can give rise to algorithmic collusion. One commonly studied class consists of reinforcement-learning algorithms, especially Q-learning, which has been shown to converge to supra-competitive outcomes in repeated
oligopoly pricing environments \citep{calvano2020artificial,klein2021autonomous,hettich2021algorithmic,abada2024collusion,asker2022artificial}.
In many of these models, collusive behavior emerges through reward-and-punishment dynamics: agents learn to coordinate on cooperative price levels and to implement credible punishments following deviations. Such mechanisms typically require extensive training, careful state representation, and the capacity to learn complex, history-dependent strategies. In contrast, we show that much simpler pricing pipelines can systematically produce supra-competitive prices without any punishment logic or learned threats.

Closer to our work, a line of research studies firms that fit a misspecified linear demand model by OLS and post the implied optimal price \citep{cooper2015learning,lin2025competition,light2026conjectural,yang2026driven}.
Across these prior works, supra-competitive outcomes arise only with an explicit source of cross-firm coupling (e.g., correlated exploration).  A novel feature of our analysis is to show that no such explicit coupling is needed: under \emph{independent} exploration, cross-firm price correlation \emph{emerges endogenously} from the estimate-then-optimize dynamics and drives supra-competitive prices.

\paragraph{Misspecified estimate--then--optimize pricing.}

The most direct comparison to our work is \citet{cooper2015learning}, who study the same estimate-then-optimize rule under misspecification.
  In their most comparable case (both intercept and slope unknown), they characterize the range of prices that can be reached as steady states, by reverse-engineering, for each target price, a short noiseless three-period history that freezes the system at that target. It is straightforward to formalize this construction in our framework: a range of target prices below, at, and above Nash can each be made a steady state of the price-moments ODE by engineering the exploration moments (\cref{prop:cooper_analog} in Appendix~\ref{app:cooper_analog}). 
  This type of construction speaks only to which prices \textit{can} persist; it does not identify which steady state actually \textit{arises} from natural initial histories, which is our focus.

  By contrast, we analyze the estimate-then-optimize dynamics to study which steady state is \textit{selected} when the system is initialized by an independent exploration history.  Notably, the supra-competitive steady states from \cite{cooper2015learning} require positively correlated initial cross-firm covariance; under our independent exploration, we instead show that the necessary cross-firm price correlation emerges endogenously during the exploitation dynamics.

\citet{lin2025competition} study related OLS dynamics under misspecified linear demand.
Their main result derives a stability condition that guarantees convergence to the competitive equilibrium; this is in contrast to our paper's main result, which is to identify conditions that lead to supra-competitive prices.
They also show that supra-competitive prices can be caused by correlation in exploration across firms, or asymmetry in the response time of algorithmic price updates. 

Two recent papers, \citet{light2026conjectural} and \citet{yang2026driven}, also study misspecified OLS demand dynamics but with ongoing independent price perturbations in each period, rather than a one-shot exploration phase as in our paper. Both find that supra-competitive prices require additional cross-firm coupling (correlated experimentation in \citealp{light2026conjectural}, price imitation in \citealp{yang2026driven}), while independent perturbations alone converge to Nash.

\paragraph{Pricing algorithms without demand estimation.}

A separate strand studies bandit and reinforcement-learning algorithms that ignore competitors' prices, but do so without fitting any demand model at all \citep{hansen2021frontiers,banchio2022artificial,douglas2024naive,bichler2024online}.
\citet{hansen2021frontiers} analyze a UCB bandit
policy in a noiseless setting; because both firms run the same deterministic algorithm, their internal learning states coincide after exactly two rounds, so their subsequent actions become fully synchronized in the third round.
\citet{douglas2024naive} study a similar noiseless setting and generalize this result from UCB to all symmetric deterministic bandit algorithms.
\citet{banchio2022artificial} study reinforcement learners (e.g., $\epsilon$-greedy Q-learning) in a two-action prisoner's dilemma and show that when learning is sufficiently slow, payoff estimates can become coupled across firms, sustaining collusive outcomes, although this coupling can disappear under alternative learning-rate specifications. 

On the other side, \citet{bichler2024online} show that a class of mean-based bandit algorithms (e.g., Exp3, MWU) provably converge to Nash equilibrium in repeated price competition.
In contrast to these works, we analyze a natural and widely used estimate--then--optimize rule in a continuous action space with explicit demand estimation. Our results do not rely on exact synchronization of algorithms across firms, nor on specific algorithmic parameters.

\paragraph{Other literature on algorithmic collusion.}

A complementary line of work studies algorithmic collusion in settings where algorithms are explicitly designed to sustain collusive outcomes. This literature dates back to the imperfect-monitoring collusion framework of \citet{green1984noncooperative} and the Edgeworth-cycle model of \citet{maskin1988theory}. 

Recent work in this direction includes \citet{meylahn2022learning,loots2023data,aouad2021algorithmic}. \citet{yang2023regulating} study how fairness regulations on prices interact with collusion in a repeated-game setting. \citet{arunachaleswaran2024algorithmic} view deploying an algorithm as a commitment and study the environment as a leader-follower game. They show that if the leader uses an arbitrary no-regret algorithm, then supra-competitive prices arise if the follower approximately optimizes within this environment.
Recent papers also evaluate whether large language models can exhibit similar collusive behavior \citep{fish2024algorithmic,keppo2025ai}.

On the empirical side, there is growing evidence that pricing algorithms can lead to elevated prices in practice \citep{chen2016empirical, abada2023artificial,calder2024algorithmic,assad2024algorithmic}.
We build on the empirical analysis of \citet{calder2024algorithmic} of the U.S. multifamily rental market by using their calibrated demand model as a more realistic environment than our stylized framework, allowing us to assess the robustness of our results.
\section{Model}
\label{sec:model}

There are $N$ symmetric firms competing over several periods under a linear demand model with idiosyncratic shocks. Firms use a myopic estimate-then-optimize pricing rule: each fits a misspecified, monopoly-style demand curve to its own price and sales history and posts the price that maximizes predicted profit. Because the fitted curve omits competitors' prices, each firm treats variation in demand caused by competitors as unobserved noise.

We study these dynamics over two phases. During the \textit{exploration phase} of $K$ periods, firms post random prices around an initial mean. Thereafter, during the \textit{exploitation phase} of $T$ periods, firms apply the estimate-then-optimize rule to their accumulating price-sales history, updating their fitted demand curves each period. We view the exploration phase as a model of the initial price-sales history that \textit{initializes the exploitation dynamics} (e.g., from legacy manual pricing), rather than a coordinated pricing algorithm that firms deliberately adopt. Any estimate-then-optimize procedure must begin from some initial data; the exploration phase is a parsimonious way to model it.
We now make this setup precise.

\subsection{True demand model}
\label{sec:true_demand_model}

Let $P_{i,t}\in\mathbb{R}_{+}$ denote the price posted by firm $i\in[N]$ in period $t\in[K+T]$. %
Firm $i$ realizes quantity $Q_{i,t}\in\mathbb{R}$, which depends on the full vector of current-period prices $(P_{j,t})_{j\in[N]}$ through the linear demand system
\begin{equation}
    Q_{i,t}
    =
    a-bP_{i,t}
    +
    \frac{c}{N-1}\sum_{j\neq i}P_{j,t}
    +
    \varepsilon_{i,t},
    \qquad i\in[N].
    \label{eq:N_demand}
\end{equation}
Here $a,b,c>0$.
The parameter $b$ captures the own-price effect, while $c$ captures the cross-price effect from competitors' prices.
We assume $b>c$, so that own-price effects dominate cross-price effects.
The demand shocks form a martingale difference sequence with respect to the filtration $\{\mathcal{F}_t\}_{t\ge 0}$ generated by the history of prices and quantities:
\[
    \mathbb{E}\!\left[\varepsilon_{i,t+1}\mid \mathcal{F}_t\right]=0,
    \qquad
    \mathbb{E}\!\left[\varepsilon_{i,t+1}^2\mid \mathcal{F}_t\right]\le \sigma_{\mathrm{env}}^2,
\]
for all $i\in[N]$ and $t\in\{0,\dots,K+T-1\}$.
Thus the shocks may be history-dependent, but they are conditionally mean zero and have uniformly bounded conditional second moments.

\paragraph{Constraints on price and quantity.}
Prices are restricted to the interval $[P_{\min},P_{\max}]$, where $P_{\min}>0$.
We assume
$
    \frac{a}{2b-c}<P_{\max}\le \frac{a}{b}.
$
The lower bound ensures that the price cap does not force firms to remain below the competitive Nash price, which we define in \cref{sec:price_benchmarks}.
The upper bound ensures that conditional expected demand is nonnegative over the feasible price range.
Since $Q_{i,t}$ represents realized sales, we also assume $Q_{i,t} > 0$ almost surely for all $i\in[N]$ and $t\in[K+T]$.

\subsection{Pricing dynamics}
\label{sec:explore_then_exploit_dynamics}

We next describe how firms choose prices, recalling that the first $K$ periods correspond to the \emph{exploration} phase, and the remaining $T$ periods correspond to the \emph{exploitation} phase.

\paragraph{Exploration $(t\le K)$.}
During exploration, firms randomize prices according to a fixed distribution $\mathcal{D}_{\mathrm{exp}}$ on $\mathbb{R}^N$:
\[
    (P_{1,t},\dots,P_{N,t})
    \overset{\mathrm{i.i.d.}}{\sim}
    \mathcal{D}_{\mathrm{exp}},
    \qquad t=1,\dots,K.
\]

We assume prices across firms are independent.
Let $\mu\in[P_{\min},P_{\max}]^N$ denote the mean of $\mathcal{D}_{\mathrm{exp}}$, and let 
$\Sigma_{\exp}:=\operatorname{diag}(\sigma_{1,\exp}^2,\dots,\sigma_{N,\exp}^2)$ denote its covariance matrix, with $\sigma_{i,\exp}^2>0$ for all $i\in[N]$.
We assume $\mathcal{D}_{\mathrm{exp}}$ is supported on $[P_{\min},P_{\max}]^N$.
The exploration phase generates the initial price--quantity histories from which firms begin estimating demand in the exploitation phase.

\paragraph{Exploitation $(t>K)$.}
During exploitation, each firm follows a myopic \emph{estimate--then--optimize} rule.
At each time $t\ge K$, firm $i$ uses its own history $\{(P_{i,s},Q_{i,s})\}_{s=1}^{t}$ to fit the linear demand curve
\begin{equation}
    q(p)=\alpha+\beta p.
\end{equation}
This model is misspecified because the true demand model \eqref{eq:N_demand} also depends on competitors' prices.
Equivalently, firm $i$ treats the variation in demand caused by competitors' prices as unobserved noise.
Let $(\hat\alpha_{i,t},\hat\beta_{i,t})$ denote the ordinary least squares estimates of $(\alpha,\beta)$ based on firm $i$'s history through period $t$.
The fitted demand curve is
\[
    \hat q_{i,t}(p)
    :=
    \hat\alpha_{i,t}+\hat\beta_{i,t}p.
\]
Firm $i$ then chooses the next period's price by maximizing predicted profit over the feasible price interval:
\begin{equation}
    P_{i,t+1}
    :=
    \argmax_{p\in[P_{\min},P_{\max}]}
    \;
    p\bigl(\hat\alpha_{i,t}+\hat\beta_{i,t}p\bigr),
    \qquad t=K,\dots,K+T-1.
    \label{eq:pricing_rule}
\end{equation}

This problem has a unique maximizer.\footnote{More specifically, this problem has a unique maximizer for every fitted demand curve generated by an admissible history; in the nonnegative-slope case, the realized-demand and price constraints \(Q_{i,s}>0\) and \(P_{i,s}\in[P_{\min},P_{\max}]\) make \(P_{\max}\) the unique endpoint maximizer; see Lemma~\ref{lem:ols_price_from_moments} in Appendix~\ref{app:terminal_moment_state}}
To summarize the feedback loop: prices generate data, the data update the fitted demand curve, and the updated fitted demand curve determines the next price.

\begin{remark}[Bayesian reformulation of the exploration phase]
    \label{rem:bayesian}
    The exploration phase can be replaced by an equivalent Bayesian formulation that encodes the same initial information through a prior rather than a random sample. Specifically, each firm runs a Bayesian linear regression for its misspecified demand curve $q(p)=\alpha+\beta p$, starting from a Gaussian prior on the intercept and slope $(\alpha,\beta)$ whose sufficient statistics match $(\mu,\Sigma_{\exp})$.
    Each period, the firm updates this prior into a posterior using the new price-sales observation, and posts the price that maximizes posterior-mean predicted profit. The resulting Bayesian price dynamics coincide with the exploitation dynamics above.
    Thus $(\mu,\Sigma_{\exp})$ admits two readings: as the moments of an initial random sample, or as the parameters of an initial Bayesian prior. See Appendix~\ref{app:bayesian_interpretation} for the derivation.
\end{remark}

\subsection{Price benchmarks}
\label{sec:price_benchmarks}

Lastly, we define two natural price benchmarks: the competitive Nash price and the monopoly price.

\paragraph{Competitive benchmark.}
For firm $i$ and price vector $p=(p_1,\dots,p_N)$, let $\bar p_{-i}:=\frac{1}{N-1}\sum_{j\neq i}p_j$
denote the average price of its competitors.
Suppose firms compete under the correct demand model \eqref{eq:N_demand} and observe each other's posted prices.
Then firm $i$'s profit-maximization problem depends on its competitors' prices only through $\bar p_{-i}$.
The best-response price for firm $i$ is
\begin{equation}
    \BR(\bar p_{-i})
    =
    \frac{a+c\,\bar p_{-i}}{2b}.
    \label{eq:best_response}
\end{equation}
The symmetric Nash equilibrium is the fixed point at which every firm charges its best response to the common price charged by its competitors.
Thus the unique \textit{Nash price}, denoted $\pNE$, is
\begin{equation}
    \pNE
    =
    \frac{a}{2b-c}.
\end{equation}
We also refer to $\pNE$ as the competitive price, thus labelling prices above $\pNE$ as \textit{supra-competitive}.

\paragraph{Monopoly benchmark.}
The \textit{monopoly price}, denoted $\pMNP$, is the symmetric price that would be chosen if firms jointly maximized total industry profits under the same demand system.
It is given by
\begin{equation}
    \pMNP
    =
    \frac{a}{2(b-c)}.
\end{equation}

The Nash price $\pNE$ and the monopoly price $\pMNP$ therefore form natural benchmarks for contextualizing the prices generated by the learning dynamics.
We assume $\pNE\ge P_{\min}$, but do not necessarily assume $\pMNP\le P_{\max}$. When $c\le b/2$, the monopoly price satisfies $\pMNP\le a/b$.

\section{Convergence to Supra-Competitive Prices: Analytical Results}
\label{sec:convergence}

We now present the main analytical results. The first step is a convergence result for the stochastic system: in a fluid scaling where the exploration and exploitation horizons grow proportionally, terminal prices converge to a deterministic limit characterized by an ODE that tracks running price means and accumulated covariances (\cref{sec:terminal_convergence_price_moments}).

In \cref{sec:supra_competitive_limiting_prices}, we use this ODE to establish two main results. 
First, we give a sufficient condition for supra-competitive limiting prices, expressed through best-response cones in the space of exploration means (\cref{thm:supra_sufficient}). Second, under symmetric exploration, we characterize the limiting price as a function of the exploration mean, quantifying the gap above Nash (\cref{thm:symmetric_explore_zero_noise}).
Finally, we discuss the geometry and scope of these cones, showing that they have a simple duopoly interpretation and arise with nontrivial probability under clustered exploration profiles (\cref{sec:cone_discussion}).

\subsection{Terminal convergence and the price-moments ODE}
\label{sec:terminal_convergence_price_moments}

For each pair $(K,T)$, the pricing dynamics in \cref{sec:model} induce a random terminal price vector $(P_{1,K+T},\dots,P_{N,K+T})$.
We show that when $K,T\to\infty$ with $(K+T)/K\to\tau$, this random vector converges to the solution of a deterministic \textit{price-moments ODE}, defined as follows.

\begin{definition}[Price-moments ODE]
\label{def:price_moments_ode}
Fix an exploration-mean vector $\mu=(\mu_1,\dots,\mu_N)$ and covariance matrix $\Sigma_{\exp}$. The price-moments ODE is the system for $(U(\tau),V(\tau))\in\mathbb{R}^N\times\mathbb{R}^{N\times N}$, $\tau\ge 1$, with $U(1)=\mu$, $V(1)=\Sigma_{\exp}$, and
\[
\dot U=\frac{P-U}{\tau}, \qquad \dot V=(P-U)(P-U)^\top,
\]
where $P(\tau)=P(\tau;\mu,\Sigma_{\exp})\in\mathbb{R}^N$ is the vector of posted prices, defined componentwise. Writing $\bar U_{-i}:=\tfrac{1}{N-1}\sum_{j\ne i}U_j$ and $\bar V_{i,-i}:=\tfrac{1}{N-1}\sum_{j\ne i}V_{ij}$,
\begin{equation}\label{eq:posted_price_formula}
    \widetilde P_i := \frac{(a+c\,\bar U_{-i})V_{ii}-c\,U_i\bar V_{i,-i}}{2(bV_{ii}-c\bar V_{i,-i})},
    \qquad
    P_i := \begin{cases}
        [\widetilde P_i]_{[P_{\min},P_{\max}]}, & \text{if } -bV_{ii}+c\bar V_{i,-i}<0,\\
        P_{\max}, & \text{otherwise.}
    \end{cases}
\end{equation}
\end{definition}

The ODE uses a scaled horizon $\tau \geq 1$, where 
$\tau$ corresponds to $(K+t)/K$ in the stochastic model ($\tau=1$ marks the start of exploitation).
The vector $U(\tau)$ tracks running price means, and the matrix $V(\tau)$ accumulates price covariances, where $V_{ii}$ captures the accumulated variance of firm $i$'s own prices and $V_{ij}$ the accumulated covariance between firms $i$ and $j$.
Each $P_i(\tau)$ is firm $i$'s posted price at time $\tau$; \cref{eq:posted_price_formula} is firm $i$'s misspecified OLS price, expressed in terms of the moments $(U, V)$ (see Appendix~\ref{app:price_moments_identification} for the derivation).

This formula \eqref{eq:posted_price_formula} also makes the source of misspecification bias transparent. When $\bar V_{i,-i}=0$, the posted price reduces to the true best-response $\widetilde P_i=\BR(\bar U_{-i})=(a+c\bar U_{-i})/(2b)$. 
When $\bar V_{i,-i}>0$ (representing positive correlation in firm $i$'s historical prices with its competitors'), this creates an upward omitted-variable bias, pushing $\widetilde P_i$ above the true best-response.

We abbreviate $P^{\mathrm{ODE}}(\tau;\mu,\Sigma_{\exp}):=P(\tau;\mu,\Sigma_{\exp})$ (or $P^{\mathrm{ODE}}(\tau)$ when $\mu$ and $\Sigma_{\exp}$ are fixed) for the ODE-implied terminal price at scaled horizon $\tau$. 
The following theorem states that this deterministic limit is the fluid limit of the original stochastic pricing process (proof in Appendix~\ref{app:proof_terminal_convergence}).

\begin{theorem}[Convergence to the price-moments ODE]
\label{thm:terminal_convergence}
Let \(\{(K_m,T_m)\}_{m\in\mathbb{N}}\) satisfy
\(K_m\to\infty\), \(T_m\to\infty\), and
$(K_m+T_m)/K_m\to \tau\in[1,\infty)$.
Then,
\[
    (P_{1,K_m+T_m},\dots,P_{N,K_m+T_m})\;
    \xrightarrow{\mathbb P}\;
    P^{\mathrm{ODE}}(\tau;\mu,\Sigma_{\exp})
    \qquad\text{as }m\to\infty.
\]
\end{theorem}

We note that a wide range of prices, both below and above Nash, can be made steady states of the price-moments ODE when exploration is allowed to be correlated across firms; this recovers, in our framework, the type of multiplicity result shown by \citet{cooper2015learning} (\cref{prop:cooper_analog} in Appendix~\ref{app:cooper_analog}).

\subsection{Supra-competitive limiting prices}
\label{sec:supra_competitive_limiting_prices}

The previous subsection reduces terminal prices to the deterministic limit
$P^{\mathrm{ODE}}(\tau;\mu,\Sigma_{\exp})$. We now state two results of this limit.
First, we give a sufficient condition on the exploration-mean vector $\mu$ under which the limiting prices are supra-competitive (\cref{thm:supra_sufficient}).
Second, under symmetric exploration, we characterize the limiting price as a function of the exploration mean (\cref{thm:symmetric_explore_zero_noise}).

We begin by defining the regions of exploration mean $\mu$ that enter the sufficient condition. 

\begin{definition}[Best-response cones]
\label{def:best_response_cones}
For an exploration-mean vector $\mu\in[P_{\min},P_{\max}]^N$, write
    $\bar\mu_{-i}:=\frac{1}{N-1}\sum_{j\ne i}\mu_j$.
The upper and lower best-response cones are
\begin{align}
\mathcal{C}^+
&:=
\bigl\{
\mu\in[P_{\min},P_{\max}]^N:
\mu_i>\BR(\bar\mu_{-i}) \ \forall i\in[N]
\bigr\},
\label{eq:upper_cone}\\
\mathcal{C}^-
&:=
\bigl\{
\mu\in[P_{\min},P_{\max}]^N:
\mu_i<\BR(\bar\mu_{-i}) \ \forall i\in[N]
\bigr\}.
\label{eq:lower_cone}
\end{align}
\end{definition}

Membership in $\mathcal{C}^+$ (resp., $\mathcal{C}^-$) means that every firm explores above (resp., below) its best response to its competitors' average exploration mean. 
The left side of \cref{fig:br_cones_intro} in the introduction illustrates the cones for $N=2$.
We defer a detailed discussion and interpretation of these cones to \cref {sec:cone_discussion}.
We now state our main result, whose proof is in \cref{sec:supracomp_proof}.
\begin{theorem}[Supra-competitive limiting prices]
\label{thm:supra_sufficient}
Under the same assumptions as \cref{thm:terminal_convergence},
\begin{enumerate}
    \item[(a)] if $\mu\in\mathcal{C}^+$, then for every $\tau\in[1,\infty)$,
    $
        P^{\mathrm{ODE}}(\tau;\mu,\Sigma_{\exp}) \succ \pNE\mathbf{1};
    $
    \item[(b)] if $\mu\in\mathcal{C}^-$, then there exists $\tau_0\in[1,\infty)$ such that for all $\tau>\tau_0$,
    $
        P^{\mathrm{ODE}}(\tau;\mu,\Sigma_{\exp}) \succ \pNE\mathbf{1}.
    $
\end{enumerate}
\end{theorem}

\cref{thm:supra_sufficient} identifies simple sufficient conditions on the mean exploration prices $\mu$ under which the long-run outcome is supra-competitive. The two cones have slightly different timing implications. If $\mu\in\mathcal{C}^-$, then prices begin below the Nash price, so the exploitation dynamics must first push prices upward; the threshold $\tau_0$ is the waiting period required for the deterministic dynamics to enter the supra-competitive region. In contrast, if $\mu\in\mathcal{C}^+$, the supra-competitive conclusion holds for every $\tau\in[1,\infty)$.

While \cref{thm:supra_sufficient} establishes that limiting prices can exceed the Nash price, it does not specify by how much. Under symmetric exploration, where $\mu=s\mathbf{1}$ and $\Sigma_{\exp}=\sigma_{\exp}^2 I_N$, we obtain a sharp characterization 
in a limit where the exploitation horizon grows and the exploration noise vanishes.

\begin{theorem}[Symmetric exploration]
\label{thm:symmetric_explore_zero_noise}
Suppose $\mu=s\mathbf{1}$ for some $s\in[P_{\min},P_{\max}]$ and $\Sigma_{\exp}=\sigma_{\exp}^2 I_N$. Let $\barpMNP:=\min\{\pMNP,P_{\max}\}$. Then
\[
    \lim_{\sigma_{\exp}\to0}\;\lim_{\tau\to\infty}
    {P^{\mathrm{ODE}}(\tau;\mu, \Sigma_{\exp})}
    =
    \begin{cases}
    s\,\mathbf{1}, & s\in[\pNE,\barpMNP],\\[3pt]
    \barpMNP\,\mathbf{1}, & s<\pNE \ \ \text{or}\ \ s>\barpMNP .
    \end{cases}
\]
\end{theorem}

The proof is in Appendix~\ref{app:symmetric_explore_zero_noise_proof}.
\cref{thm:symmetric_explore_zero_noise} reveals a stark threshold structure, illustrated in the right plot of \cref{fig:br_cones_intro}. When $s\in[\pNE,\barpMNP]$, the limiting price locks in at the exploration mean $s$ itself. Outside this range, either below Nash or above the capped monopoly price, prices converge all the way to $\barpMNP$. Thus the price gap above Nash can be substantial, potentially reaching the monopoly benchmark when $\pMNP\le P_{\max}$.\footnote{
We show that under symmetric histories, the pairwise price correlation $\rho$ also admits a conduct-parameter interpretation, mapping $\rho=0$ to the Nash price and $\rho=1$ to the capped monopoly price; see Appendix~\ref{app:correlated_history_conduct} for details.
}

\subsection{Coverage of best-response cones}
\label{sec:cone_discussion}

The sets $\mathcal{C}^+$ and $\mathcal{C}^-$ consist of exploration profiles where, if each firm were to best-respond to its competitors' average exploration price, all firms would adjust their prices in the same direction  (a single round of fictitious play).
Concretely, $\mu\in\mathcal{C}^+$ means each firm $i$ explores at a price $\mu_i$ exceeding its best-response to the average exploration mean of its competitors; $\mu\in\mathcal{C}^-$ is the analogous region where every firm explores below that best-response. 
This is only a characterization of the cones, not what firms actually do: under the misspecified demand model, firms do not compute best responses. The actual pricing rule in \cref{def:price_moments_ode} can be decomposed into a best-response term plus a covariance-driven bias.

We interpret the cones in two parts. 
First, in the duopoly case, they admit a simple geometric description and also arise endogenously from the dynamics.
Second, for general $N$, we introduce a \textit{random interval prior} that models clustered exploration profiles and show that the probability of landing in a cone is at least $1/4$ and often much higher, for any $N$ and any demand parameters.

\subsubsection{Cone coverage for $N=2$}
\label{sec:cone_duopoly}

\begin{figure}[h]
\centering
\begin{tikzpicture}[x=3.4cm,y=3.4cm]
\pgfmathsetmacro{\xMax}{4/3}
\pgfmathsetmacro{\yMax}{4/3}
\pgfmathsetmacro{\pNEtikz}{2/3}
\pgfmathsetmacro{\xCut}{5/6}
\pgfmathsetmacro{\xZero}{1/2}
\pgfmathsetmacro{\yZero}{1/2}

\definecolor{green}{RGB}{0,120,0}
\definecolor{navy}{RGB}{0,0,180}

\fill[red!15] (\pNEtikz,\pNEtikz) -- (\xMax,\xCut) -- (\xMax,\yMax) -- (\xCut,\yMax) -- cycle;
\fill[red!5] (0,0) -- (\xZero,0) -- (\pNEtikz,\pNEtikz) -- (0,\yZero) -- cycle;

\draw[->] (0,0) -- (\xMax+0.08,0) node[right] {$\mu_1$};
\draw[->] (0,0) -- (0,\yMax+0.08) node[above] {$\mu_2$};

\draw[thick, navy] (0,\yZero) -- (\xMax,\xCut);
\node[below right, navy] at (0.88,0.78) {$\mu_2=\BR(\mu_1)$};

\draw[thick, green] (\xZero,0) -- (\xCut,\yMax);
\node[above left, green] at (0.83,1.12) {$\mu_1=\BR(\mu_2)$};

\fill (\pNEtikz,\pNEtikz) circle[radius=0.02];
\node[above left] at (\pNEtikz,\pNEtikz) {$\bigl(\pNE,\pNE\bigr)$};

\node at (1.05,1.05) {$\mathcal{C}^+$};
\node at (0.25,0.25) {$\mathcal{C}^-$};

\draw[gray!55] (\xMax,0) -- (\xMax,\yMax) -- (0,\yMax);
\end{tikzpicture}
\caption{Best-response cones in the duopoly case, shown in the $(\mu_1,\mu_2)$ plane for $a=b=1$ and $c=1/2$. The two best-response lines intersect at $(\pNE,\pNE)$; the upper-right and lower-left wedges are $\mathcal{C}^+$ and $\mathcal{C}^-$, respectively.}
\label{fig:exp_cones_c_half}
\end{figure}
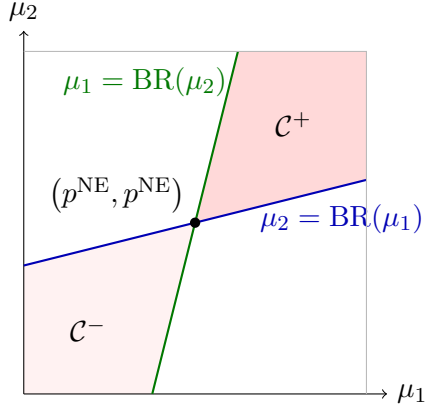

For $N=2$, the cones in \cref{def:best_response_cones} simplify because $\bar\mu_{-1}=\mu_2$ and $\bar\mu_{-2}=\mu_1$:
\begin{align*}
\mathcal{C}^+
&=
\bigl\{
(\mu_1,\mu_2)\in[P_{\min},P_{\max}]^2:
\mu_1>\BR(\mu_2)\ \text{and}\ \mu_2>\BR(\mu_1)
\bigr\},\\
\mathcal{C}^-
&=
\bigl\{
(\mu_1,\mu_2)\in[P_{\min},P_{\max}]^2:
\mu_1<\BR(\mu_2)\ \text{and}\ \mu_2<\BR(\mu_1)
\bigr\}.
\end{align*}
\cref{fig:exp_cones_c_half} illustrates these regions in the $(\mu_1,\mu_2)$ plane. The two best-response lines $\mu_2=\BR(\mu_1)$ and $\mu_1=\BR(\mu_2)$ intersect at $(\pNE,\pNE)$, and $\mathcal{C}^+$ and $\mathcal{C}^-$ correspond to the upper-right and lower-left wedges. Informally, the cones capture settings where both firms explore at similar prices and where the first best-response adjustment points in the same direction for both firms.

Beyond this geometric interpretation, the cones also arise endogenously from the dynamic:
for any exploration price vector \(\mu\), the duopoly ODE price converges, and the limit
must lie in \(\overline{\mathcal{C}^+}\cup\overline{\mathcal{C}^-}\), where
\(\overline{\mathcal{C}^\pm}\) denotes the closure of \(\mathcal{C}^\pm\) (allowing equality in the
defining inequalities \eqref{eq:upper_cone}-\eqref{eq:lower_cone}).

\begin{proposition}[Duopoly convergence to the closed best-response cones]
\label{prop:duopoly_limit_points_cones}
Take \(N=2\). Fix a diagonal covariance matrix \(\Sigma_{\exp}\) with positive diagonal entries
and an exploration-mean vector \(\mu\in [P_{\min},P_{\max}]^2\). Then $\displaystyle 
    P^\infty(\mu,\Sigma_{\exp})
    := \lim_{\tau\to\infty} P^{\mathrm{ODE}}(\tau;\mu,\Sigma_{\exp})$
exists, and $\displaystyle P^\infty(\mu,\Sigma_{\exp})
\in \overline{\mathcal{C}^+}\cup \overline{\mathcal{C}^-}.$
\end{proposition}

The proof is in Appendix~\ref{app:duopoly_limit_points_cone_proof}. To see why this matters,
suppose firms' exploration prices are themselves the output of an earlier round of
estimate-then-optimize pricing, rather than arbitrary points in the feasible set; this is natural if
real-world prices inherit structure from past pricing rather than being drawn uniformly. \cref{prop:duopoly_limit_points_cones} guarantees that such ``recycled'' exploration prices
lie in \(\overline{\mathcal{C}^{+}}\cup\overline{\mathcal{C}^{-}}\). \cref{thm:supra_sufficient} then applies after restarting the dynamic from these prices. The cones are thus not an
arbitrary corner of the parameter space but a typical destination of the dynamic.

\subsubsection{Cone coverage for general $N$}
\label{sec:cone_coverage}

We now ask, for general $N$, how often a randomly drawn exploration profile $\mu\in[P_{\min},P_{\max}]^N$ lands in a best-response cone. 
In practice, exploratory prices tend to cluster within a common local band, since firms typically tie their experiments to recently observed market prices or industry-typical price levels. We introduce the following \emph{random interval prior} to formalize this clustering.

\begin{definition}[Random interval prior]
\label{def:random_interval_prior}
Fix an outer band $[\underline P,\overline P]$
containing $\pNE$. Two firms draw exploration means uniformly from this band,
\[
    x,y\stackrel{\mathrm{iid}}{\sim}\operatorname{Unif}[\underline P,\overline P],
    \qquad
    \ell:=\min\{x,y\},
    \qquad
    u:=\max\{x,y\},
\]
and we set $\mu_1=x$ and $\mu_2=y$. The remaining $N-2$ firms draw exploration means uniformly from the random interval $[\ell,u]$:
\[
    \mu_i\mid(\ell,u)\stackrel{\mathrm{iid}}{\sim}\operatorname{Unif}[\ell,u],
    \qquad i=3,\dots,N.
\]
\end{definition}

Two firms act as anchors by drawing from the full outer band, forming a random subinterval $[\ell,u]$; the remaining $N-2$ firms then cluster within that subinterval.
The only input to this construction is the outer band $[\underline P,\overline P]$. Letting $r:=c/b$ denote the ratio of the cross-price effect to the own-price effect, the following result gives a universal lower bound on cone membership (proof in Appendix~\ref{app:random_interval_cone_probability_proof}).

\begin{proposition}[Cone membership under the random interval prior]
\label{prop:random_interval_cone_probability}
Under the random interval prior, for every $N\ge2$ and $r\in(0,1)$,
\[
    \Pr\bigl(\mu\in\mathcal{C}^+\cup\mathcal{C}^-\bigr)
    \ge
    \frac{(N-1)(2-r)}{4(N-1)-r(N-2)}
    \ge
    \frac{1}{4}.
\]
\end{proposition}

Under the random interval prior, the probability of landing in a cone is at least $1/4$ for \emph{any} number of firms $N$ and \emph{any} demand parameters; \cref{tab:random_interval_cone_probability} reports the bound for representative values, and the bound converges to $(2-r)/(4-r)$ as $N\to\infty$.

Perhaps surprisingly, the bound \emph{increases} in $N$ for each fixed $r$. This runs counter to a naive volume intuition: under a uniform prior on the hypercube $[P_{\min},P_{\max}]^N$, the best-response cones shrink exponentially as $N$ grows. Under the random interval prior, however, the inner interval $[\ell,u]$ is the same for all firms, so additional firms only reinforce the alignment that places $\mu$ inside a cone.

\begin{table}[!htbp]
\centering
\renewcommand{\arraystretch}{1.15}
\begin{tabular}{c|cccc}
\hline
$N$ & $r=0.25$ & $r=0.50$ & $r=0.75$ & $r\to1.00$ \\
\hline
$2$      & $0.4375$ & $0.3750$ & $0.3125$ & $0.2500$ \\
$3$      & $0.4518$ & $0.4006$ & $0.3461$ & $0.2877$ \\
$5$      & $0.4591$ & $0.4143$ & $0.3647$ & $0.3094$ \\
$10$     & $0.4633$ & $0.4222$ & $0.3756$ & $0.3225$ \\
$\infty$ & $0.4667$ & $0.4286$ & $0.3846$ & $0.3333$ \\
\hline
\end{tabular}
\caption{Lower bounds on the probability of landing in $\mathcal{C}^+\cup\mathcal{C}^-$ under the random interval prior, for representative values of $N$ and $r=c/b$. The column $r\to1.00$ is the limiting boundary case.}
\label{tab:random_interval_cone_probability}
\end{table}

\section{Computational Results via the ODE}\label{sec:computational_simulations}

The previous section shows that terminal prices are characterized by the deterministic price-moments ODE and gives sufficient conditions for supra-competitive limits. 
We now use the ODE to computationally evaluate prices beyond the analytical cases, where we focus on the duopoly case ($N=2$) for ease of illustration.
In this section, we compute the ODE terminal prices over a grid of exploration means, illustrating how the supra-competitive region changes with $\mu$, $\tau$ and $\sigma_{\exp}$ (\cref{sec:duopoly_ode_sims}). 
Then, we compare these ODE values to the original finite-sample stochastic system from \cref{sec:model}, showing how the deterministic heatmaps emerge as the exploration sample size grows (\cref{sec:stochastic_eval}).

\subsection{ODE-based duopoly simulations}
\label{sec:duopoly_ode_sims}

\begin{figure}[h]
\centering
    \includegraphics[width=1.00\textwidth]{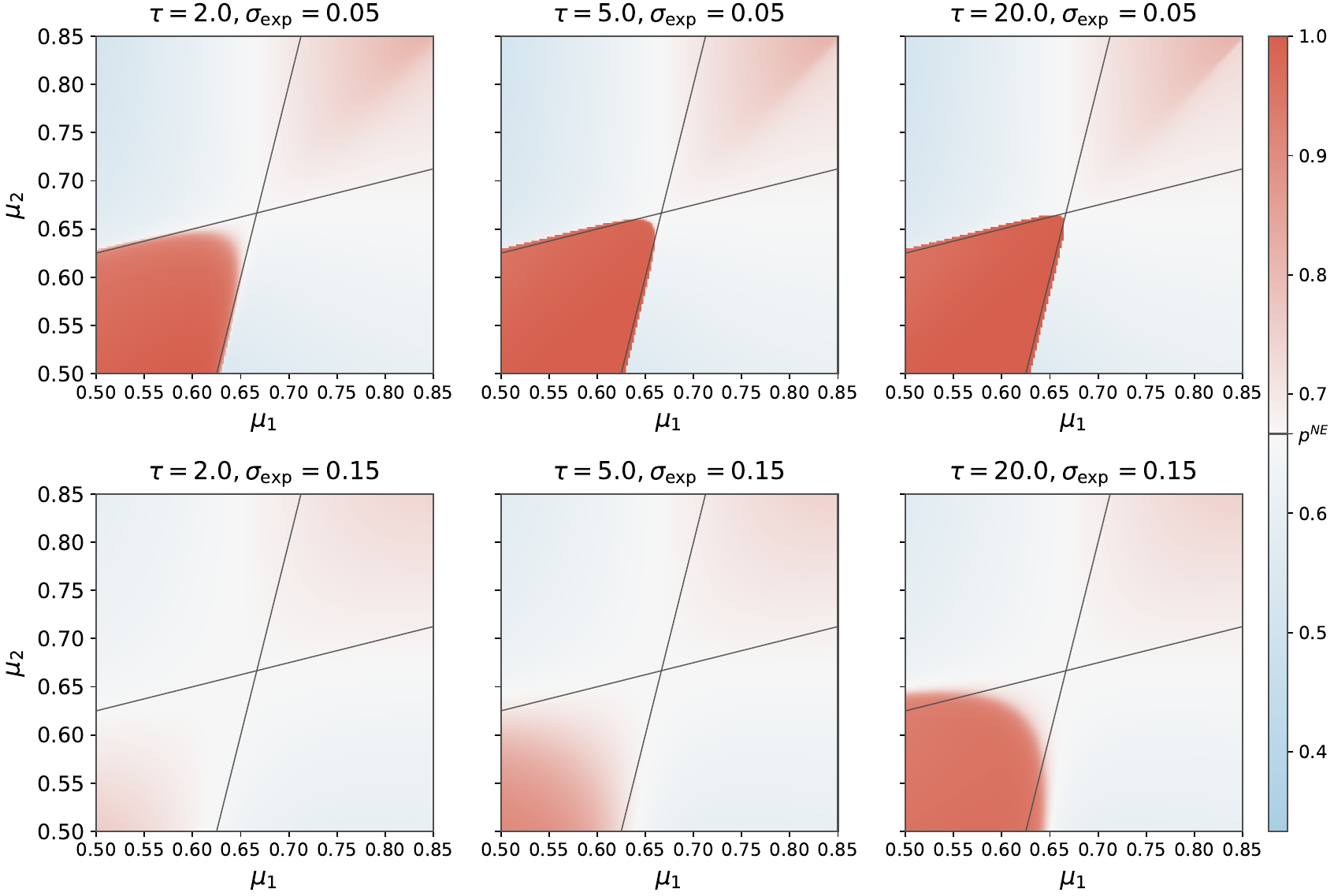}
    \caption{ODE-implied terminal price \(P^{\mathrm{ODE}}_1(\tau;\mu,\sigma_{\exp}^2 I_{2\times 2})\) in the duopoly case. Each panel fixes \((\tau,\Sigma_{\exp}=\sigma_{\exp}^2 I_{2\times2})\) and varies the exploration means $(\mu_1,\mu_2)$. White marks the Nash price $\pNE=2/3$, while red and blue indicate terminal prices above and below Nash. Thin lines mark the best-response boundaries defining the two cones.}
    \label{fig:ode_heatmap}
\end{figure}

We numerically evaluate the price-moments ODE for $N=2$.
We fix $a=b=1$ and $c=1/2$, so that $\pNE=\frac{2}{3}$ and $\pMNP=1$.
We set the covariance during exploration to be $\Sigma_{\exp}=\sigma_{\exp}^2 I_{2\times 2}$.
\cref{fig:ode_heatmap} plots the ODE-implied terminal price $P^{\mathrm{ODE}}_1(\tau;\mu,\Sigma_{\exp})$ over a grid of exploration means $(\mu_1,\mu_2)$, varying both the scaled horizon $\tau$ and the exploration noise $\sigma_{\exp}$.

\paragraph{Horizons sharpen cone-like regions.} As the horizon $\tau$ increases (left to right), the heatmaps show convergence that is qualitatively similar to the best-response cones defined in \cref{def:best_response_cones}, though the supra-competitive region is in fact larger than the cones.
At larger $\sigma_{\exp}$ (bottom plots), this convergence is slower, so the cone-like regions are less sharply separated at the plotted horizons.
The diagonal line ($\mu_1 = \mu_2$) corresponds to symmetric exploration, and the heatmap shows values consistent with the symmetric exploration result of \cref{thm:symmetric_explore_zero_noise}.
Specifically, the terminal price is close to the monopoly price when $\mu_1 < \pNE$, while terminal prices increase with $\mu_1$ when $\mu_1 \geq \pNE$.

\paragraph{Exploration noise pulls prices toward Nash.} 
When the exploration noise $\sigma_{\exp}$ is higher (bottom plots), prices are closer to Nash.
Because supra-competitive prices arise from correlation in prices across firms, more exploration noise weakens this effect: it gives each firm more uncorrelated own-price variation, diluting the cross-firm correlation that drives the bias.
Beyond the duopoly case, Appendix~\ref{app:additional_computations} evaluates the ODE for general $N$ and shows that supra-competitive prices arise robustly under clustered exploration profiles.

\subsection{Comparing the ODE to stochastic simulations}
\label{sec:stochastic_eval}

\begin{figure}[h]
\centering
    \includegraphics[width=1.00\textwidth]{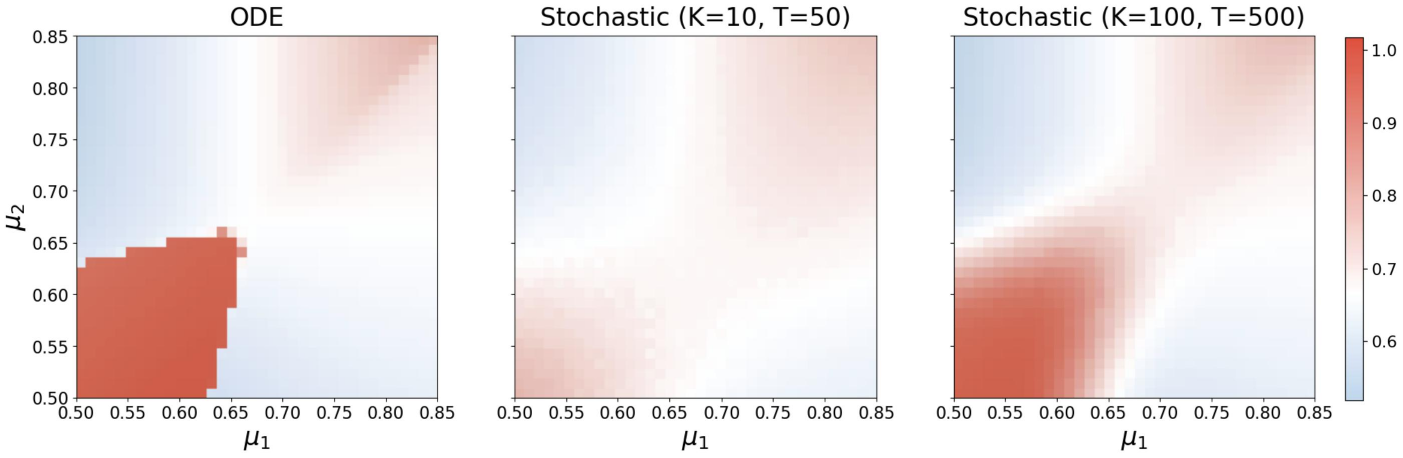}
    \caption{ODE and stochastic mean terminal-price heatmaps. The left panel is the deterministic ODE prediction, while the middle and right panels average terminal prices across $2500$ stochastic simulation runs. For small $K$, finite-sample exploration noise blurs the sharp ODE transition regions. As $K$ increases from $10$ to $100$, the stochastic heatmap becomes visibly closer to the deterministic ODE profile.}
    \label{fig:stochastic_heatmap}
\end{figure}

We compare the ODE predictions (assuming an asymptotic scaling) to the original discrete-time stochastic system from \cref{sec:model} for small, finite horizons.
We use the same parameters as \cref{sec:duopoly_ode_sims} for the demand ($a=b=1$ and $c=1/2$, so that $\pNE=\frac{2}{3}$ and $\pMNP=1$).
We evaluate the stochastic system for $(K, T) \in \{(10, 50), (100, 500)\}$, then run the ODE for the corresponding $\tau = (K+T)/K = 6$.
We fix $\sigma_{\exp}=0.05$ and set the standard deviation of the random demand shock to $0.05$. 
For each exploration-mean pair $(\mu_1,\mu_2)\in[0.5,0.85]^2$, we run $2500$ independent simulation pipelines.
\cref{fig:stochastic_heatmap} compares the mean terminal price of firm~1 to the ODE-implied value $P^{\mathrm{ODE}}_1(\tau;\mu,\Sigma_{\exp})$.

\paragraph{Finite samples blur transitions.} The stochastic heatmaps are blurred versions of the ODE heatmap, with the blurring diminishing as $K$ grows. This is consistent with \cref{thm:terminal_convergence}, which predicts convergence to the ODE profile as $K\to\infty$ with $\tau$ fixed. When $K$ is small, the exploration phase provides a noisy estimate of the moments that initialize exploitation, so nearby exploration profiles can lead to different dynamic regimes. As $K$ increases, these moments concentrate around their population values and the sharp transitions in the ODE map reappear.

\begin{figure}[h]
\centering
    \includegraphics[width=1.00\textwidth]{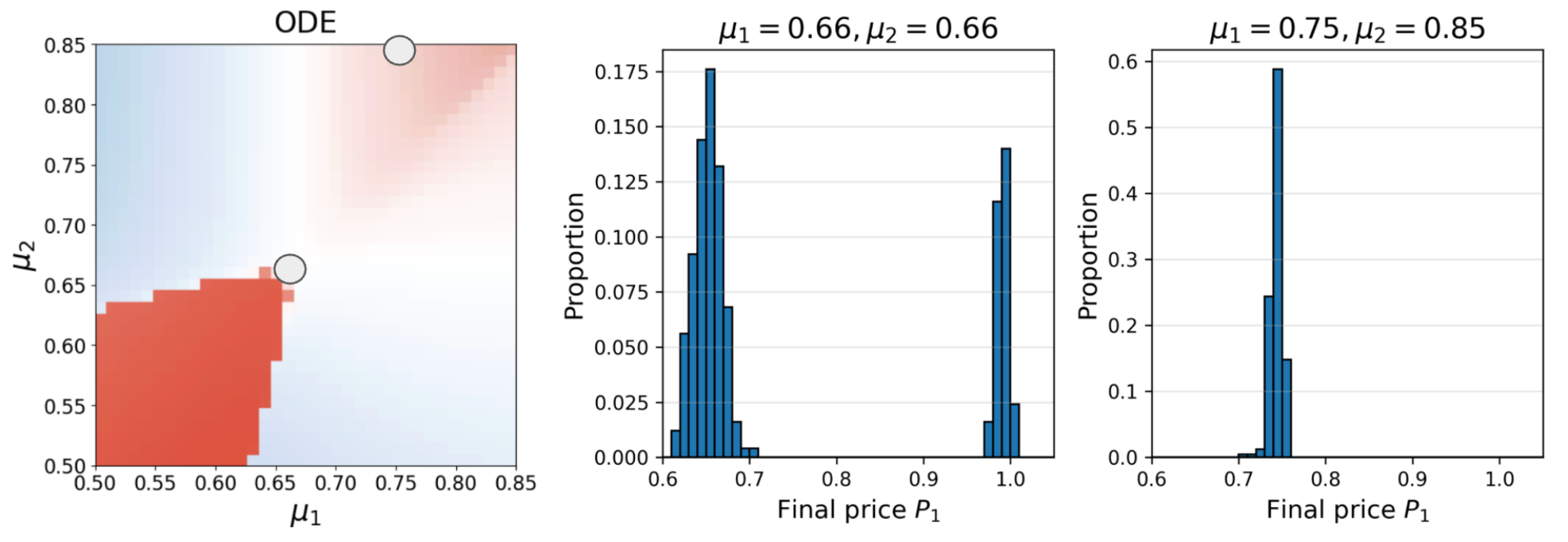}
    \caption{ODE map and terminal-price histograms from stochastic simulations at $\tau=100$, $\sigma_{\exp}=0.05$, and demand-shock standard deviation $0.05$ ($2500$ runs; $K=100$, $K+T=10000$). The two marked profiles have similar mean terminal prices, but different distributions: $(0.66,0.66)$ lies near a sharp ODE transition and produces a mixture of Nash-like and monopoly-like outcomes, while $(0.75,0.85)$ lies in a smoother region and produces a concentrated distribution.}
    \label{fig:stochastic_histograms}
\end{figure}

\paragraph{Boundary points create mixtures.} As the heatmap only presents the mean price over 2500 runs, we also examine the distribution of prices across runs.
In \cref{fig:stochastic_histograms}, we pick two exploration means $(\mu_1, \mu_2) = (0.66, 0.66)$ and $(\mu_1, \mu_2) = (0.75, 0.85)$, and we plot the distribution of terminal prices across runs.
The first point $(0.66, 0.66)$ corresponds to a sharp boundary between Nash-like and monopoly-like outcomes in the ODE heatmap.
Finite-sample noise therefore pushes different runs to different sides of the boundary, producing a bimodal distribution: roughly one quarter of runs remain near Nash, while the rest move close to $\pMNP$. By contrast, the second point $(0.75,0.85)$ lies in a smoother region of the heatmap, where the stochastic noise does not lead to regime switching.

\FloatBarrier
\section{Empirically calibrated simulations: Boston multifamily rentals}\label{sec:sim_design}

We complement our analysis with simulations calibrated to a large multifamily rental market, adapting the heterogeneous-logit demand system in \citet{calder2024algorithmic}. We test whether an estimate-then-optimize pipeline initialized by exploration can generate systematically elevated prices, and evaluate whether the patterns from our theory persist in this richer setting. Relative to our stylized model, this environment features differentiated firms, heterogeneous customers, and nonlinear logit demand.

\subsection{Market setup and calibration}

\paragraph{Data.}
We instantiate a market calibrated to the Boston Core multifamily rentals (PUMAs 08001--08003) using ACS microdata. We exclude rentals below the within-bedroom-count 5th percentile of rent and restrict to adult households with income $\ge\$30$k. After filtering, our sample comprises $N=801$ representative rentals and $H=920$ representative households, each weighted by its ACS survey weight.

We denote by $p_0$ the vector of observed rents and by $s_0$ the corresponding observed shares (constructed from normalized survey weights). We model each of the $N$ representative rentals as an independent firm offering one product, and thus refer to ``rentals'' and ``firms'' interchangeably.

\paragraph{Demand-side model.}
In each period $t=1,\ldots,K+T$, each household $h \in \{1, \dots, H\}$ chooses among rentals $j\in\{1,\ldots,N\}$ (plus an outside option). Given the price vector $P_t=(P_{1,t},\dots,P_{N,t}) \in \mathbb{R}^N$, the total share (demand) for rental $j$ at time $t$ is
\[
s_{j,t}(P_t; \xi)
=\sum_{h=1}^{H} \tilde w_h\,
\frac{\exp\!\left(\alpha_{h}\, P_{j,t} + x_{j}'\beta_{h} + \xi_j\right)}
{1+\sum_{k=1}^N \exp\!\left(\alpha_{h}\, P_{k,t} + x_{k}'\beta_{h} + \xi_k\right)},
\qquad
\tilde w_h:=\frac{w_h}{\sum_{\ell=1}^{H} w_\ell},
\]
where $x_j$ are observed rental characteristics, $\xi_j$ is a rental vertical-differentiation fixed-effect term, and $(\alpha_h, \beta_h)$ encode renter-specific price sensitivity and characteristic preferences. We adopt the parameterization of $(\alpha_h, \beta_h)$ from \citet{calder2024algorithmic} and calibrate $\{\xi_j\}_{j=1}^N$ to match observed shares $s_{0}$ at the observed prices $p_0$ ($s_{j}(p_0;\xi)=s_{j0}$); see Appendix~\ref{app:semiempirical_demand} for details.

\paragraph{Supply-side model.}
We assume that each rental $j$ has a product-specific shadow cost $\lambda_j > 0$ capturing the intertemporal opportunity cost of renting today, so that its period-$t$ objective is
\[
\pi_{j,t}(P_t,\xi)=\bigl(P_{j,t}-\lambda_j\bigr)s_{j,t}(P_t,\xi).
\]
We choose $\lambda=(\lambda_j)_{j=1}^N$ so that the observed rent vector $p_0$ is the Nash equilibrium of the calibrated static game. This pins down each shadow cost uniquely via the firm's first-order condition; the explicit formula is in Appendix~\ref{app:semiempirical_lambda}. The fitted values are reasonable: the 25th and 75th percentiles of $\lambda_j/p_{j0}$ are $0.817$ and $0.877$, respectively.

\paragraph{Nash and monopoly benchmarks.}
The Nash benchmark is the observed rent vector $p_0$ by construction, since $\lambda$ was calibrated to make $p_0$ the Nash equilibrium. The monopoly benchmark $p^{\mathrm{M}}$ is defined by joint profit maximization across all rentals; in our calibration, the monopoly markup is relatively uniform, with the 25th and 75th percentiles of $p_j^{\mathrm{M}}/p_{j0}$ at $1.236$ and $1.263$, respectively.

\subsection{Estimate-then-optimize pricing pipeline}

We apply an analogous estimate-then-optimize pricing pipeline at the rental level.
During exploration ($t=1,\ldots,K$), each rental $j$ posts a perturbed price $P_{j,t}=\mu_j(1+v_{j,t})$ with $v_{j,t}\overset{\text{i.i.d.}}{\sim}\mathcal{N}(0,\sigma^2)$ around an exploration mean $\mu_j$, with price clipped to \([0,\bar P]\), where $\bar P$ is a large exogenous upper bound. During exploitation ($t \ge K$), each rental fits a naive binary logit on its own price-share history (treating competitors' prices as unobserved) and sets its next-period rent by myopically maximizing predicted profit. Full equations are in Appendix~\ref{app:semiempirical_dynamic}.

\paragraph{Specifying exploration means $\mu$.}
To mirror the comparative statics in the stylized model, we vary two features of exploration means: their overall level relative to the Nash benchmark and the dispersion of prices across rentals. Specifically, for parameters $m, \sigma_\nu \ge 0$, we set
\[
\mu_j = m\,p_{j0}(1+\nu_j), \qquad \nu_j \overset{\text{i.i.d.}}{\sim} \text{Unif}(-\sqrt{3}\sigma_{\nu},\sqrt{3}\sigma_{\nu}).
\]
Thus $m$ shifts the average exploration price vs. Nash, while $\sigma_\nu$ controls cross-rental dispersion.

\subsection{Results and Discussion}
\label{subsec:semiempirical-results}

We report two experiments. First, we vary the exploration-price parameters $m$, $\sigma_{\nu}$, and $\sigma$, which respectively control the average exploration level relative to Nash, the cross-rental dispersion in exploration means, and the within-rental exploration noise (Subsection~\ref{subsec:semiempirical-price-comparative-statics}). Second, we vary the time parameters: the exploitation horizon $T$ and the exploration length $K$ (Subsection~\ref{subsec:semiempirical-time-dynamics}).

For rental $j$, define the terminal percentage change relative to Nash by
\[
    \Delta_{j,T}
    :=
    100\left(\frac{P_{j,K+T}}{p_{j0}} - 1\right).
\]
The figures plot the 10th, 50th, and 90th percentiles of $\Delta_{j,T}$ across rentals.

\subsubsection{Supra-competitive prices across exploration-price designs}
\label{subsec:semiempirical-price-comparative-statics}

\begin{figure}[!ht]
    \centering
    \includegraphics[width=0.9\textwidth]{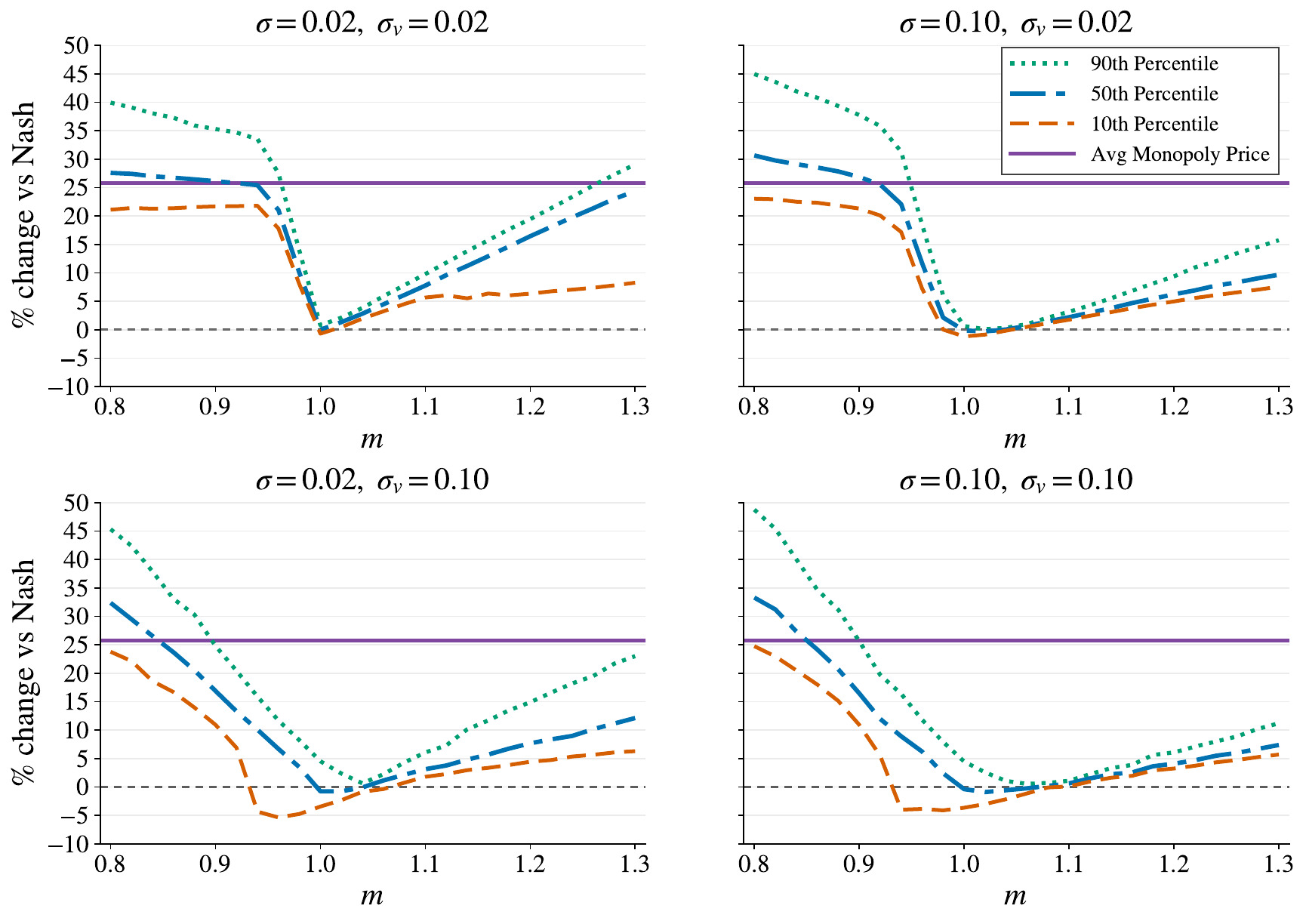}
    \caption{Terminal rent changes relative to Nash as the exploration mean multiplier $m$ varies. Each panel fixes the within-rental exploration noise $\sigma$ and cross-rental dispersion $\sigma_{\nu}$, with $K=50$ and $T=450$. The percentile curves summarize the cross-section of rentals, while the horizontal benchmark marks the average monopoly price. The main pattern is the same U-shape predicted by the stylized model: prices are elevated when exploration is clustered away from Nash.
    }
    \label{fig:semiempirical-m-sigma-sigmav}
\end{figure}

We fix $K=50$ and $T=450$ ($\tau=10$), placing the experiment close to the converged regime. We vary the exploration mean multiplier $m \in \{0.80,0.82,\ldots,1.30\}$ across four combinations of within-rental exploration noise $\sigma$ and cross-rental dispersion $\sigma_\nu$, each taking values in $\{0.02, 0.10\}$. \cref{fig:semiempirical-m-sigma-sigmav} plots the 10th, 50th, and 90th percentiles of $\Delta_{j,T}$ across rentals as a function of $m$, with the calibrated average monopoly price as a benchmark.

\paragraph{Main pattern is consistent with theory.}
The main finding in \cref{fig:semiempirical-m-sigma-sigmav} is that terminal rents exceed Nash across a wide range of $m$ values and all four panels: supra-competitive outcomes are the norm, not the exception. The median rent change is positive for most of the plotted range, and the 10th percentile lies above Nash except near $m=1$. The pattern of when prices are most elevated is consistent with the analytical results. Terminal rents are closest to Nash near $m=1$ and rise when exploration is clustered away from Nash on either side, forming the U-shape predicted by \cref{thm:symmetric_explore_zero_noise}. The low-$m$ side is especially steep, while the high-$m$ side rises more gradually.

\paragraph{Variance drives attenuation to Nash.}
Cross-rental dispersion $\sigma_\nu$ controls how clustered exploration is across rentals. 
When $\sigma_{\nu}$ is small (top panels), exploration means are close together, and the terminal price profile as a function of $m$ closely matches the characterization in \cref{thm:symmetric_explore_zero_noise}: prices are near Nash when $m \approx 1$ and rise toward the monopoly benchmark as $m$ moves away from Nash in either direction.
When $\sigma_\nu$ is larger (bottom panels), this correspondence weakens as exploration means spread across rentals, though supra-competitive outcomes remain over a broad range of $m$.
Within-rental noise $\sigma$ attenuates supra-competitiveness, especially when $m>1$. Larger $\sigma$ induces more price variation for each rental, 
reducing the role of endogenous correlation during exploitation. This attenuation is weaker for $m<1$, where prices move more during exploitation and therefore generate more correlated price variation after exploration.

\subsubsection{Effect of time horizon}
\label{subsec:semiempirical-time-dynamics}

\begin{figure}[!ht]
    \centering
    \includegraphics[width=0.75\textwidth]{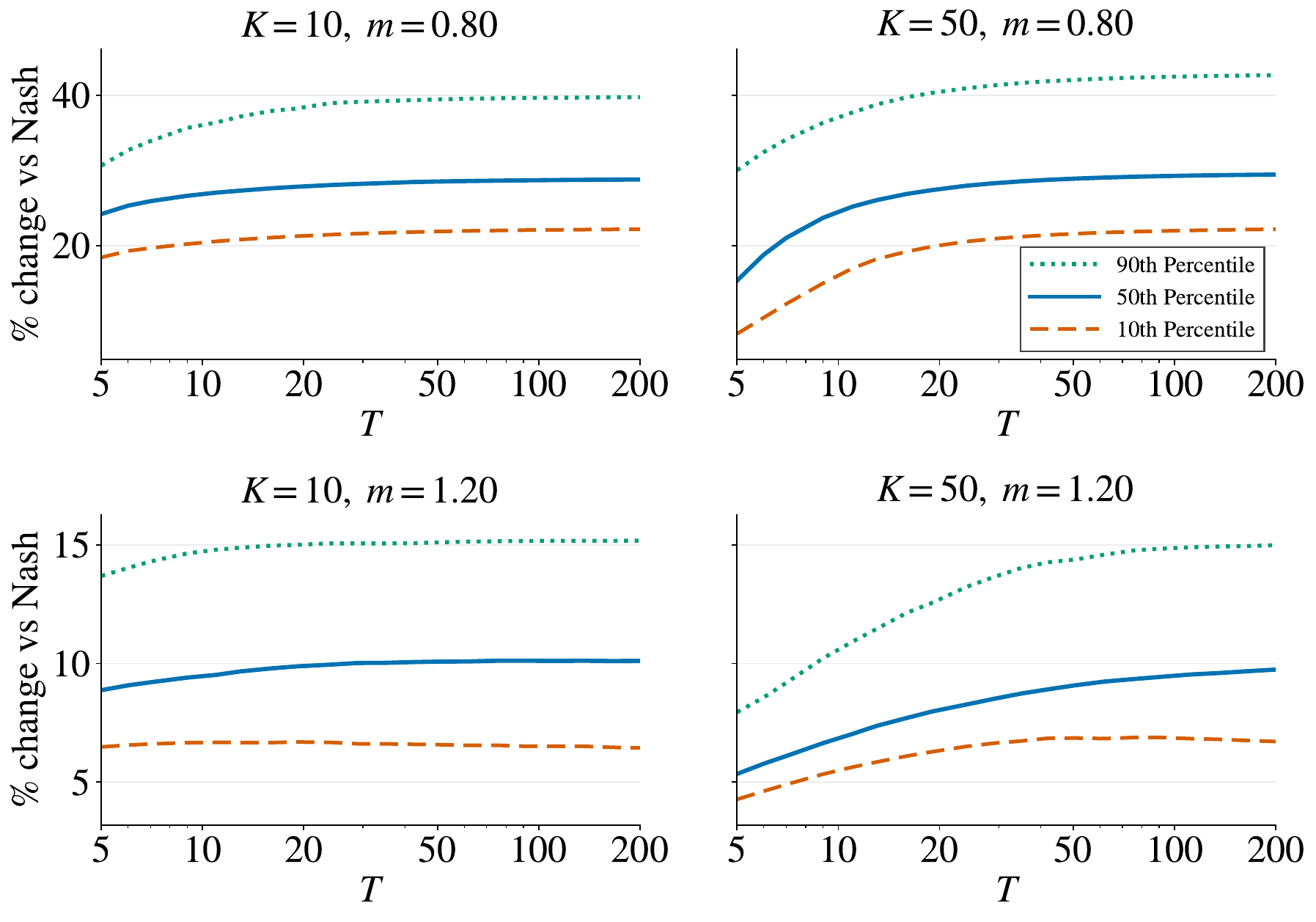}
    \caption{Finite-time dynamics of terminal rent changes relative to Nash. We fix $\sigma=\sigma_{\nu}=0.05$ and vary the exploitation horizon $T$ on a log scale, comparing $K\in\{10,50\}$ and $m\in\{0.80,1.20\}$. }
    \label{fig:semiempirical-time-dynamics}
\end{figure}

We now examine how the time horizon affects terminal prices. We fix $\sigma=\sigma_{\nu}=0.05$ and vary the exploitation horizon $T \in [5, 200]$, comparing two exploration levels ($m \in \{0.80,1.20\}$, placing exploration below and above Nash) and two exploration lengths ($K \in \{10,50\}$). \cref{fig:semiempirical-time-dynamics} plots the percentile curves of $\Delta_{j,T}$ as a function of $T$.

\FloatBarrier

\paragraph{Results show fast convergence.}
\cref{fig:semiempirical-time-dynamics} shows that supra-competitive rents emerge quickly and do not require long exploitation horizons. The percentile curves are already above Nash at $T=5$ in all four panels, and the curves flatten by moderate horizons such as $T=50$. The strongest case is $K=10$, $m=0.80$: even at the smallest plotted horizons, nearly all rentals are already substantially supra-competitive, with the 10th percentile well above Nash. Across all panels, convergence to a supra-competitive steady state is rapid: most of the price increase relative to Nash occurs within the first $T=50$ exploitation periods, and further exploitation yields diminishing additional movement.\footnote{The comparison across $K$ should be interpreted through the normalized clock $\tau=(K+T)/K$. Holding $T$ fixed, the $K=10$ process has a larger $\tau$ than the $K=50$ process, so it is farther along the ODE trajectory. This is why the $K=10$ curves tend to be closer to their apparent long-run levels at the same value of $T$, while the $K=50$, $m=0.80$ panel starts lower but catches up quickly.}

Taken together, these simulations confirm that supra-competitive outcomes arise robustly in a realistic rental market, persisting across a wide range of exploration parameters, finite time horizons, heterogeneous products, and nonlinear demand.
\section{Proof of Theorem~\ref{thm:supra_sufficient}: Convergence to Supra-competitive Prices}
\label{sec:supracomp_proof}

We outline the proof of Theorem~\ref{thm:supra_sufficient}, that limiting prices are strictly supra-competitive, by analyzing the price-moments ODE from Definition~\ref{def:price_moments_ode}.
Throughout, let $(U(\tau),V(\tau))$ denote the ODE state and $P(\tau)=P(U(\tau),V(\tau))$ the corresponding posted-price vector\footnote{To keep the main-text proof focused, we state the intuition in the interior case, where the posted price is not clipped at $P_{\min}$ or $P_{\max}$; the appendices prove the corresponding statements for the full capped price map.}, so that the terminal price in Theorem~\ref{thm:supra_sufficient} is $P^{\mathrm{ODE}}(\tau;\mu,\Sigma_{\exp})=P(\tau;\mu,\Sigma_{\exp})$.

The argument has three steps.
First, positive historical price correlation creates an omitted-variable bias that pushes the misspecified OLS price above the true best response.
Second, the best-response cones initialize common-sign movements relative to running means, and this common-sign pattern is forward invariant, so the induced positive cross-covariances persist through the dynamics.
Third, the persistent correlation bias yields prices greater than Nash.
The core of the proof is the forward-invariance argument in Step~2.

\subsection{Step 1: Correlation bias raises OLS prices above best responses}
\label{sec:supracomp_corr_bias}

The mechanism here is derived from omitted-variable bias.
Firm $i$ fits a one-dimensional demand curve in its own price, omitting the competitor term $c\bar P_{-i}$ from true demand \eqref{eq:N_demand}.
Positive cross-firm price correlation therefore biases the fitted slope toward less price-sensitivity, so firm $i$ prices above the true best response.
In the ODE notation, this correlation is captured by $\displaystyle 
    \bar V_{i,-i}:=\frac{1}{N-1}\sum_{j\neq i}V_{ij}.$ The following lemma states the bias implication precisely.

\begin{lemma}[Correlation bias]
\label{lem:corr_bias_jackie}
Fix $i\in[N]$ and hold $U$ and $V_{ii}$ fixed, with $bV_{ii}-c\,\bar V_{i,-i}>0$.
Then:
\begin{enumerate}[label=(\arabic*),leftmargin=*]
    \item if $\bar V_{i,-i}=0$, then $\widetilde P_i(U,V)=\BR(\bar U_{-i})$;
    \item $\widetilde P_i(U,V)$ is strictly increasing in $\bar V_{i,-i}$.
\end{enumerate}
\end{lemma}

Thus zero cross-covariance gives exactly the best response to the historical average competitor price, while positive cross-covariance pushes the price strictly above it (proof in Appendix~\ref{app:corr_bias_proof}). Conversely, without the correlation bias, the running-mean dynamics would reduce to iterated best responses to historical average competitor prices. Based on the classical fictitious-play convergence result of \citet{robinson1951iterative}, this leads to convergence to the symmetric Nash fixed point $\pNE\mathbf{1}$. 

\subsection{Step 2: Best-response cones generate forward-invariant co-movements}
\label{sec:supracomp_cone_initialization}
\label{sec:supracomp_step1}
\label{sec:supracomp_forward_invariance}

We start by recalling the definition of best-response cones: for a set of exploration prices in a best-response cone, all firms move in the same direction relative to their running means at the start of exploitation.

\paragraph{Initialization of positive covariance.} Independent exploration across firms leads to $\bar V_{i,-i}(1)=0$, so by Lemma~\ref{lem:corr_bias_jackie}, $P_i(1)=\BR(\bar\mu_{-i})$. Hence, on $\mathcal{C}^+$, the cone condition gives
$U_i(1)=\mu_i>\BR(\bar\mu_{-i})=P_i(1)$, so $P_i(1)-U_i(1)<0$ for every $i$. Whereas for $\mathcal{C}^-$, the inequalities reverse, so $P_i(1)-U_i(1)>0$ for every $i$. In either case, all firms initially move in the same direction relative to their exploration means.
Since $\dot V_{ij}=(P_i-U_i)(P_j-U_j)$, these common-sign gaps give $\dot V_{ij}(1)>0$ for every $i\neq j$, and hence $\bar V_{i,-i}(\tau)>0$ for $\tau$ slightly above 1.

\paragraph{Forward invariance.}
The following lemma presents the key step, which is that the co-movement relative to running means is self-sustaining.
We interpret this as follows: starting from $\mathcal{C}^+$, once all firms price below their running means, no firm can be the first to cross above its running mean; likewise, for $\mathcal{C}^-$, the analogous statement holds.

\begin{lemma}[Forward invariance of common-sign gap]
\label{lem:drift_orthant_invariant_jackie}
Fix $\tau_0\geq1$.
If $P-U\preceq0$ at time $\tau_0$, then $P-U\preceq0$ for all $\tau\geq \tau_0$.
Likewise, if $P-U\succeq0$ at time $\tau_0$, then $P-U\succeq0$ for all $\tau\geq \tau_0$.
\end{lemma}

We give the intuition for the $\mathcal{C}^+$ case; the $\mathcal{C}^-$ case is the same with signs reversed.
Suppose firm $i$ reaches the boundary $P_i=U_i$ at time $\tau=\tau^\star$, while every other firm still has $P_j-U_j\leq0$.
We want to show that firm $i$'s price does not move upward at this moment, so the gap $P_i-U_i$ cannot rise above zero.
At this boundary,
\begin{align} 
    \dot U_i(\tau^\star)=0,\qquad
    \dot V_{ii}(\tau^\star)=0,\qquad
    \dot{\bar V}_{i,-i}(\tau^\star)=0,
    \label{eq:derivatives_zero}
\end{align}
because each derivative carries the factor $P_i-U_i$.
Thus firm $i$'s own running mean, own variance, and average cross-covariance are momentarily frozen.

Recall that firm $i$ fits the linear demand curve $Q_i\approx\widehat\alpha_i+\widehat\beta_iP_i$.
In terms of the ODE moments, the fitted coefficients are
\[
    \widehat\alpha_i
    =
    a+c\bar U_{-i}
    -
    cU_i\frac{\bar V_{i,-i}}{V_{ii}},
    \qquad
    \widehat\beta_i
    =
    -b+c\frac{\bar V_{i,-i}}{V_{ii}}.
\]
Using \eqref{eq:derivatives_zero}, we have
\[
    \dot{\widehat\alpha}_i(\tau^\star)
    =
    c\,\dot{\bar U}_{-i}(\tau^\star),
    \qquad
    \dot{\widehat\beta}_i(\tau^\star)=0.
\]
Since $\widehat\beta_i(\tau^\star)<0$, the price $\widetilde P_i=-\widehat\alpha_i/(2\widehat\beta_i)$ moves in the same direction as the fitted intercept $\widehat\alpha_i$.
The other firms have $P_j-U_j\leq0$, so their running means satisfy $\dot U_j\leq0$ and therefore $\dot{\bar U}_{-i}(\tau^\star)\leq0$.
Thus $\dot{\widetilde P}_i(\tau^\star)\leq0$.
Combined with $\dot U_i(\tau^\star)=0$, this implies that $P_i-U_i$ cannot cross from nonpositive to positive at the boundary.
In the $\mathcal{C}^-$ case, the same argument gives $\dot{\bar U}_{-i}(\tau^\star)\geq0$ and prevents $P_i-U_i$ from crossing from nonnegative to negative.
The formal proof is in Appendices~\ref{app:drift_orthant_invariant_proof} and~\ref{app:negative_orthant_invariant_proof}.

\paragraph{Consequence for correlation.}
Because the common-sign gap persists for all $\tau$, the covariance equation $\dot V_{ij}=(P_i-U_i)(P_j-U_j)$ keeps the average cross-covariances $\bar V_{i,-i}(\tau)$ nonnegative throughout; after the strict initial common-sign movement, they are strictly positive for all $\tau>1$.

\subsection{Step 3: Correlation bias gives a Nash lower bound}
\label{sec:supracomp_step2}
\label{sec:supracomp_nash_limit_bound}

Combining Steps 1 and 2, we conclude that as the average cross-covariances remain nonnegative, each firm's price is at least its best response to the historical average competitor price. The following lemma then formalizes the dynamics to connect this observation to the limiting price vector.

\begin{lemma}[Nash lower bound from nonnegative correlation bias]
\label{lem:weak_supra_from_bias_jackie}
Along any solution of the price-moments ODE, assume (i) $\bar V_{i,-i}(\tau)\geq0$ for all $i$ and $\tau\geq1$, and (ii) $U$ is componentwise monotone in $\tau$.
Then the limit $U^\infty:=\lim_{\tau\to\infty}U(\tau)$ exists and satisfies
$U^\infty\succeq\pNE\mathbf{1}$.
\end{lemma}

The argument is by contradiction.
If the lowest limiting component were below Nash, then that firm's true best response to the limiting historical average of its competitors would exceed its own limiting mean.
The nonnegative correlation bias only strengthens this push, so for all large $\tau$ the firm would have a persistent positive gap $P_i(\tau)-U_i(\tau)$, contradicting convergence of $U_i(\tau)$.

For $\mu\in\mathcal{C}^+$ this completes Theorem~\ref{thm:supra_sufficient}(a): since $\mu\succ\pNE\mathbf{1}$ and the sign-reversed argument gives $P-U\preceq0$ with positive cross-covariances for all $\tau>1$, Lemmas~\ref{lem:corr_bias_jackie} and~\ref{lem:weak_supra_from_bias_jackie} yield $P_i(\tau)>\pNE$ for every $i$ and every $\tau\geq1$ (Appendix~\ref{app:upper_cone_strictness_proof}), hence
\[
    P^{\mathrm{ODE}}(\tau;\mu,\Sigma_{\exp})\succ\pNE\mathbf{1}
    \qquad\text{for every }\tau\in[1,\infty).
\]
For $\mu\in\mathcal{C}^-$, Lemma~\ref{lem:weak_supra_from_bias_jackie} gives only the weak bound $U^\infty\succeq\pNE\mathbf{1}$, and it remains to exclude convergence exactly to Nash.
We prove this by contradiction, by showing that convergence to Nash would leave the cross-covariances at strictly positive limits. The formal argument is in Appendix~\ref{app:omitted_pf_supra_sufficient}.

\qed

\section{Conclusion}

This paper demonstrates that simple estimate-then-optimize pricing algorithms, relying on misspecified demand models, can lead to supra-competitive prices.
The mechanism does not rely on punishment, communication, or explicit coordination. Instead, independent exploratory histories can generate price movements that become positively correlated during exploitation; this correlation creates omitted-variable bias, making firms perceive demand as less price-sensitive and leading them to choose prices above the competitive benchmark.

We close with a brief discussion of our modeling assumptions. 
Our results rest on one core assumption: each firm fits a \emph{misspecified} demand model that ignores competitors' prices. 
Several other assumptions are for tractability rather than substance: linear demand and symmetric firms simplify the analysis, and the empirically calibrated simulations of Section~\ref{sec:sim_design} show that supra-competitive outcomes persist under nonlinear (logit) demand, heterogeneous products, and asymmetric costs. The common exploration length $K$ across firms is a modeling normalization for summarizing initial histories, not an assumption that firms coordinate on phase timing. The fluid scaling regime ($K, T \to \infty$ with $(K+T)/K \to \tau$) is what yields the deterministic ODE limit, but the underlying stochastic dynamics show the same qualitative behavior at moderate finite horizons (Section~\ref{sec:stochastic_eval}). Finally, we deliberately impose \textit{independent} exploration across firms, so the supra-competitive outcomes we establish emerge endogenously and not from any coordinated experimentation, in contrast to \citet{lin2025competition}.

Future work could extend the paper in several directions. One direction is a theoretical extension to nonlinear demand or asymmetric firms (which our simulations already suggest is robust). Another is to consider dynamics with persistent demand shocks or ongoing experimentation. Lastly, characterizing when analogous biases arise in other supermodular games would show how broadly the supra-competitive channel extends beyond multi-firm pricing dynamics.

\bibliographystyle{abbrvnat}
\bibliography{references} 

\newpage

\appendix  %

\section*{Appendix Table of Contents}

\startcontents[appendix]
\printcontents[appendix]{}{1}{\setcounter{tocdepth}{1}}
\vspace{1cm}

\section{Derivation of the Bayesian Interpretation (Section~\ref{sec:explore_then_exploit_dynamics})}
\label{app:bayesian_interpretation}

This appendix formalizes the Bayesian interpretation of initialization. It first derives estimate-then-optimize from Bayesian linear regression, then shows that the exploration-implied prior generates the price-moments ODE.

\subsection{Estimate-then-optimize as Bayesian linear regression}
\label{app:bayesian_update}

The estimate-then-optimize step can be read as a certainty-equivalent Bayesian rule.
Fix firm \(i\), and write the misspecified own-price demand model as
\[
    Q_i \approx \alpha_i+\beta_iP_i,\qquad
    \theta_i:=(\alpha_i,\beta_i)^\top .
\]
For an observation \(s\), define the design vector \(x_{i,s}:=(1,P_{i,s})^\top\).
We write the Bayesian regression in normalized information units, suppressing a common observation-noise scale since only posterior mean coefficients enter the pricing rule.

\begin{proposition}[Bayesian form of estimate-then-optimize]
\label{prop:bayesian_update_app}
Suppose firm \(i\) has Gaussian prior \(\theta_i\sim N(m_i^0,(\Lambda_i^0)^{-1})\).
After observing \(\{(P_{i,s},Q_{i,s})\}_{s\le t}\), its posterior mean is
\[
    \widehat\theta_{i,t}
    =
    \begin{pmatrix}
        \widehat\alpha_{i,t}\\
        \widehat\beta_{i,t}
    \end{pmatrix}
    =
    \left(\Lambda_i^0+\sum_{s\le t}x_{i,s}x_{i,s}^{\top}\right)^{-1}
    \left(\Lambda_i^0m_i^0+\sum_{s\le t}x_{i,s}Q_{i,s}\right).
\]
If firm \(i\) posts the posterior-mean optimal price,
\[
    P_{i,t+1}\in\argmax_{p\in[P_{\min},P_{\max}]}
    p(\widehat\alpha_{i,t}+\widehat\beta_{i,t}p),
\]
then the policy is the same estimate-then-optimize rule as in the main model, with the fitted demand curve interpreted as a posterior mean demand curve.
\end{proposition}

\begin{myproof}
The likelihood under the normalized Gaussian linear model is proportional to
\[
    \exp\left\{
        -\frac12\sum_{s\le t}(Q_{i,s}-x_{i,s}^{\top}\theta_i)^2
    \right\}.
\]
The Gaussian prior density is proportional to
\[
    \exp\left\{
        -\frac12(\theta_i-m_i^0)^\top\Lambda_i^0(\theta_i-m_i^0)
    \right\}.
\]
Multiplying likelihood and prior gives a posterior density proportional to
\[
    \exp\left\{
        -\frac12
        \left[
        \theta_i^\top\left(\Lambda_i^0+\sum_{s\le t}x_{i,s}x_{i,s}^{\top}\right)\theta_i
        -2\theta_i^\top\left(\Lambda_i^0m_i^0+\sum_{s\le t}x_{i,s}Q_{i,s}\right)
        \right]
    \right\}.
\]
Thus the posterior precision is
\[
    \Lambda_{i,t}=\Lambda_i^0+\sum_{s\le t}x_{i,s}x_{i,s}^{\top},
\]
and the posterior information vector is
\[
    r_{i,t}=\Lambda_i^0m_i^0+\sum_{s\le t}x_{i,s}Q_{i,s}.
\]
The posterior mean is therefore \(\widehat\theta_{i,t}=\Lambda_{i,t}^{-1}r_{i,t}\), as claimed.

Given this posterior mean, the posterior-mean predicted demand curve is
\[
    \widehat q_{i,t}(p)=\widehat\alpha_{i,t}+\widehat\beta_{i,t}p.
\]
A certainty-equivalent Bayesian firm maximizes posterior-mean predicted profit \(p\widehat q_{i,t}(p)\) over the feasible interval.
This is exactly the estimate-then-optimize step in the main model.
With the improper flat prior \(\Lambda_i^0=0\), the posterior mean reduces to the ordinary least-squares estimate based only on the observed data.
\end{myproof}

\subsection{The exploration-implied prior and the price-moments ODE}
\label{app:bayesian_prior_ode}

We now choose the prior so that its sufficient statistics match the population moments generated by the exploration phase.
This shows that the same price-moments ODE can be obtained without explicitly referring to a finite exploration sample.

For each firm \(i\), let \(\sigma_{i,\exp}^2:=(\Sigma_{\exp})_{ii}\), and set
\[
    m_i^0
    :=
    \begin{pmatrix}
        a+c\bar\mu_{-i}\\
        -b
    \end{pmatrix},
    \qquad
    \Lambda_i^0
    :=
    \begin{pmatrix}
        1 & \mu_i\\
        \mu_i & \mu_i^2+\sigma_{i,\exp}^2
    \end{pmatrix}.
\]
Thus the prior mean curve is \(q_i^0(p)=a-bp+c\bar\mu_{-i}\): firm \(i\)'s own-price model treats competitors' prices as fixed at their prior reference levels.
The precision matrix makes \(\mu_i\) the price at which the prior demand curve is most tightly pinned down.
Indeed,
\[
    (1,p)(\Lambda_i^0)^{-1}(1,p)^\top
    =
    1+\frac{(p-\mu_i)^2}{\sigma_{i,\exp}^2}.
\]
So \(\mu_i\) is a prior price center, not a prior optimal price.

\begin{proposition}[Bayesian representation of the price-moments ODE]
\label{prop:bayesian_ode_app}
Suppose each firm \(i\) has prior \(\theta_i\sim N(m_i^0,(\Lambda_i^0)^{-1})\), with \(m_i^0\) and \(\Lambda_i^0\) defined above.
Normalize the prior mass to one, and let firms update their Bayesian sufficient statistics continuously while posting the posterior-mean optimal price.
Then the induced continuous-time dynamics reduce to the price-moments ODE:
\[
    U(1)=\mu,\qquad V(1)=\Sigma_{\exp},\qquad
    \dot U(t)=\frac{P(t)-U(t)}{t},\qquad
    \dot V(t)=(P(t)-U(t))(P(t)-U(t))^\top,
\]
with the same posted-price map as in Definition~\ref{def:price_moments_ode}.
\end{proposition}

\begin{myproof}
Let
\[
    M(t)=\bigl(M^P(t),M^Q(t),M^{PP}(t),M^{PQ}(t)\bigr)
\]
denote the normalized posterior moment state.
Here \(M^P\) and \(M^Q\) are price and quantity means, \(M^{PP}\) is the raw price second-moment matrix, and \(M^{PQ}\) is the vector of own price-quantity moments.

The prior induces the initial moments
\[
    M_i^P(1)=\mu_i,\qquad
    M_{ij}^{PP}(1)=\mu_i\mu_j+(\Sigma_{\exp})_{ij}.
\]
Since \(\Sigma_{\exp}\) is diagonal in the main model, the prior also gives
\[
    M_i^Q(1)=a-b\mu_i+c\bar\mu_{-i},
    \qquad
    M_i^{PQ}(1)=\mu_iM_i^Q(1)-b(\Sigma_{\exp})_{ii}.
\]
These are exactly the population moments of the exploration distribution, now read as prior sufficient statistics.

Given a posted price vector \(P(t)\), write true mean demand as
\[
    d_i(P(t)):=a-bP_i(t)+c\bar P_{-i}(t).
\]
The continuous-time Bayesian sufficient-statistic updates are the running-average equations
\[
    \dot M_i^P=\frac{P_i-M_i^P}{t},
    \qquad
    \dot M_i^Q=\frac{d_i(P)-M_i^Q}{t},
\]
\[
    \dot M_{ij}^{PP}=\frac{P_iP_j-M_{ij}^{PP}}{t},
    \qquad
    \dot M_i^{PQ}=\frac{P_id_i(P)-M_i^{PQ}}{t}.
\]
Define
\[
    U(t):=M^P(t),
    \qquad
    V(t):=t\left(M^{PP}(t)-U(t)U(t)^\top\right).
\]
Then \(U(1)=\mu\) and \(V(1)=\Sigma_{\exp}\).
The first running-average equation gives
\[
    \dot U=\frac{P-U}{t}.
\]
For \(V\), differentiate \(V=t(M^{PP}-UU^\top)\). Using
\[
    \dot M^{PP}=\frac{PP^\top-M^{PP}}{t},
    \qquad
    \dot U=\frac{P-U}{t},
\]
we obtain
\[
    \dot V=(P-U)(P-U)^\top.
\]
Thus the state dynamics match the price-moments ODE.

It remains to recover the posted-price map from \((U,V)\).
First,
\[
    M_i^Q(t)=a-bU_i(t)+c\bar U_{-i}(t).
\]
This holds because both sides agree at \(t=1\), and differentiating the right-hand side gives the same running-average equation as \(\dot M_i^Q=(d_i(P)-M_i^Q)/t\).

Next,
\[
    M_i^{PQ}(t)
    =
    U_i(t)M_i^Q(t)
    +
    \frac{-bV_{ii}(t)+c\bar V_{i,-i}(t)}{t}.
\]
To see this, let
\[
    C_i(t):=M_i^{PQ}(t)-U_i(t)M_i^Q(t).
\]
Then
\[
    \dot C_i
    =
    \frac{(P_i-U_i)(d_i(P)-M_i^Q)-C_i}{t}.
\]
Using \(d_i(P)-M_i^Q=-b(P_i-U_i)+c(\bar P_{-i}-\bar U_{-i})\), this becomes
\[
    \dot C_i
    =
    \frac{-b(P_i-U_i)^2+c(P_i-U_i)(\bar P_{-i}-\bar U_{-i})-C_i}{t}.
\]
The quantity \((-bV_{ii}+c\bar V_{i,-i})/t\) satisfies the same differential equation and has the same initial value, so the identity follows.

The posterior mean slope is therefore
\[
    \widehat\beta_i(t)
    =
    \frac{M_i^{PQ}(t)-M_i^P(t)M_i^Q(t)}
    {M_{ii}^{PP}(t)-(M_i^P(t))^2}
    =
    -b+c\frac{\bar V_{i,-i}(t)}{V_{ii}(t)}.
\]
The posterior mean intercept is
\[
    \widehat\alpha_i(t)
    =
    M_i^Q(t)-\widehat\beta_i(t)U_i(t)
    =
    a+c\bar U_{-i}(t)
    -
    cU_i(t)\frac{\bar V_{i,-i}(t)}{V_{ii}(t)}.
\]
On the interior branch, the posterior-mean optimal price is
\[
    \widetilde P_i(t)
    =
    -\frac{\widehat\alpha_i(t)}{2\widehat\beta_i(t)}
    =
    \frac{(a+c\bar U_{-i}(t))V_{ii}(t)-cU_i(t)\bar V_{i,-i}(t)}
    {2(bV_{ii}(t)-c\bar V_{i,-i}(t))}.
\]
Adding the same clipping and nonnegative-slope branch as in Definition~\ref{def:price_moments_ode} gives exactly the posted-price map in the price-moments ODE.
This proves the proposition.
\end{myproof}

\section{Discussion of multiplicity of steady states in \citet{cooper2015learning}}
\label{app:cooper_analog}

\citet{cooper2015learning} show that estimate-then-optimize dynamics admit a range of steady-state prices, spanning sub-competitive to supra-competitive, depending on initial conditions. The following proposition establishes the analog within our framework: for a range of target prices spanning both below and above Nash, there exist initial conditions $(\mu, \Sigma_{\exp})$ under which the price-moments ODE is stationary at $p^*$ (with $P(t) = U(t) = p^*$ for all $t \ge 1$), provided $\Sigma_{\exp}$ is allowed to be non-diagonal.

\begin{proposition}[Cooper-style multiplicity]
\label{prop:cooper_analog}
For any target $p^*\in [p^*_{\min}, \min(\pMNP,P_{\max})]$, where
\[
p^*_{\min} := \frac{a(N-1)}{(N-1)(2b-c) + c},
\]
the following is a steady-state initialization of the price-moments ODE: $\mu_i = p^*$ for all $i$, and $V_{ii}(1) = \sigma^2$, $V_{ij}(1) = \rho^*\sigma^2$ for $i\ne j$, where $\rho^* := (2b-c)(p^* - \pNE)/(c\,p^*)$ and $\sigma^2 > 0$ is arbitrary. The resulting dynamics satisfy $P_i(t) = U_i(t) = p^*$ for all $i$ and all $t\ge 1$.
\end{proposition}

\begin{myproof}[Proof of \cref{prop:cooper_analog}.]
Substituting into \eqref{eq:posted_price_formula} gives $\widetilde P_i = (a+cp^*(1-\rho^*))/(2(b-c\rho^*)) = p^*$ by the definition of $\rho^*$. The range $p^*\in[p^*_{\min},\min(\pMNP,P_{\max})]$ corresponds to $\rho^*\in[-1/(N-1),1]$, which makes $\Sigma_{\exp}$ positive semi-definite. Hence $P_i(1)=p^*=U_i(1)$, so $\dot U(1)=\dot V(1)=0$ and the state remains at $(p^*\mathbf{1},\Sigma_{\exp})$ for all $t\ge 1$.
\end{myproof}

Note that $p^*_{\min} < \pNE$: the ratio $p^*_{\min}/\pNE = (N-1)(2b-c)/[(N-1)(2b-c) + c]$ is strictly less than 1, approaching 1 as $N\to\infty$. Hence the achievable range always contains a strictly sub-Nash region.
Under the independent exploration of our model ($\Sigma_{\exp}$ diagonal, so $\rho^* = 0$), the only steady-state initialization is the symmetric Nash profile $\mu = \pNE\mathbf{1}$; \cref{prop:cooper_analog} shows that other steady-state initializations exist only by allowing correlated initial covariance.
\section{Supplemental general-\texorpdfstring{$N$}{N} computational simulations (Section~\ref{sec:duopoly_ode_sims})}
\label{app:additional_computations}

The duopoly heatmaps in Section~\ref{sec:duopoly_ode_sims} are useful because, when $N=2$, the exploration-mean space is two-dimensional and the full terminal-price map can be swept directly. For general $N$, the terminal-price map is a function on an $N$-dimensional hypercube, so an exhaustive visualization is no longer available. This appendix therefore uses the same price-moments ODE to study two clustered exploration designs for general $N$. These designs are not intended to characterize the full hypercube. Rather, they evaluate the ODE on broad and interpretable families of exploration profiles in which firms experiment around a common local price band or a common anchor price.

Throughout, unless otherwise noted, we use the same primitives as in Section~\ref{sec:duopoly_ode_sims}: $a=b=1$ and $c=1/2$, so that $\pNE=2/3$ and $\pMNP=1$. The simulations in this appendix restrict the exploration covariance to the scalar form
\[
    \Sigma_{exp}=\sigma_{exp}^2 I_N,
\]
where $\sigma_{exp}$ is the common exploration standard deviation. For $N>2$, when reporting a scalar terminal price, we use the cross-firm average
\[
    \bar P^{\mathrm{ODE}}(\tau;\mu,\Sigma_{exp})
    :=
    \frac{1}{N}\sum_{i=1}^N P_i^{\mathrm{ODE}}(\tau;\mu,\Sigma_{exp}).
\]
Under the scalar covariance specification above, this is
\[
    \bar P^{\mathrm{ODE}}(\tau;\mu,\sigma_{exp}^2 I_N)
    =
    \frac{1}{N}\sum_{i=1}^N P_i^{\mathrm{ODE}}(\tau;\mu,\sigma_{exp}^2 I_N).
\]

\subsection{Interval sampling}
\label{app:generalN_interval_sampling}

\paragraph{Motivation.}
The first design is the general-$N$ analogue of the ``similar exploration prices'' emphasized by the cone analysis. Instead of attempting to sample uniformly from the full $N$-dimensional exploration-mean space, we draw firms' exploration means from a common interval. This focuses attention on clustered profiles, where firms explore within the same local price band.

\paragraph{Design.}
For each boundary pair $(\ell,u)$ with $0.40\le \ell<u\le1.00$, we fix two exploration means at the endpoints and draw the remaining firms inside the interval:
\[
    \mu_1=\ell,\qquad
    \mu_2=u,\qquad
    \mu_i\mid(\ell,u)\stackrel{\mathrm{iid}}{\sim}\operatorname{Unif}[\ell,u],
    \quad i=3,\ldots,N.
\]
This is the computational analogue of the random-interval prior in \cref{def:random_interval_prior}: two firms determine a local price band, and the remaining firms explore within that same band. For each sampled profile, we solve the price-moments ODE with $\Sigma_{exp}=\sigma_{exp}^2 I_N$ and record the mean terminal price $\bar P^{\mathrm{ODE}}(\tau;\mu,\sigma_{exp}^2 I_N)$. The figure uses $N=10$ and $\sigma_{exp}=0.10$. Each boundary pair averages over $100$ independent draws of the interior $N-2$ exploration means. For visualization, we symmetrize the heatmap: cells with \(\mu_2<\mu_1\) are filled using the value from the corresponding flipped pair \((\mu_2,\mu_1)\).

\begin{figure}[!htbp]
\centering
\includegraphics[width=0.60\textwidth,keepaspectratio]{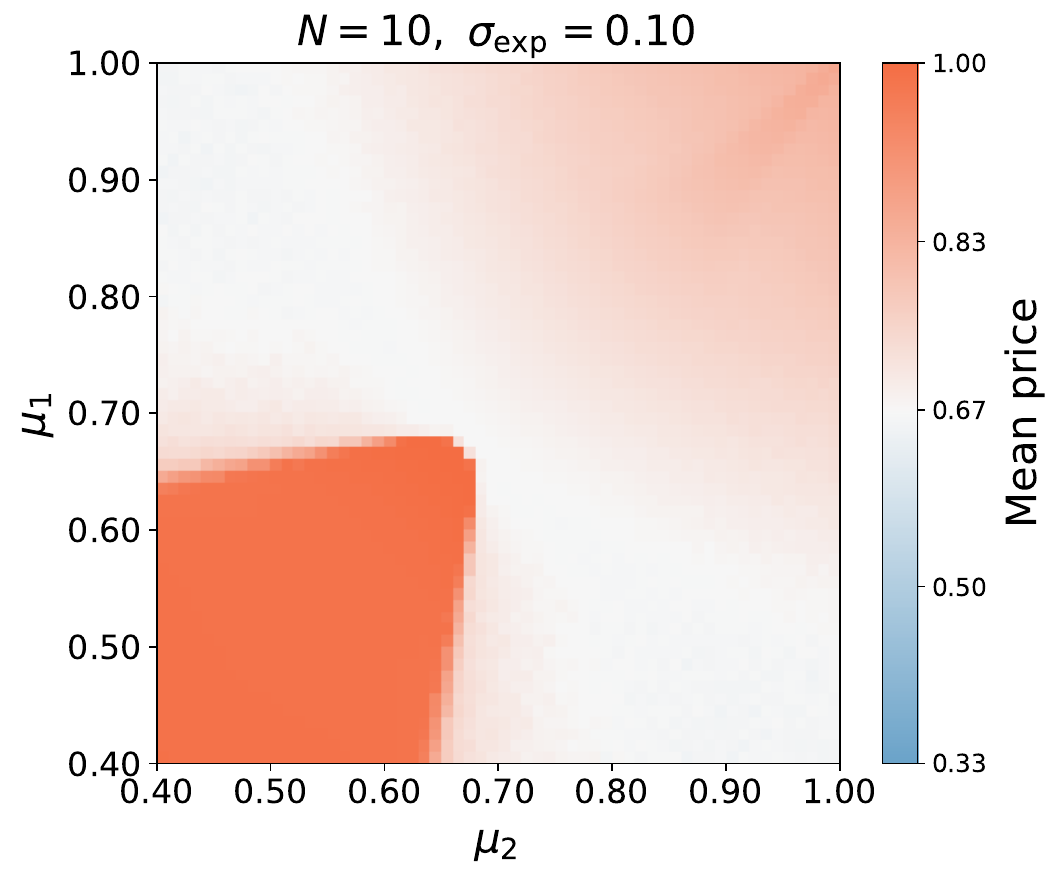}
\caption{ODE-implied mean terminal price under interval sampling. For each boundary pair $0.40\le \ell<u\le1.00$ on a grid with increments of $0.01$, two exploration means are fixed at $\ell$ and $u$, and the remaining $N-2$ means are drawn uniformly from $[\ell,u]$. Each boundary pair averages over $100$ draws, with $N=10$ and scalar exploration covariance $\Sigma_{exp}=0.10^2 I_N$. White is centered at the Nash price; red and blue indicate supra-competitive and sub-competitive mean prices, respectively. The heatmap is then symmetrized by assigning cells with \(\mu_2<\mu_1\) the value from the flipped pair.}
\label{fig:sec_5_Neq10_interval_ODE_simuls}
\end{figure}

\paragraph{Results.}
\cref{fig:sec_5_Neq10_interval_ODE_simuls} shows that the cone calculation is conservative. Across boundary pairs, $78.6\%$ have mean terminal price above Nash. At the simulation level, $74.2\%$ of draws have total profit above the Nash total-profit benchmark, and $70.2\%$ of boundary pairs have average total profit above that benchmark. The overall mean terminal price is $0.758$, compared with $\pNE=0.667$, and the range of mean terminal prices is $[0.652,1.000]$. Thus, within this family of clustered general-$N$ exploration profiles, sub-competitive outcomes remain close to Nash, while supra-competitive outcomes can reach the monopoly benchmark.

\subsection{Center--dispersion sampling}
\label{app:generalN_center_sampling}

\paragraph{Motivation.}
The second design parameterizes clustered exploration profiles by a common anchor price and a dispersion level. This separates the location of the exploration cluster from the amount of cross-firm heterogeneity. Small dispersion approximates symmetric exploration, while larger dispersion tests how robust the qualitative pattern is when firms' exploration means are no longer nearly identical.

\paragraph{Design.}
For an anchor price $s$ and dispersion parameter $\nu$, draw
\[
    \mu_i\stackrel{\mathrm{iid}}{\sim}
    \operatorname{Unif}[s-\sqrt{3}\nu,\ s+\sqrt{3}\nu],
    \qquad i=1,\ldots,N.
\]
Thus $s$ controls the location of the exploration cluster and $\nu$ controls cross-firm heterogeneity. The factor $\sqrt{3}$ normalizes $\nu$ as the standard deviation of the cross-firm exploration means. We solve the price-moments ODE with $N=10$ and compare $\nu\in\{0.02,0.10\}$ and scalar exploration covariances $\Sigma_{exp}=\sigma_{exp}^2 I_N$ with $\sigma_{exp} \in\{0.02,0.10\}$.

\begin{figure}[!htbp]
\centering
\includegraphics[width=1.0\textwidth,keepaspectratio]{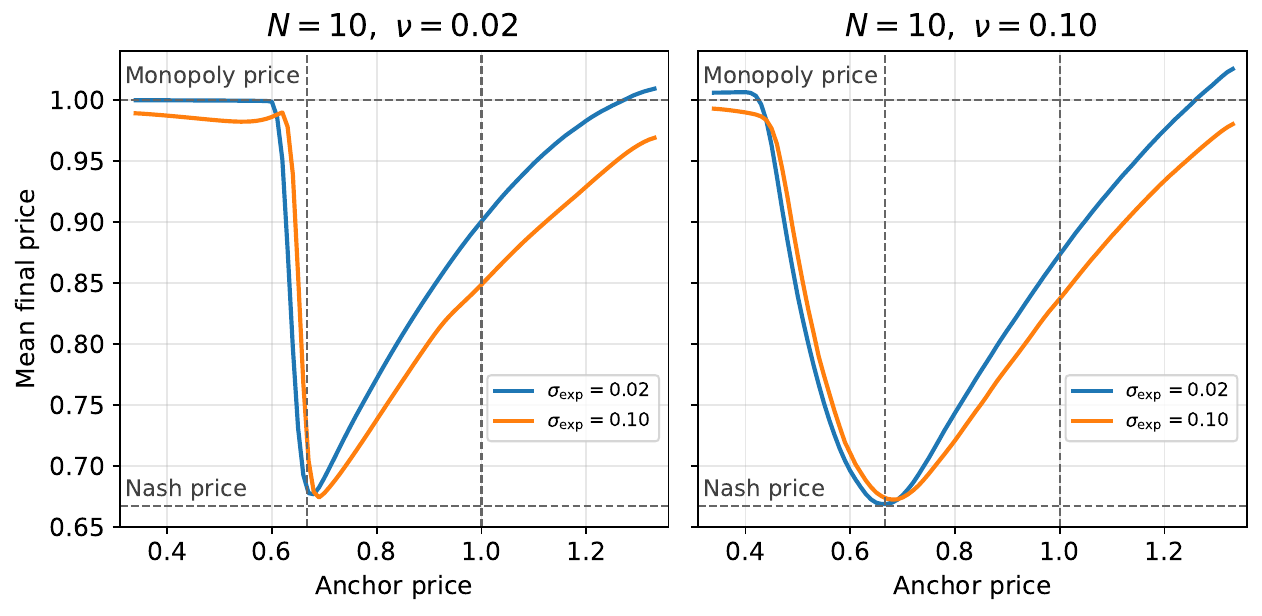}
\caption{Mean terminal price under center--dispersion sampling. For each anchor price $s$, exploration means are drawn independently from $\operatorname{Unif}[s-\sqrt{3}\nu,\ s+\sqrt{3}\nu]$. The panels compare cross-firm dispersion levels $\nu\in\{0.02,0.10\}$, and the lines compare scalar exploration covariances $\Sigma_{exp}=\sigma_{exp}^2 I_N$ with $\sigma_{exp}\in\{0.02,0.10\}$, using $N=10$. Dashed horizontal and vertical lines mark the Nash and monopoly benchmarks.}
\label{fig:sec_5_Neq10_center_ODE_simuls}
\end{figure}

\paragraph{Results.}
The pattern in \cref{fig:sec_5_Neq10_center_ODE_simuls} mirrors the symmetric-exploration benchmark. Prices are closest to Nash when the anchor is near $\pNE$, and they rise substantially when the anchor is below Nash or above the monopoly benchmark. Increasing either the cross-firm dispersion $\nu$ or the common exploration standard deviation $\sigma_{exp}$ attenuates the supra-competitive effect, but the qualitative shape remains stable.

Taken together, these general-$N$ simulations are not an exhaustive statement about the full exploration-mean space. They instead evaluate the ODE on broad families of clustered profiles. Within these families, supra-competitive outcomes arise more broadly than the analytical cone certificate alone would imply.
\FloatBarrier
\section{Details for the Empirically Calibrated Simulations (Section~\ref{sec:sim_design})}
\label{app:semiempirical_design}

This appendix collects deferred details for the empirically calibrated simulation design described in \cref{sec:sim_design}. The demand and observed rents are calibrated from data, while the pricing dynamics are simulated counterfactually on top of that calibrated environment.

\subsection{Demand-side parameterization}
\label{app:semiempirical_demand}

Recall the random-coefficients logit demand system from \cref{sec:sim_design}:
\[
s_{j,t}(P_t; \xi)
=\sum_{h=1}^{H} \tilde w_h\,
\frac{\exp\!\left(\alpha_{h}\, P_{j,t} + x_{j}'\beta_{h} + \xi_j\right)}
{1+\sum_{k=1}^N \exp\!\left(\alpha_{h}\, P_{k,t} + x_{k}'\beta_{h} + \xi_k\right)}.
\]
This demand system converts any rent vector into predicted market shares by aggregating heterogeneous household choice probabilities.
The covariates $x_j$ are observed rental characteristics (e.g., bedrooms, new-building indicators, and quality proxies). The renter-specific coefficients $(\alpha_h, \beta_h)$ are set as in \citet{calder2024algorithmic}: $\alpha_h<0$ is a decreasing linear function of $\log$ income, and $\beta_h$ uses their estimated interaction structure (Table~13 of \citealp{calder2024algorithmic}).

The vertical-differentiation fixed effects $\{\xi_j\}_{j=1}^N$ are calibrated so that at observed rents $p_0$, the model matches the observed shares $s_0$:
\[
s_{j}(p_0;\xi)=s_{j0} \quad \text{for each } j=1, \dots, N.
\]
These fixed effects absorb product-specific vertical quality not captured by observed characteristics, so the calibrated model exactly matches baseline shares.

\subsection{Calibration of the shadow costs}
\label{app:semiempirical_lambda}

We calibrate $\lambda=(\lambda_j)_{j=1}^N$ so that the observed rent vector $p_0$ is the Nash equilibrium of the calibrated static game. These shadow costs are not observed accounting costs; they are supply-side primitives chosen to rationalize observed rents as Nash. Formally, firm $j$ solves
\[
\max_{p_j}(p_j-\lambda_j)\,s_j\bigl((p_j,p_{-j,0});\xi\bigr),
\]
taking rivals' rents as fixed at $p_{-j,0}$. We set $\lambda_j$ so that the first-order condition holds at $p_{j0}$:
\[
0=s_j(p_0;\xi)+\bigl(p_{j0}-\lambda_j\bigr)\frac{\partial s_j}{\partial p_j}(p_0;\xi), \qquad j=1,\dots,N.
\]
Because the demand side is calibrated so that $s_j(p_0;\xi)=s_{j0}$, this pins down a unique $\lambda_j$ for each rental:
\[
\lambda_j=p_{j0}+\frac{s_{j0}}{\partial s_j/\partial p_j(p_0;\xi)}.
\]

\subsection{Explore-then-exploit pipeline: full equations}
\label{app:semiempirical_dynamic}

We provide the full equations for the explore-then-exploit dynamics.

\paragraph{Exploration.}
Fix an exploration mean $\mu_j$ for each rental $j$. During $t=1,\ldots,K$, rental $j$ posts a perturbed price
\[
P_{j,t}=\mu_j(1+v_{j,t}),\qquad v_{j,t}\overset{\text{i.i.d.}}{\sim}\mathcal{N}(0,\sigma^2),
\]
clipped to $[0,\bar P]$, where $\bar P$ is a large exogenous upper bound.

\paragraph{Exploitation.}
For $t\ge K$, each rental $j$ estimates a naive binary logit using only its own price-share history $\{(P_{j,r},s_{j,r})\}_{r\le t}$, treating competitors' prices as unobserved:
\[
y_{j,r}:=\log\!\Big(\frac{s_{j,r}}{1-s_{j,r}}\Big)\approx \eta_{j,t}+\theta_{j,t}P_{j,r},\qquad r\le t.
\]
Here ``naive'' means that the fitted logit is one-product and own-price only, even though true shares depend on all rents.
This yields $(\hat\eta_{j,t},\hat\theta_{j,t})$ and a predicted share $\hat s_{j,t}(p)=\Lambda(\hat\eta_{j,t}+\hat\theta_{j,t}p)$, where $\Lambda(z)=(1+e^{-z})^{-1}$ is the logistic function. The next-period rent is set by myopically maximizing predicted profit:
\[
P_{j,t+1}\in\arg\max_{p\in[0,\bar P]}\; (p-\lambda_j)\hat s_{j,t}(p),
\qquad t=K,\ldots,K+T-1.
\]
\section{Proof of Theorem~\ref{thm:terminal_convergence}: convergence to the price-moments ODE (Section~\ref{sec:terminal_convergence_price_moments})}
\label{app:proof_terminal_convergence}

This appendix proves \cref{thm:terminal_convergence}. The proof is organized in three steps. First, we write the empirical price--quantity moments as a stochastic approximation recursion. Second, we show that this recursion tracks a deterministic mean-field ODE on the fluid time scale. Third, we identify the posted-price coordinate of that ODE with the price-moments ODE in \cref{def:price_moments_ode}.

The argument is written for the diagonal exploration covariance matrix $\Sigma_{\exp}$ from the model section, with positive diagonal entries. Proofs of the auxiliary lemmas stated below are collected in Appendix~\ref{app:omitted_pf_terminal_convergence}.

The main point is that the stochastic process and the ODE are written in two different coordinate systems. We first prove convergence in a raw empirical-moment state, where the stochastic approximation is transparent, and only afterward change variables to the centered price moments used in the main text.

\subsection{Empirical moments and the OLS price map}
\label{app:terminal_moment_state}

We begin by defining a state that contains exactly the information used by each firm's own-price OLS regression and by the conditional mean of the next observation.

For each period $t$, let
\[
    P_t=(P_{1,t},\dots,P_{N,t})^\top,
    \qquad
    Q_t=(Q_{1,t},\dots,Q_{N,t})^\top,
\]
and define the one-period moment vector
\[
    X_t:=
    \bigl(P_t,Q_t,P_tP_t^\top,P_t\odot Q_t\bigr)
    \in
    \mathbb{R}^N\times\mathbb{R}^N\times\mathbb{R}^{N\times N}\times\mathbb{R}^N,
\]
where $\odot$ denotes componentwise multiplication. Let
\[
    \bar X_t
    :=
    \frac1t\sum_{s=1}^t X_s
    =
    \bigl(\bar P_t,\bar Q_t,\overline{PP}_t,\overline{PQ}_t\bigr).
\]
Thus $\bar P_t$ and $\bar Q_t$ are running price and quantity means, $\overline{PP}_t$ is the raw price second-moment matrix, and $\overline{PQ}_t$ is the vector of own price--quantity moments.

We keep raw second moments in this state because they update linearly as running averages. The centered variances and covariances needed for OLS can then be recovered by subtracting products of first moments.

For a generic state
\[
    x=(x^P,x^Q,x^{PP},x^{PQ})
    \in
    \mathbb{R}^N\times\mathbb{R}^N\times\mathbb{R}^{N\times N}\times\mathbb{R}^N,
\]
define the local own-price variance and own price--quantity covariance terms
\[
    R_i(x):=x^{PP}_{ii}-(x^P_i)^2,
    \qquad
    S_i(x):=x^{PQ}_i-x^P_i x^Q_i.
\]
On states with $R_i(x)>0$, the OLS slope and intercept in firm $i$'s own-price regression are
\[
    \hat\beta_i(x)=\frac{S_i(x)}{R_i(x)},
    \qquad
    \hat\alpha_i(x)=x^Q_i-\hat\beta_i(x)x^P_i.
\]
Substituting these coefficients into the myopic one-dimensional pricing problem gives a deterministic price as a function of the empirical moments. This is the key Markovian reduction: the full past history enters future prices only through $\bar X_t$.

The induced OLS pricing map is
\[
\pi_i(x):=
\begin{cases}
\left[
\dfrac{x^P_i x^{PQ}_i-x^{PP}_{ii}x^Q_i}
      {2\bigl(x^{PQ}_i-x^P_i x^Q_i\bigr)}
\right]_{[P_{\min},P_{\max}]},
& \text{if } S_i(x)<0,\\[12pt]
P_{\max}, & \text{if } S_i(x)\ge 0.
\end{cases}
\]
Let $\pi(x):=(\pi_1(x),\dots,\pi_N(x))^\top$.

\begin{lemma}[OLS price from empirical moments]
\label{lem:ols_price_from_moments}
For every exploitation period $t\ge K$, $P_{t+1}=\pi(\bar X_t)$. Moreover, on the finite-horizon neighborhoods used in the ODE-tracking argument below, the map $\pi$ is continuous and Lipschitz.
\end{lemma}

After this lemma, proving terminal convergence is reduced to proving convergence of the empirical moment vector $\bar X_t$ and then applying the continuous price map $\pi$.

\subsection{Stochastic approximation recursion}

We next compute the conditional drift of the moment state during exploitation. Conditional on the current history, the next posted price is already fixed by \cref{lem:ols_price_from_moments}; only the demand shocks remain random.

For a posted-price vector $p\in\mathbb{R}^N$, define expected demand by
\[
    d_i(p):=a-bp_i+\frac{c}{N-1}\sum_{j\ne i}p_j,
    \qquad
    d(p):=(d_1(p),\dots,d_N(p))^\top.
\]
Given the empirical moment state $x$, the conditional mean of the next observation is
\[
    f(x):=
    \bigl(\pi(x),d(\pi(x)),\pi(x)\pi(x)^\top,\pi(x)\odot d(\pi(x))\bigr).
\]
Define the drift
\begin{equation}
    h(x):=f(x)-x.
    \label{eq:raw_moment_drift}
\end{equation}
Thus $f(x)$ is the population one-period moment vector induced by the price chosen from state $x$, and $h(x)$ is the gap between this new target observation and the current running average.

\begin{lemma}[Conditional mean update]
\label{lem:conditional_mean_update}
For every exploitation period $t\ge K$,
\[
    \mathbb{E}[X_{t+1}\mid\mathcal{F}_t]=f(\bar X_t).
\]
\end{lemma}

The running-average identity then gives the stochastic approximation recursion.

The factor $1/(t+1)$ below is the usual running-average step size. It is this slowly vanishing step size that produces the multiplicative, or fluid, time scale used in the limiting ODE.

\begin{lemma}[SA recursion and start condition]
\label{lem:sa_recursion_start}
Let
\[
    \xi_{t+1}:=X_{t+1}-f(\bar X_t).
\]
Then, for all $t\ge K$,
\[
    \bar X_{t+1}
    =
    \bar X_t+\frac{1}{t+1}
    \bigl(h(\bar X_t)+\xi_{t+1}\bigr),
    \qquad
    \mathbb{E}[\xi_{t+1}\mid\mathcal{F}_t]=0.
\]
Moreover, there is a constant $\sigma_\xi<\infty$ such that
\[
    \mathbb{E}\bigl[\|\xi_{t+1}\|_2^2\mid\mathcal{F}_t\bigr]\le \sigma_\xi^2.
\]
Under exploration, letting $\mu_X:=\mathbb{E}[X_1]$ and $\Sigma_X:=\operatorname{Var}(X_1)$,
\[
    \mathbb{E}[\bar X_K]=\mu_X,
    \qquad
    \mathbb{E}\bigl[\|\bar X_K-\mu_X\|_2^2\bigr]
    =
    \frac{\operatorname{tr}(\Sigma_X)}{K}.
\]
\end{lemma}

This lemma gives the two ingredients needed for the fluid limit: the noise is a controlled martingale difference, and the exploration phase initializes the recursion near the deterministic point $\mu_X$.

The exploration mean $\mu_X$ has the following explicit coordinates. Define
\[
    \bar\mu_{-i}:=\frac{1}{N-1}\sum_{j\ne i}\mu_j,
    \qquad
    \bar\Sigma_{\exp,i,-i}:=\frac{1}{N-1}\sum_{j\ne i}(\Sigma_{\exp})_{ij}.
\]
Writing
\[
    \mu_X=
    \bigl(\mu^P,\mu^Q,\mu^{PP},\mu^{PQ}\bigr),
\]
we have
\[
    \mu^P_i=\mu_i,
    \qquad
    \mu^Q_i=a-b\mu_i+c\bar\mu_{-i},
\]
\[
    \mu^{PP}=\mu\mu^\top+\Sigma_{\exp},
    \qquad
    \mu^{PQ}_i
    =
    \mu_i\mu^Q_i-b(\Sigma_{\exp})_{ii}+c\bar\Sigma_{\exp,i,-i}.
\]
These formulas make explicit how the exploration distribution enters the limiting initial condition. In particular, the covariance matrix $\Sigma_{\exp}$ affects the initial price--quantity moments and therefore the initial OLS slopes.

\subsection{Mean-field ODE and ODE tracking}

Dropping the martingale term from the stochastic approximation recursion gives the deterministic mean-field dynamics. The time variable is normalized so that $\tau=1$ corresponds to the end of exploration.

The deterministic ODE associated with the recursion in \cref{lem:sa_recursion_start} is
\begin{equation}
    \dot M(\tau)=\frac1\tau h(M(\tau)),
    \qquad
    M(1)=\mu_X,
    \qquad \tau\ge1.
    \label{eq:raw_moment_ode}
\end{equation}
Write
\[
    M(\tau)=\bigl(M^P(\tau),M^Q(\tau),M^{PP}(\tau),M^{PQ}(\tau)\bigr),
\]
where $M^{PP}(\tau)$ is an $N\times N$ matrix. If
\[
    P(\tau):=\pi(M(\tau)),
    \qquad
    q(\tau):=d(P(\tau)),
\]
then \eqref{eq:raw_moment_ode} has the coordinate form
\[
    \dot M^P_i(\tau)
    =
    \frac{P_i(\tau)-M^P_i(\tau)}{\tau},
\]
\[
    \dot M^Q_i(\tau)
    =
    \frac{q_i(\tau)-M^Q_i(\tau)}{\tau},
\]
\[
    \dot M^{PP}_{ij}(\tau)
    =
    \frac{P_i(\tau)P_j(\tau)-M^{PP}_{ij}(\tau)}{\tau},
    \qquad i,j\in[N],
\]
\[
    \dot M^{PQ}_i(\tau)
    =
    \frac{P_i(\tau)q_i(\tau)-M^{PQ}_i(\tau)}{\tau}.
\]
Each coordinate has the same running-average interpretation: the current mean-field moment is pulled toward the population moment generated by the current deterministic price $P(\tau)$.

The next lemma is the fluid-limit step.

\begin{lemma}[Mean-field ODE convergence]
\label{lem:mean_field_convergence}
Let $M(\cdot)$ solve \eqref{eq:raw_moment_ode}. Let $\{K_m\}_{m\ge1}$ satisfy $K_m\to\infty$, and let $\{n_m\}_{m\ge1}$ satisfy $n_m\ge K_m$ and $n_m/K_m\to\tau\in[1,\infty)$. Then
\[
    \bar X_{n_m}\xrightarrow{\mathbb P} M(\tau).
\]
\end{lemma}

The proof of \cref{lem:mean_field_convergence} uses a localized Lipschitz argument. Although the OLS pricing rule contains ratios of empirical moments and is not globally Lipschitz, the mean-field path stays in a region where expected demand is bounded away from zero and own-price variance is bounded below. This gives a fixed tube around the path on which the drift is regular.

Thus the possible singularities in the OLS ratios are avoided on every fixed fluid-time window, and the martingale error in \cref{lem:sa_recursion_start} vanishes in the limit.

\subsection{Identification with the price-moments ODE}
\label{app:price_moments_identification}

The convergence lemma is stated in the raw moment coordinates $M$. The ODE in \cref{def:price_moments_ode}, however, is written only in terms of running price means and centered price co-movements, so we now translate between the two descriptions.

It remains to rewrite the mean-field ODE in centered price-moment coordinates. Define
\[
    U(\tau):=M^P(\tau),
    \qquad
    V(\tau):=\tau\bigl(M^{PP}(\tau)-U(\tau)U(\tau)^\top\bigr).
\]
Thus $U(\tau)$ is the running price mean and $V(\tau)$ is the scaled centered price second-moment matrix.

The scaling by $\tau$ records accumulated price variation rather than the centered second moment itself. This normalization is what gives $V$ the simple evolution $\dot V=(P-U)(P-U)^\top$.

\begin{lemma}[Mean-field ODE in price-moments coordinates]
\label{lem:mean_field_to_price_moments}
Let $M(\cdot)$ solve \eqref{eq:raw_moment_ode}, and define $(U,V)$ as above. Then $(U,V)$ solves
\[
    U(1)=\mu,\qquad V(1)=\Sigma_{\exp},
    \qquad
    \dot U=\frac{P-U}{\tau},
    \qquad
    \dot V=(P-U)(P-U)^\top,
\]
where $P(\tau)=\pi(M(\tau))$ is given componentwise as follows. For each $i\in[N]$, let
\[
    \bar U_{-i}:=\frac{1}{N-1}\sum_{j\ne i}U_j,
    \qquad
    \bar V_{i,-i}:=\frac{1}{N-1}\sum_{j\ne i}V_{ij}.
\]
Then
\[
    P_i=
    \begin{cases}
    [\widetilde P_i]_{[P_{\min},P_{\max}]},
        & \text{if } -bV_{ii}+c\bar V_{i,-i}<0,\\
    P_{\max},
        & \text{otherwise,}
    \end{cases}
\]
with
\[
    \widetilde P_i
    =
    \frac{(a+c\bar U_{-i})V_{ii}-cU_i\bar V_{i,-i}}
    {2(bV_{ii}-c\bar V_{i,-i})}.
\]

Equivalently, the mean-field coordinates are recovered from $(U,V)$ by
\[
    M^P_i(\tau)=U_i(\tau),
    \qquad
    M^Q_i(\tau)=a-bU_i(\tau)+c\bar U_{-i}(\tau),
\]
\[
    M^{PP}(\tau)=U(\tau)U(\tau)^\top+\frac1\tau V(\tau),
\]
and
\[
    M^{PQ}_i(\tau)
    =
    U_i(\tau)M^Q_i(\tau)
    +
    \frac1\tau\bigl(-bV_{ii}(\tau)+c\bar V_{i,-i}(\tau)\bigr).
\]
\end{lemma}

The recovery formulas show that no information relevant for the posted-price coordinate is lost: once $(U,V)$ is known, the quantity moments appearing in the raw OLS state are pinned down by the linear demand model. These dynamics and the posted-price coordinate $P(\tau)$ coincide with the price-moments ODE in \cref{def:price_moments_ode}.

This completes the deterministic identification step: the stochastic approximation is tracked in the full moment state, but its limiting posted prices can be read from the price-moments coordinates of \cref{def:price_moments_ode}.

\subsection{Terminal convergence}

\begin{myproof}[Proof of \cref{thm:terminal_convergence}]
We now combine the preceding pieces. The empirical moment state converges to the mean-field state, the terminal price is a continuous function of that state, and the resulting deterministic price is the posted-price coordinate of the price-moments ODE.

Let $\{(K_m,T_m)\}_{m\ge1}$ satisfy
\[
    K_m\to\infty,
    \qquad
    T_m\to\infty,
    \qquad
    \frac{K_m+T_m}{K_m}\to\tau\in[1,\infty).
\]
Set $n_m:=K_m+T_m-1$. Since
\[
    \frac{n_m}{K_m}
    =
    \frac{K_m+T_m}{K_m}
    -
    \frac1{K_m}
    \to\tau,
\]
\cref{lem:mean_field_convergence} gives
\[
    \bar X_{n_m}\xrightarrow{\mathbb P}M(\tau).
\]
This is convergence of the full empirical state through the last period whose data are used to compute the terminal exploitation price.

The terminal exploitation price is computed from the empirical moments through period $K_m+T_m-1$. Therefore, by \cref{lem:ols_price_from_moments},
\[
    P_{K_m+T_m}=\pi(\bar X_{n_m}).
\]
Because $\pi$ is continuous in a neighborhood of the limiting path, the continuous mapping theorem implies
\[
    P_{K_m+T_m}
    =
    \pi(\bar X_{n_m})
    \xrightarrow{\mathbb P}
    \pi(M(\tau)).
\]
The remaining task is only notational: the limit $\pi(M(\tau))$ is still expressed in the raw moment coordinates, while the theorem states the limit in the price-moments notation.

By \cref{lem:mean_field_to_price_moments}, the vector $\pi(M(\tau))$ is exactly $P(\tau)$, the posted-price coordinate of the price-moments ODE initialized at
\[
    (U(1),V(1))=(\mu,\Sigma_{\exp}).
\]
In the notation of \cref{def:price_moments_ode}, this gives
\[
    \pi(M(\tau))=P^{\mathrm{ODE}}(\tau;\mu,\Sigma_{\exp}).
\]
Therefore
\[
    (P_{1,K_m+T_m},\dots,P_{N,K_m+T_m})
    \xrightarrow{\mathbb P}
    P^{\mathrm{ODE}}(\tau;\mu,\Sigma_{\exp}),
\]
as claimed.

This proves the desired terminal convergence and completes the reduction from the finite-sample explore--then--exploit process to the deterministic price-moments ODE.
\end{myproof}

\section{Omitted Lemmas in Proof of Theorem~\ref{thm:terminal_convergence} (Section~\ref{sec:terminal_convergence_price_moments})}
\label{app:omitted_pf_terminal_convergence}

\subsection{Proof of Lemma~\ref{lem:ols_price_from_moments}}
\label{app:proof_price_from_moments}

\begin{myproof}
For each firm $i$, the OLS regression uses only the four own-firm empirical moments
\[
\bar{P}_{i,t}:=\frac1t\sum_{s=1}^t P_{i,s},\qquad
\bar{Q}_{i,t}:=\frac1t\sum_{s=1}^t Q_{i,s},\qquad
\overline{P^2}_{i,t}:=\frac1t\sum_{s=1}^t P_{i,s}^2,\qquad
\overline{PQ}_{i,t}:=\frac1t\sum_{s=1}^t P_{i,s}Q_{i,s}.
\]
In the full moment state
\[
\bar X_t=(\bar P_t,\bar Q_t,\overline{PP}_t,\overline{PQ}_t),
\]
these are the local coordinates
\[
\bar X_{i,t}^{\mathrm{loc}}
:=
\bigl(
\bar P_{i,t},\bar Q_{i,t},\overline{PP}_{ii,t},\overline{PQ}_{i,t}
\bigr)
=
\bigl(
\bar P_{i,t},\bar Q_{i,t},\overline{P^2}_{i,t},\overline{PQ}_{i,t}
\bigr).
\]

Define the clipping operator
\[
[x]_{[P_{\min},P_{\max}]}:=\min\{\max\{x,P_{\min}\},\,P_{\max}\}.
\]
We first show that, for every exploitation period $t\ge K$,
\[
P_{i,t+1}
=
f_P(\bar X_{i,t}^{\mathrm{loc}}),
\]
where
\[
f_P(\bar X_{i,t}^{\mathrm{loc}})
=
\begin{cases}
\displaystyle
\left[\,
\frac{\bar{P}_{i,t}\,\overline{PQ}_{i,t}-\overline{P^2}_{i,t}\,\bar{Q}_{i,t}}
{2\big(\overline{PQ}_{i,t}-\bar{P}_{i,t}\,\bar{Q}_{i,t}\big)}
\,\right]_{[P_{\min},P_{\max}]},
& \text{if }\ \overline{PQ}_{i,t}-\bar{P}_{i,t}\,\bar{Q}_{i,t}<0,\\[1.1em]
P_{\max}, & \text{otherwise.}
\end{cases}
\]
Equivalently, in the notation of Appendix~\ref{app:proof_terminal_convergence},
\[
    \pi_i(\bar X_t)=f_P(\bar X_{i,t}^{\mathrm{loc}}),
    \qquad
    P_{t+1}=\pi(\bar X_t).
\]

Fix a firm $i\in[N]$ and a time $t\ge K$. Define
\[
S_{i,t}:=\overline{PQ}_{i,t}-\bar{P}_{i,t}\,\bar{Q}_{i,t},
\qquad
R_{i,t}:=\overline{P^2}_{i,t}-\bigl(\bar{P}_{i,t}\bigr)^2.
\]
By nondegenerate exploration, $R_{i,t}>0$ a.s.\ for all $t\ge K$, so the OLS slope is well-defined.
Since $Q_{i,s}>0$ a.s.\ for all $s\le t$, we also have $\bar{Q}_{i,t}>0$ a.s.

The OLS normal equations, with an intercept, for regressing $Q_{i,s}$ on $(1,P_{i,s})$ over
$s\le t$ are
\[
\hat\alpha_{i,t}+\hat\beta_{i,t}\,\bar{P}_{i,t}=\bar{Q}_{i,t},
\qquad
\hat\alpha_{i,t}\,\bar{P}_{i,t}+\hat\beta_{i,t}\,\overline{P^2}_{i,t}=\overline{PQ}_{i,t}.
\]
Subtracting $\bar{P}_{i,t}$ times the first equation from the second gives
$\hat\beta_{i,t}R_{i,t}=S_{i,t}$, i.e.
\[
\hat\beta_{i,t}=\frac{S_{i,t}}{R_{i,t}},
\qquad
\hat\alpha_{i,t}=\bar{Q}_{i,t}-\hat\beta_{i,t}\bar{P}_{i,t}.
\]

Let $\hat\pi_{i,t}(p):=p(\hat\alpha_{i,t}+\hat\beta_{i,t}p)$ for
$p\in[P_{\min},P_{\max}]$. By the exploitation rule,
\[
P_{i,t+1}=\arg\max_{p\in[P_{\min},P_{\max}]}\hat\pi_{i,t}(p).
\]

\paragraph{Case 1: $S_{i,t}<0$ (equivalently $\hat\beta_{i,t}<0$).}
Then $\hat\pi_{i,t}$ is strictly concave, so its unique unconstrained maximizer is
$p^\star=-\hat\alpha_{i,t}/(2\hat\beta_{i,t})$, and the constrained maximizer is
$P_{i,t+1}=[p^\star]_{[P_{\min},P_{\max}]}$.
Using $\hat\alpha_{i,t}=\bar{Q}_{i,t}-\hat\beta_{i,t}\bar{P}_{i,t}$ and
$\hat\beta_{i,t}R_{i,t}=S_{i,t}$,
\[
p^\star
=-\frac{\bar{Q}_{i,t}-\hat\beta_{i,t}\bar{P}_{i,t}}{2\hat\beta_{i,t}}
=\frac{\bar{P}_{i,t}S_{i,t}-\bar{Q}_{i,t}R_{i,t}}{2S_{i,t}}
=\frac{\bar{P}_{i,t}\,\overline{PQ}_{i,t}-\overline{P^2}_{i,t}\,\bar{Q}_{i,t}}
{2\bigl(\overline{PQ}_{i,t}-\bar{P}_{i,t}\,\bar{Q}_{i,t}\bigr)}.
\]
This is the first branch of the claimed formula.

\paragraph{Case 2: $S_{i,t}\ge 0$ (equivalently $\hat\beta_{i,t}\ge 0$).}
Then $\hat\pi_{i,t}$ is convex, or linear, on $[P_{\min},P_{\max}]$, so maximizers lie at the endpoints.
Moreover,
\begin{align*}
\hat\pi_{i,t}(P_{\max})-\hat\pi_{i,t}(P_{\min})
&=(P_{\max}-P_{\min})\bigl(\hat\alpha_{i,t}+\hat\beta_{i,t}(P_{\max}+P_{\min})\bigr)\\
&=(P_{\max}-P_{\min})\Bigl(\bar{Q}_{i,t}
+\hat\beta_{i,t}(P_{\max}+P_{\min}-\bar{P}_{i,t})\Bigr)\;>\;0,
\end{align*}
since $P_{\max}>P_{\min}$, $\bar{Q}_{i,t}>0$, $\hat\beta_{i,t}\ge 0$, and
$\bar{P}_{i,t}\le P_{\max}$ implies
$P_{\max}+P_{\min}-\bar{P}_{i,t}\ge P_{\min}>0$. Hence the unique maximizer is
$P_{\max}$, giving the second branch.

Combining the two cases proves the stated deterministic mapping from the local empirical moments
to $P_{i,t+1}$, and hence proves $P_{t+1}=\pi(\bar X_t)$.

The continuity and Lipschitz statement used in the finite-horizon ODE-tracking argument follows
from Lemmas~\ref{lem:price_lip_appN} and~\ref{lem:tube_lip_appN} below. Those lemmas show that,
on the fixed tube around the mean-field trajectory, each local block
\[
(x_i^P,x_i^Q,x_{ii}^{PP},x_i^{PQ})
\]
lies in a domain where the OLS denominator is bounded away from zero and the pricing map is
Lipschitz. Therefore $\pi$ is continuous and Lipschitz on the finite-horizon neighborhoods used
in Lemma~\ref{lem:mean_field_convergence}.
\end{myproof}

\subsection{Proof of Lemma~\ref{lem:conditional_mean_update}}
\label{app:proof_state_update}

\begin{myproof}
Recall that
\[
    X_t=(P_t,Q_t,P_tP_t^\top,P_t\odot Q_t),
    \qquad
    \bar X_t=\frac1t\sum_{s=1}^t X_s.
\]
Let
\[
    \F_t:=\sigma\big((P_{j,s},Q_{j,s})_{j\in[N],\,1\le s\le t}\big)
\]
be the history filtration.

Fix $t\ge K$. By Lemma~\ref{lem:ols_price_from_moments},
\[
    P_{t+1}=\pi(\bar X_t),
\]
so $P_{t+1}$ is $\F_t$-measurable. Write $p:=\pi(\bar X_t)$. Then, by the demand model and
$\E[\varepsilon_{i,t+1}\mid\F_t]=0$,
\[
\E[Q_{i,t+1}\mid \F_t]
= a - bp_i + \frac{c}{N-1}\sum_{j\ne i}p_j
= d_i(p).
\]
Since $P_{t+1}$ is $\F_t$-measurable, we also have
\[
\E[P_{t+1}\mid\F_t]=p,
\qquad
\E[P_{t+1}P_{t+1}^\top\mid\F_t]=pp^\top,
\]
and, for each $i$,
\[
\E[P_{i,t+1}Q_{i,t+1}\mid\F_t]
=
p_i\,\E[Q_{i,t+1}\mid\F_t]
=
p_i d_i(p).
\]
Therefore
\[
\E[X_{t+1}\mid\F_t]
=
\bigl(
    \pi(\bar X_t),\
    d(\pi(\bar X_t)),\
    \pi(\bar X_t)\pi(\bar X_t)^\top,\
    \pi(\bar X_t)\odot d(\pi(\bar X_t))
\bigr)
=
f(\bar X_t),
\]
which proves the claim.
\end{myproof}

\subsection{Proof of Lemma~\ref{lem:sa_recursion_start} (SA recursion and start condition)}
\label{app:proof_SA_N}

\begin{myproof}
Recall
\[
X_t:=(P_t,Q_t,P_tP_t^\top,P_t\odot Q_t),
\qquad
\bar X_t:=\frac1t\sum_{s=1}^t X_s.
\]
Let
\[
\F_t:=\sigma\big((P_{j,s},Q_{j,s})_{j\in[N],\,1\le s\le t}\big)
\]
be the history filtration.

\paragraph{Step 1: Running-mean identity and SA decomposition.}
For every $t\ge 1$,
\[
\bar X_{t+1}
=\frac{1}{t+1}\sum_{s=1}^{t+1}X_s
=\frac{t}{t+1}\bar X_t+\frac{1}{t+1}X_{t+1}
=\bar X_t+\frac{1}{t+1}\big(X_{t+1}-\bar X_t\big).
\]
For $t\ge K$, define
\[
\xi_{t+1}:=X_{t+1}-f(\bar X_t),
\qquad
h(\bar X_t):=f(\bar X_t)-\bar X_t.
\]
Then
\[
\bar X_{t+1}
=
\bar X_t+\frac{1}{t+1}\big(h(\bar X_t)+\xi_{t+1}\big),
\qquad t\ge K.
\]
By Lemma~\ref{lem:conditional_mean_update},
\[
\E[\xi_{t+1}\mid\F_t]
=
\E[X_{t+1}\mid\F_t]-f(\bar X_t)
=
0.
\]

\paragraph{Step 2: Componentwise form of the noise.}
Fix $t\ge K$. By Lemma~\ref{lem:ols_price_from_moments},
\[
P_{t+1}=\pi(\bar X_t),
\]
hence $P_{t+1}$ is $\F_t$-measurable and satisfies
$P_{\min}\le P_{i,t+1}\le P_{\max}$ for every $i$.
By the demand model and $\E[\varepsilon_{i,t+1}\mid\F_t]=0$,
\[
\E[Q_{i,t+1}\mid \F_t]
= a - bP_{i,t+1} + \frac{c}{N-1}\sum_{j\ne i}P_{j,t+1}.
\]
Moreover, since $P_{t+1}$ is $\F_t$-measurable,
\[
\E[P_{t+1}P_{t+1}^\top\mid\F_t]
=
P_{t+1}P_{t+1}^\top,
\]
and
\[
\E[P_{i,t+1}Q_{i,t+1}\mid \F_t]
=
P_{i,t+1}\E[Q_{i,t+1}\mid \F_t].
\]
Thus the noise coordinates satisfy
\[
\xi_{t+1}^{P}=0,
\qquad
\xi_{t+1}^{PP}=0_{N\times N},
\]
and, for each $i$,
\[
\xi_{t+1}^{Q_i}
=
Q_{i,t+1}-\E[Q_{i,t+1}\mid\F_t]
=
\varepsilon_{i,t+1},
\qquad
\xi_{t+1}^{PQ_i}
=
P_{i,t+1}\varepsilon_{i,t+1}.
\]

\paragraph{Step 3: Uniform conditional second-moment bound.}
Using the preceding identities,
\[
\|\xi_{t+1}\|_2^2
=
\sum_{i=1}^N\Big((\varepsilon_{i,t+1})^2+(P_{i,t+1}\varepsilon_{i,t+1})^2\Big)
\le
(1+P_{\max}^2)\sum_{i=1}^N \varepsilon_{i,t+1}^2.
\]
Taking conditional expectation and using
$\E[\varepsilon_{i,t+1}^2\mid \F_t]\le \sigma_{\mathrm{env}}^2$ yields
\[
\E[\|\xi_{t+1}\|_2^2\mid \F_t]
\le
(1+P_{\max}^2)\sum_{i=1}^N \E[\varepsilon_{i,t+1}^2\mid \F_t]
\le
N(1+P_{\max}^2)\sigma_{\mathrm{env}}^2.
\]
Thus we may take
\[
\sigma_\xi^2:=N(1+P_{\max}^2)\sigma_{\mathrm{env}}^2.
\]

\paragraph{Step 4: Exploration start condition.}
Under the exploration law, $(X_1,\dots,X_K)$ are i.i.d.\ with mean
$\mu_X:=\E[X_1]$ and covariance $\Sigma_X:=\Var(X_1)$. Hence
\[
\E[\bar X_K]
=
\E\Big[\frac1K\sum_{s=1}^K X_s\Big]
=
\mu_X,
\]
and
\[
\Var(\bar X_K)
=
\Var\Big(\frac1K\sum_{s=1}^K X_s\Big)
=
\frac{1}{K^2}\sum_{s=1}^K \Var(X_s)
=
\frac{1}{K}\Sigma_X.
\]
Finally,
\[
\E\big[\|\bar X_K-\mu_X\|_2^2\big]
=
\tr\big(\Var(\bar X_K)\big)
=
\frac{\tr(\Sigma_X)}{K}.
\]
\end{myproof}

\subsection{Proof of Lemma~\ref{lem:mean_field_convergence} (Mean-field ODE convergence)}
\label{app:proof_ode_limit}

\begin{myproof}
Fix $\bar\tau\in[1,\infty)$ and a sequence $\{K_m\}_{m\ge1}$ with $K_m\to\infty$ and
$\{n_m\}_{m\ge1}$ satisfying $n_m\ge K_m$ and $n_m/K_m\to\bar\tau$.
Fix $\tau_+>\bar\tau$. For all sufficiently large $m$, $n_m/K_m\le \tau_+$; throughout we work
on the finite window $[1,\tau_+]$.

Let $M:[1,\infty)\to
\R^N\times\R^N\times\R^{N\times N}\times\R^N$ solve the mean ODE
\begin{equation}\label{eq:mean_ode_app}
\dot M(\tau)=\frac{1}{\tau}\,h(M(\tau)),\qquad M(1)=\mu_X.
\end{equation}
We will show $\bar X_{n_m}\xrightarrow{\P}M(\bar\tau)$ by comparing the SA recursion to an Euler
scheme and then to $M$, using a fixed Lipschitz tube around the ODE trajectory on
$[1,\tau_+]$.

\begin{lemma}[{Mean-field bounds on $M(\tau)$ on $[1,\tau_+\rbrack$}]
\label{lem:ode_bounds_appN}
Define
\[
Q_{\min}:=a-bP_{\max}+cP_{\min}>0,
\qquad
Q_{\max}:=a-bP_{\min}+cP_{\max}.
\]
Then for all $\tau\in[1,\tau_+]$ and all $i,j\in[N]$,
\[
P_{\min}\le M_i^P(\tau)\le P_{\max},
\qquad
Q_{\min}\le M_i^Q(\tau)\le Q_{\max},
\]
\[
P_{\min}^2\le M_{ij}^{PP}(\tau)\le P_{\max}^2,
\qquad
P_{\min}Q_{\min}\le M_i^{PQ}(\tau)\le P_{\max}Q_{\max}.
\]
\end{lemma}

\begin{myproof}[Proof of Lemma~\ref{lem:ode_bounds_appN}]
For any state $x$, the induced next-period price vector is $\pi(x)\in[P_{\min},P_{\max}]^N$, and the induced expected-demand vector is $d(\pi(x))\in\R^N$. Since $\pi_i(x)\in[P_{\min},P_{\max}]$ for all $i$, we have
$d_i(\pi(x))\in[Q_{\min},Q_{\max}]$ for all $i$.

Writing the mean ODE \eqref{eq:mean_ode_app} coordinatewise, for each $i,j\in[N]$,
\[
\dot M_i^P(\tau)=\frac{1}{\tau}\big(\pi_i(M(\tau))-M_i^P(\tau)\big),
\qquad
\dot M_i^Q(\tau)=\frac{1}{\tau}\big(d_i(\pi(M(\tau)))-M_i^Q(\tau)\big),
\]
\[
\dot M_{ij}^{PP}(\tau)
=
\frac{1}{\tau}\big(\pi_i(M(\tau))\pi_j(M(\tau))-M_{ij}^{PP}(\tau)\big),
\qquad
\dot M_i^{PQ}(\tau)
=
\frac{1}{\tau}\big(\pi_i(M(\tau))\,d_i(\pi(M(\tau)))-M_i^{PQ}(\tau)\big).
\]
Multiplying by $\tau$ and integrating from $1$ to $\tau$ gives the identities
\[
M_i^P(\tau)
=
\frac{1}{\tau}\left(M_i^P(1)+\int_1^\tau \pi_i(M(s))\,ds\right),
\qquad
M_i^Q(\tau)
=
\frac{1}{\tau}\left(M_i^Q(1)+\int_1^\tau d_i(\pi(M(s)))\,ds\right),
\]
\[
M_{ij}^{PP}(\tau)
=
\frac{1}{\tau}\left(M_{ij}^{PP}(1)+\int_1^\tau \pi_i(M(s))\pi_j(M(s))\,ds\right),
\]
\[
M_i^{PQ}(\tau)
=
\frac{1}{\tau}\left(M_i^{PQ}(1)+\int_1^\tau \pi_i(M(s))d_i(\pi(M(s)))\,ds\right).
\]
Now use the bounds $\pi_i(M(s))\in[P_{\min},P_{\max}]$ and
$d_i(\pi(M(s)))\in[Q_{\min},Q_{\max}]$ for all $s$, so that
\[
\pi_i(M(s))\pi_j(M(s))\in[P_{\min}^2,P_{\max}^2],
\qquad
\pi_i(M(s))d_i(\pi(M(s)))\in[P_{\min}Q_{\min},P_{\max}Q_{\max}].
\]
The initial condition $M(1)=\mu_X$ satisfies the same bounds, because it is the corresponding
exploration moment vector. Each display above therefore implies that the corresponding coordinate
of $M(\tau)$ remains within the stated interval for all $\tau\ge 1$, and in particular for all
$\tau\in[1,\tau_+]$.
\end{myproof}

\begin{lemma}[Own-price variance floor along the mean ODE]
\label{lem:var_floor_appN}
For each $i$, define
\[
R_i^M(\tau):=M_{ii}^{PP}(\tau)-\big(M_i^{P}(\tau)\big)^2.
\]
Then $\tau R_i^M(\tau)$ is nondecreasing and
\[
R_i^M(\tau)\ge \frac{R_i^M(1)}{\tau}=\frac{(\Sigma_{\exp})_{ii}}{\tau}\qquad\text{for all }\tau\ge 1.
\]
In particular, for all $\tau\in[1,\tau_+]$, $R_i^M(\tau)\ge (\Sigma_{\exp})_{ii}/\tau_+$.
\end{lemma}

\begin{myproof}[Proof of Lemma~\ref{lem:var_floor_appN}]
From the ODE coordinates,
\[
\dot M_i^{P}(\tau)=\frac{1}{\tau}\big(\pi_i(M(\tau))-M_i^P(\tau)\big),
\qquad
\dot M_{ii}^{PP}(\tau)=\frac{1}{\tau}\big(\pi_i(M(\tau))^2-M_{ii}^{PP}(\tau)\big).
\]
Differentiate $R_i^M(\tau)=M_{ii}^{PP}(\tau)-(M_i^P(\tau))^2$:
\begin{align*}
\dot R_i^M(\tau)
&=\dot M_{ii}^{PP}(\tau)-2M_i^{P}(\tau)\dot M_i^{P}(\tau)\\
&=\frac{1}{\tau}\Big(\pi_i(M(\tau))^2-M_{ii}^{PP}-2M_i^P(\pi_i(M(\tau))-M_i^P)\Big)\\
&=\frac{1}{\tau}\Big((\pi_i(M(\tau))-M_i^P)^2 - (M_{ii}^{PP}-(M_i^P)^2)\Big)\\
&=\frac{1}{\tau}\Big((\pi_i(M(\tau))-M_i^P)^2 - R_i^M(\tau)\Big).
\end{align*}
Therefore
\[
\frac{d}{d\tau}\big(\tau R_i^M(\tau)\big)
=
\tau\dot R_i^M(\tau)+R_i^M(\tau)
=
(\pi_i(M(\tau))-M_i^P(\tau))^2
\ge 0,
\]
so $\tau R_i^M(\tau)$ is nondecreasing and $\tau R_i^M(\tau)\ge R_i^M(1)$.
Finally,
\[
R_i^M(1)=M_{ii}^{PP}(1)-(M_i^P(1))^2
=
\Var_{\exp}(P_{i,1})
=
(\Sigma_{\exp})_{ii}.
\]
\end{myproof}

\begin{lemma}[Lipschitz pricing map under $(Q,\Var(P))$ floors]
\label{lem:price_lip_appN}
Fix constants $\eta_Q>0$ and $\eta_V>0$, and define the block-domain
\begin{align*}
\mathcal{S}(\eta_Q,\eta_V) \coloneqq \bigg\{ x=(x^P,x^Q,x^{PP},x^{PQ}) \in \mathbb{R}^4 :
&\ 0\le x^P\le P_{\max}+1,\ \ \eta_Q\le x^Q\le Q_{\max}+1,\\
&\ 0\le x^{PP}\le (P_{\max}+1)^2,\ \ 0\le x^{PQ}\le (P_{\max}+1)(Q_{\max}+1),\\
&\ x^{PP}-(x^P)^2\ge \eta_V \bigg\}.
\end{align*}
Then $f_P:\mathcal S(\eta_Q,\eta_V)\to[P_{\min},P_{\max}]$ is globally Lipschitz: there exists
$L_P=L_P(P_{\min},P_{\max},Q_{\max},\eta_Q,\eta_V)<\infty$ such that for all
$x,y\in\mathcal S(\eta_Q,\eta_V)$,
\[
|f_P(x)-f_P(y)|\le L_P\,\|x-y\|_2.
\]
\end{lemma}

\begin{myproof}[Proof of Lemma~\ref{lem:price_lip_appN}]
Fix $x\in\mathcal S(\eta_Q,\eta_V)$ and define
\[
v(x):=x^{PP}-(x^P)^2\ (\ge \eta_V),\qquad
s(x):=x^{PQ}-x^P x^Q,
\]
\[
\beta(x):=\frac{s(x)}{v(x)},
\qquad
\alpha(x):=x^Q-\beta(x)x^P.
\]
Set the shorthand $P_+:=P_{\max}+1$ and $Q_+:=Q_{\max}+1$.

\paragraph{Step 1: explicit bounds and Lipschitz constants for $(\alpha,\beta)$.}
On $\mathcal S(\eta_Q,\eta_V)$ we have $0\le x^P\le P_+$,
$\eta_Q\le x^Q\le Q_+$, and $0\le x^{PQ}\le P_+Q_+$, hence
\[
|s(x)|\le |x^{PQ}|+|x^P x^Q|\le 2P_+Q_+ \eqqcolon B_s.
\]
Moreover, $\nabla v=(-2x^P,0,1,0)$ and $\nabla s=(-x^Q,-x^P,0,1)$, so
\[
\|\nabla v\|_2\le \sqrt{4P_+^2+1}\eqqcolon L_v,
\qquad
\|\nabla s\|_2\le \sqrt{Q_+^2+P_+^2+1}\eqqcolon L_s.
\]
By the mean value theorem, for all $x,y\in\mathcal S(\eta_Q,\eta_V)$,
\[
|v(x)-v(y)|\le L_v\|x-y\|_2,
\qquad
|s(x)-s(y)|\le L_s\|x-y\|_2.
\]
Using $v(\cdot)\ge \eta_V$ and $|s(\cdot)|\le B_s$, we obtain
\[
|\beta(x)-\beta(y)|
\le
\left(\frac{L_s}{\eta_V}+\frac{B_sL_v}{\eta_V^2}\right)\|x-y\|_2
\eqqcolon L_\beta\,\|x-y\|_2,
\qquad
|\beta(x)|\le \frac{B_s}{\eta_V}\eqqcolon B_\beta.
\]
Finally,
\[
|\alpha(x)|\le |x^Q|+|\beta(x)|\,|x^P|
\le Q_+ + B_\beta P_+ \eqqcolon B_\alpha,
\]
and
\[
|\alpha(x)-\alpha(y)|
\le |x^Q-y^Q| + |\beta(x)x^P-\beta(y)y^P|
\le \bigl(1+B_\beta+P_+L_\beta\bigr)\|x-y\|_2
\eqqcolon L_\alpha\,\|x-y\|_2.
\]

\paragraph{Step 2: Lipschitzness of $f_P$.}
Let $[u]_{[P_{\min},P_{\max}]}:=\min\{\max\{u,P_{\min}\},P_{\max}\}$.
With $g(\alpha,\beta):=-\alpha/(2\beta)$, the pricing map can be written as
\[
f_P(x)=
\begin{cases}
P_{\max}, & \beta(x)\ge 0,\\[0.25em]
\bigl[g(\alpha(x),\beta(x))\bigr]_{[P_{\min},P_{\max}]}, & \beta(x)<0.
\end{cases}
\]
If $\beta(x)<0$, then
\[
\alpha(x)=x^Q-\beta(x)x^P\ge x^Q\ge \eta_Q,
\]
since $x^P\ge 0$. Define
\[
\delta:=\frac{\eta_Q}{2P_{\max}}>0.
\]
If $\beta(x)\in[-\delta,0)$, then
\[
g(\alpha(x),\beta(x))
=
\frac{\alpha(x)}{2|\beta(x)|}
\ge
\frac{\eta_Q}{2\delta}
=
P_{\max},
\]
hence $f_P(x)=P_{\max}$ whenever $\beta(x)\ge -\delta$.

On the region $\{\beta\le -\delta\}$, $g$ is smooth and
\[
\left|\frac{\partial g}{\partial \alpha}\right|\le \frac{1}{2\delta},
\qquad
\left|\frac{\partial g}{\partial \beta}\right|\le \frac{B_\alpha}{2\delta^2}.
\]
Thus, by the mean value theorem, for $\beta(x),\beta(y)\le -\delta$,
\[
|g(\alpha(x),\beta(x))-g(\alpha(y),\beta(y))|
\le
\frac{1}{2\delta}|\alpha(x)-\alpha(y)|
+
\frac{B_\alpha}{2\delta^2}|\beta(x)-\beta(y)|.
\]
Since $u\mapsto [u]_{[P_{\min},P_{\max}]}$ is $1$-Lipschitz, the same bound holds for
$|f_P(x)-f_P(y)|$, giving
\[
|f_P(x)-f_P(y)|
\le
\left(\frac{L_\alpha}{2\delta}+\frac{B_\alpha L_\beta}{2\delta^2}\right)\|x-y\|_2,
\qquad\text{whenever }\beta(x),\beta(y)\le -\delta.
\]

\paragraph{Step 3: mixed-region pairs and conclusion.}
If $\beta(x)\ge -\delta$ and $\beta(y)\le -\delta$, then $f_P(x)=P_{\max}$ and
\[
P_{\max}-f_P(y)
\le
g(\alpha(y),-\delta)-g(\alpha(y),\beta(y))
\le
\frac{B_\alpha}{2\delta^2}\big(|\beta(y)|-\delta\big)
\le
\frac{B_\alpha}{2\delta^2}|\beta(x)-\beta(y)|
\le
\frac{B_\alpha L_\beta}{2\delta^2}\|x-y\|_2.
\]
The symmetric case is identical. Combining all cases, we may take
\[
L_P:=\frac{L_\alpha}{2\delta}+\frac{B_\alpha L_\beta}{2\delta^2},
\]
which proves global Lipschitzness of $f_P$ on $\mathcal S(\eta_Q,\eta_V)$.
\end{myproof}

\begin{lemma}[Fixed Lipschitz tube for $h$]
\label{lem:tube_lip_appN}
Let $\Sigma_{\exp,\min}:=\min_{i\in[N]}(\Sigma_{\exp})_{ii}$ and set
\[
\eta_V:=\frac{\Sigma_{\exp,\min}}{2\tau_+},
\qquad
\eta_Q:=\frac{Q_{\min}}{2}.
\]
Let
\[
\Gamma:=\{M(\tau):\tau\in[1,\tau_+]\}.
\]
Then there exists $R\in(0,1]$ and constants $L_h,H_h<\infty$ such that on the tube
\[
\Gamma_R:=
\left\{
x\in\R^N\times\R^N\times\R^{N\times N}\times\R^N:\dist(x,\Gamma)\le R
\right\},
\]
the following hold:
\begin{enumerate}
\item[(i)] For every $x\in\Gamma_R$ and every $i$, the local block
\[
x_i^{\mathrm{loc}}:=(x_i^P,x_i^Q,x_{ii}^{PP},x_i^{PQ})
\]
lies in $\mathcal S(\eta_Q,\eta_V)$.
\item[(ii)] $h$ is $L_h$-Lipschitz and bounded by $H_h$ on $\Gamma_R$:
\[
\|h(x)-h(y)\|_2\le L_h\|x-y\|_2\quad\forall x,y\in\Gamma_R,
\qquad
\sup_{x\in\Gamma_R}\|h(x)\|_2\le H_h.
\]
\end{enumerate}
\end{lemma}

\begin{myproof}[Proof of Lemma~\ref{lem:tube_lip_appN}]
\textbf{Step 1: margins on the path.}
By Lemma~\ref{lem:ode_bounds_appN}, for all $\tau\in[1,\tau_+]$,
\[
M_i^Q(\tau)\ge Q_{\min}=2\eta_Q.
\]
By Lemma~\ref{lem:var_floor_appN}, for all $\tau\in[1,\tau_+]$,
\[
M_{ii}^{PP}(\tau)-(M_i^P(\tau))^2 \ge \frac{(\Sigma_{\exp})_{ii}}{\tau_+}\ge 2\eta_V.
\]
Also Lemma~\ref{lem:ode_bounds_appN} gives the remaining coordinate bounds on $\Gamma$.

\textbf{Step 2: choose $R$ so margins persist.}
Define the blockwise maps
\[
q(x_i^{\mathrm{loc}}):=x_i^Q,
\qquad
v(x_i^{\mathrm{loc}}):=x_{ii}^{PP}-(x_i^P)^2.
\]
On the bounded set $0\le x_i^P\le P_{\max}+1$ we have
\[
\|\nabla v(x_i^{\mathrm{loc}})\|_2
=
\|(-2x_i^P,0,1,0)\|_2
\le
\sqrt{4(P_{\max}+1)^2+1}
=:L_v^{(1)},
\]
so $v$ is $L_v^{(1)}$-Lipschitz there, while $q$ is $1$-Lipschitz.
Choose
\[
R:=
\min\left\{
1,\ \eta_Q,\ \frac{\eta_V}{L_v^{(1)}},\ P_{\min},\ P_{\min}Q_{\min}
\right\}.
\]
If $x\in\Gamma_R$, pick $\tau\in[1,\tau_+]$ with $\|x-M(\tau)\|_2\le R$. Then for each $i$,
\[
\|x_i^{\mathrm{loc}}-M_i^{\mathrm{loc}}(\tau)\|_2\le R,
\]
hence
\[
x_i^Q
\ge
M_i^Q(\tau)-\|x_i^{\mathrm{loc}}-M_i^{\mathrm{loc}}(\tau)\|_2
\ge
2\eta_Q-R
\ge
\eta_Q,
\]
and
\[
x_{ii}^{PP}-(x_i^P)^2
\ge
\big(M_{ii}^{PP}(\tau)-(M_i^P(\tau))^2\big)
-
L_v^{(1)}\|x_i^{\mathrm{loc}}-M_i^{\mathrm{loc}}(\tau)\|_2
\ge
2\eta_V-L_v^{(1)}R
\ge
\eta_V.
\]
Moreover,
\[
x_i^P
\ge
M_i^P(\tau)-\|x_i^{\mathrm{loc}}-M_i^{\mathrm{loc}}(\tau)\|_2
\ge
P_{\min}-R
\ge
0,
\]
and
\[
x_i^P\le P_{\max}+R\le P_{\max}+1.
\]
The same coordinatewise comparison gives
\[
0\le x_{ii}^{PP}\le (P_{\max}+1)^2,
\qquad
0\le x_i^{PQ}\le (P_{\max}+1)(Q_{\max}+1),
\]
after shrinking by the above choice of $R$, because $\Gamma$ satisfies the bounds in
Lemma~\ref{lem:ode_bounds_appN} and each coordinate is $1$-Lipschitz in Euclidean distance.
This proves (i).

\textbf{Step 3: Lipschitzness of $h$ on the tube.}
By (i) and Lemma~\ref{lem:price_lip_appN}, each local block map
$x_i^{\mathrm{loc}}\mapsto f_P(x_i^{\mathrm{loc}})$ is $L_P$-Lipschitz on $\Gamma_R$.
Therefore the vector price map $\pi$ satisfies
\[
\|\pi(x)-\pi(y)\|_2^2
=
\sum_{i=1}^N |f_P(x_i^{\mathrm{loc}})-f_P(y_i^{\mathrm{loc}})|^2
\le
L_P^2\sum_{i=1}^N\|x_i^{\mathrm{loc}}-y_i^{\mathrm{loc}}\|_2^2
\le
L_P^2\|x-y\|_2^2,
\]
so
\[
\|\pi(x)-\pi(y)\|_2\le L_P\|x-y\|_2.
\]

Define the linear expected-demand map $d:[P_{\min},P_{\max}]^N\to\R^N$ by
\[
d_i(p) := a-bp_i+\frac{c}{N-1}\sum_{j\ne i}p_j.
\]
Its matrix has diagonal entries $-b$ and off-diagonal entries $c/(N-1)$, hence its induced
$1$- and $\infty$-norms are both $b+c$, so
\[
\|d(p)-d(p')\|_2\le (b+c)\|p-p'\|_2.
\]

Now define
\[
\Phi:\R^N\to\R^N\times\R^N\times\R^{N\times N}\times\R^N
\]
by
\[
\Phi(p):=\bigl(p,\ d(p),\ pp^\top,\ p\odot d(p)\bigr).
\]
Using the bounds $0\le p_i\le P_{\max}$ and $Q_{\min}\le d_i(p)\le Q_{\max}$, one checks that
for all $p,p'\in[P_{\min},P_{\max}]^N$,
\[
\|\Phi(p)-\Phi(p')\|_2
\le
L_\Phi \|p-p'\|_2,
\]
where one may take
\[
L_\Phi
:=
\sqrt{
1+(b+c)^2+(2\sqrt{N}P_{\max})^2+
\big(P_{\max}(b+c)+Q_{\max}\big)^2
}.
\]
Indeed,
\[
\|pp^\top-p'p'^\top\|_F
\le
\|(p-p')p^\top\|_F+\|p'(p-p')^\top\|_F
\le
2\sqrt{N}P_{\max}\|p-p'\|_2,
\]
and
\[
\|p\odot d(p)-p'\odot d(p')\|_2
\le
Q_{\max}\|p-p'\|_2+P_{\max}\|d(p)-d(p')\|_2.
\]
Since $f(x)=\Phi(\pi(x))$, it follows that
\[
\|f(x)-f(y)\|_2
\le
L_\Phi\|\pi(x)-\pi(y)\|_2
\le
L_\Phi L_P\|x-y\|_2.
\]
Finally,
\[
\|h(x)-h(y)\|_2
=
\|(f(x)-x)-(f(y)-y)\|_2
\le
\|f(x)-f(y)\|_2+\|x-y\|_2
\le
(L_\Phi L_P+1)\|x-y\|_2.
\]
Thus (ii) holds with
\[
L_h:=L_\Phi L_P+1.
\]
Boundedness $\sup_{x\in\Gamma_R}\|h(x)\|_2<\infty$ follows from continuity of $h$ and compactness
of $\Gamma_R$, and we denote this bound by $H_h$.
\end{myproof}

\paragraph{Step 1: log-time reparametrization and a deterministic Euler scheme.}
Define the log-time variable $\ell:=\log \tau$, so $\tau=e^\ell$, and the reparametrized mean-field trajectory
\[
M^\ell(\ell):=M(e^\ell),\qquad \ell\ge 0.
\]
Since $M(\cdot)$ solves $\dot M(\tau)=\frac{1}{\tau}h(M(\tau))$, the chain rule gives
\[
\frac{d}{d\ell}M^\ell(\ell)=h(M^\ell(\ell)),\qquad M^\ell(0)=M(1)=\mu_X.
\]
Thus $M^\ell$ solves the autonomous ODE
\begin{equation}\label{eq:logtime_ode}
\frac{d}{d\ell}M^\ell(\ell)=h(M^\ell(\ell)),\qquad M^\ell(0)=\mu_X.
\end{equation}

Fix $m$ and abbreviate $K:=K_m$ and $n:=n_m$. For calendar times $s\ge K$, define the accumulated
step size
\[
S_{K,s}:=\sum_{r=K+1}^{s}\frac1r,\qquad S_{K,K}=0.
\]
For all $s\in\{K,\dots,n\}$,
\[
S_{K,s}\le \int_{K}^{s}\frac{dx}{x}=\log\!\Big(\frac{s}{K}\Big)
\le \log\!\Big(\frac{n}{K}\Big)\le \log \tau_+,
\]
for all sufficiently large $m$. Hence $S_{K,s}\in[0,\log\tau_+]$ and
\[
M^\ell(S_{K,s})\in \Gamma:=\{M(\tau):\tau\in[1,\tau_+]\}
\qquad\text{for all }s\in\{K,\dots,n\}.
\]

Define the deterministic Euler scheme for \eqref{eq:logtime_ode} on the grid $\{S_{K,s}\}$:
\[
Y_{K,K}:=\mu_X,\qquad
Y_{K,s+1}:=Y_{K,s}+\frac{1}{s+1}h(Y_{K,s}),\qquad s\ge K.
\]
Define the Euler error
\[
e_s:=Y_{K,s}-M^\ell(S_{K,s}),\qquad s\ge K,
\]
so $e_K=0$.

We now prove the uniform bound \eqref{eq:euler_error_bound}. The argument uses only that on the
tube $\Gamma_R$ from Lemma~\ref{lem:tube_lip_appN}, $h$ is $L_h$-Lipschitz and bounded by $H_h$.

\medskip
\noindent\textbf{Step 1a (local truncation error for the exact flow).}
For each $s\ge K$,
\[
M^\ell(S_{K,s+1})-M^\ell(S_{K,s})
=
\int_{S_{K,s}}^{S_{K,s+1}} h(M^\ell(u))\,du.
\]
Write
\[
\Delta_s:=S_{K,s+1}-S_{K,s}=\frac{1}{s+1}.
\]
Add and subtract $\Delta_s\,h(M^\ell(S_{K,s}))$ to obtain
\[
M^\ell(S_{K,s+1})
=
M^\ell(S_{K,s})+\Delta_s\,h(M^\ell(S_{K,s}))+\rho_{s+1},
\]
where
\[
\rho_{s+1}
:=
\int_{S_{K,s}}^{S_{K,s+1}}
\Big(h(M^\ell(u))-h(M^\ell(S_{K,s}))\Big)\,du.
\]
Since $M^\ell(u)\in\Gamma\subseteq \Gamma_R$ and $\|h(\cdot)\|\le H_h$ on $\Gamma_R$, for
$u\in[S_{K,s},S_{K,s+1}]$,
\[
\|h(M^\ell(u))-h(M^\ell(S_{K,s}))\|_2
\le
L_h\|M^\ell(u)-M^\ell(S_{K,s})\|_2
\le
L_hH_h\,(u-S_{K,s}),
\]
and therefore
\[
\|\rho_{s+1}\|_2
\le
\int_{S_{K,s}}^{S_{K,s+1}} L_hH_h\,(u-S_{K,s})\,du
=
\frac{L_hH_h}{2}\,\Delta_s^2
=
\frac{L_hH_h}{2}\,\frac{1}{(s+1)^2}.
\]

\medskip
\noindent\textbf{Step 1b (error recursion, localized to the tube).}
Define the deterministic exit time from the tube
\[
s_{\mathrm{E}}
:=
\inf\{s\in\{K,K+1,\dots,n\}:\ Y_{K,s}\notin \Gamma_R\},
\qquad
(\inf\emptyset:=n+1).
\]
For $s<s_{\mathrm{E}}$ we have $Y_{K,s}\in\Gamma_R$ and
$M^\ell(S_{K,s})\in\Gamma\subseteq\Gamma_R$, so
\begin{align*}
e_{s+1}
&=Y_{K,s+1}-M^\ell(S_{K,s+1})\\
&=\Big(Y_{K,s}+\Delta_s h(Y_{K,s})\Big)
-\Big(M^\ell(S_{K,s})+\Delta_s h(M^\ell(S_{K,s}))+\rho_{s+1}\Big)\\
&=e_s+\Delta_s\Big(h(Y_{K,s})-h(M^\ell(S_{K,s}))\Big)-\rho_{s+1}.
\end{align*}
Taking norms and using Lipschitzness of $h$ on $\Gamma_R$ gives, for all $s<s_{\mathrm{E}}$,
\begin{equation}\label{eq:e_recursion}
\|e_{s+1}\|_2
\le
\Big(1+\frac{L_h}{s+1}\Big)\|e_s\|_2
+
\frac{L_hH_h}{2}\,\frac{1}{(s+1)^2}.
\end{equation}

\medskip
\noindent\textbf{Step 1c (solve the recursion).}
Let $a_s:=\|e_s\|_2$ and $c:=L_hH_h/2$. From \eqref{eq:e_recursion}, for all
$s<s_{\mathrm{E}}$,
\[
a_{s+1}
\le
\Big(1+\frac{L_h}{s+1}\Big)a_s+\frac{c}{(s+1)^2},
\qquad
a_K=0.
\]
Unrolling yields, for $K\le s\le s_{\mathrm{E}}\wedge n$,
\[
a_s
\le
c\sum_{r=K+1}^{s}\frac{1}{r^2}
\prod_{j=r+1}^{s}\Big(1+\frac{L_h}{j}\Big).
\]
Moreover, $\log(1+L_h/j)\le L_h/j$ implies
\[
\prod_{j=r+1}^{s}\Big(1+\frac{L_h}{j}\Big)
\le
\exp\!\left(\sum_{j=r+1}^{s}\frac{L_h}{j}\right)
\le
\left(\frac{s}{r}\right)^{L_h}
\le
\left(\frac{n}{K}\right)^{L_h}
\le
\tau_+^{L_h},
\]
for all sufficiently large $m$. Hence, for $s\le s_{\mathrm{E}}\wedge n$,
\[
\|e_s\|_2
\le
c\,\tau_+^{L_h}\sum_{r=K+1}^{n}\frac{1}{r^2}.
\]
Define
\[
C_{\mathrm{E}}:=\frac{L_hH_h}{2}\,\tau_+^{L_h}.
\]

\medskip
\noindent\textbf{Step 1d (the Euler iterates stay in the tube).}
Since $\sum_{r=K+1}^{n}r^{-2}\le \int_{K}^{\infty}x^{-2}\,dx=1/K$, we have
\[
\max_{K\le s\le s_{\mathrm{E}}\wedge n}\|e_s\|_2\le \frac{C_{\mathrm{E}}}{K}.
\]
Because $K_m\to\infty$, for all sufficiently large $m$ we have $C_{\mathrm{E}}/K\le R/2$.
Fix such an $m$. Then for all $s\le s_{\mathrm{E}}\wedge n$,
\[
\|Y_{K,s}-M^\ell(S_{K,s})\|_2=\|e_s\|_2\le R/2,
\]
and since $M^\ell(S_{K,s})\in\Gamma$, this implies
$Y_{K,s}\in\Gamma_{R/2}\subseteq\Gamma_R$ for all $s\le s_{\mathrm{E}}\wedge n$.
Thus $s_{\mathrm{E}}=n+1$, i.e., $Y_{K,s}\in\Gamma_R$ for all $s\in\{K,\dots,n\}$, and
\begin{equation}\label{eq:euler_error_bound}
\max_{K\le s\le n}\|Y_{K,s}-M^\ell(S_{K,s})\|_2
=
\max_{K\le s\le n}\|e_s\|_2
\le
\frac{C_{\mathrm{E}}}{K}.
\end{equation}

\paragraph{Step 2: SA vs.\ Euler (martingale perturbation) via localization.}
Let
\[
d_s:=\bar X_s-Y_{K,s}.
\]
Define the stopping time
\[
s_{\mathrm{SA}}
:=
\inf\{s\in\{K,K+1,\dots,n\}:\|d_s\|_2>R/2\},
\qquad
(\inf\emptyset:=\infty),
\]
and the stopped process
\[
\tilde d_s:=d_{s\wedge s_{\mathrm{SA}}}.
\]
On $\{s<s_{\mathrm{SA}}\}$ we have $\bar X_s,Y_{K,s}\in\Gamma_R$, hence
\[
\|h(\bar X_s)-h(Y_{K,s})\|_2\le L_h\|d_s\|_2.
\]

Using Lemma~\ref{lem:sa_recursion_start} and the Euler recursion for $Y_{K,s}$, for $s\ge K$ we have
\[
d_{s+1}
=
d_s+\frac{1}{s+1}\big(h(\bar X_s)-h(Y_{K,s})\big)
+
\frac{1}{s+1}\xi_{s+1}.
\]
Taking conditional expectation of $\|\tilde d_{s+1}\|_2^2$ given $\F_s$ and using
$\E[\xi_{s+1}\mid\F_s]=0$ yields
\[
\E\big[\|\tilde d_{s+1}\|_2^2\mid\F_s\big]
\le
\left(1+\frac{2L_h}{s+1}+\frac{L_h^2}{(s+1)^2}\right)\|\tilde d_s\|_2^2
+
\frac{1}{(s+1)^2}\E\big[\|\xi_{s+1}\|_2^2\mid\F_s\big].
\]
Using $\E[\|\xi_{s+1}\|_2^2\mid\F_s]\le \sigma_\xi^2$ and
$(s+1)^{-2}\le (s+1)^{-1}$, there is a constant $C_1$ such that
\[
\E\|\tilde d_{s+1}\|_2^2
\le
\left(1+\frac{C_1}{s+1}\right)\E\|\tilde d_s\|_2^2
+
\frac{\sigma_\xi^2}{(s+1)^2}.
\]
Iterating and using
\[
\sum_{r=K+1}^{n} \frac1r \le \log\tau_+,
\qquad
\sum_{r=K+1}^{n}\frac{1}{r^2}\le \frac{1}{K},
\]
gives
\begin{equation}\label{eq:stopped_L2}
\E\|\tilde d_{n}\|_2^2
\le
C_2\left(\E\|d_{K}\|_2^2+\frac{1}{K}\right),
\end{equation}
where $C_2<\infty$ depends only on $C_1,\sigma_\xi^2,\tau_+$. Since
$d_K=\bar X_K-\mu_X$, Lemma~\ref{lem:sa_recursion_start} gives
\[
\E\|d_K\|_2^2=O(1/K),
\]
hence
\[
\E\|\tilde d_n\|_2^2=O(1/K).
\]
Therefore
\[
\Pr(s_{\mathrm{SA}}\le n)
\le
\Pr\!\left(\|\tilde d_n\|_2\ge \frac{R}{2}\right)
\le
\frac{4}{R^2}\E\|\tilde d_n\|_2^2
\longrightarrow 0.
\]
On $\{s_{\mathrm{SA}}>n\}$ we have $d_n=\tilde d_n$, so
\[
d_n\xrightarrow{\P}0.
\]

\paragraph{Step 3: conclude $\bar X_{n_m}\to M(\bar\tau)$.}
By \eqref{eq:euler_error_bound} and $d_n\xrightarrow{\P}0$,
\[
\bar X_{n}-M^\ell(S_{K,n})
=
\underbrace{(\bar X_{n}-Y_{K,n})}_{d_n}
+
\underbrace{(Y_{K,n}-M^\ell(S_{K,n}))}_{e_n}
\xrightarrow{\P}0.
\]
It remains to identify the limit of $M^\ell(S_{K,n})$. Using the integral test for harmonic sums,
\[
\int_{K}^{n}\frac{dx}{x}
\le
\sum_{r=K+1}^{n}\frac1r
\le
\int_{K}^{n}\frac{dx}{x}+\frac{1}{K},
\]
so
\[
S_{K,n}
=
\log\!\left(\frac{n}{K}\right)+O\!\left(\frac{1}{K}\right),
\quad\text{hence}\quad
e^{S_{K,n}}
=
\frac{n}{K}\cdot \exp\!\left(O\!\left(\frac{1}{K}\right)\right)
\longrightarrow \bar\tau.
\]
By continuity of $M(\cdot)$,
\[
M^\ell(S_{K,n})=M(e^{S_{K,n}})\to M(\bar\tau).
\]
Therefore
\[
\bar X_{n_m}\xrightarrow{\P}M(\bar\tau),
\]
proving Lemma~\ref{lem:mean_field_convergence}.
\end{myproof}

\subsection{Proof of Lemma~\ref{lem:mean_field_to_price_moments}}
\label{app:ode_conversion_proof}

\begin{myproof}
Let $M(\cdot)$ be the mean-field ODE solution from Lemma~\ref{lem:mean_field_convergence}, and write
\[
M(\tau)=\bigl(M^P(\tau),M^Q(\tau),M^{PP}(\tau),M^{PQ}(\tau)\bigr).
\]
By definition of the mean-field moment coordinates, for $\tau\ge 1$ the price mean
$M^P(\tau)\in\R^N$ and raw price second-moment matrix $M^{PP}(\tau)\in\R^{N\times N}$ satisfy
\[
\dot M^P=\frac{P-M^P}{\tau},
\qquad
\dot M^{PP}=\frac{PP^\top-M^{PP}}{\tau},
\]
where $P(\tau)=\pi(M(\tau))$. Define
\[
U(\tau):=M^P(\tau),
\qquad
\Sigma_P(\tau):=M^{PP}(\tau)-U(\tau)U(\tau)^\top,
\qquad
V(\tau):=\tau\,\Sigma_P(\tau).
\]

\paragraph{Deriving the $(U,V)$ dynamics.}
From $U=M^P$ we immediately have
\[
\dot U=\frac{P-U}{\tau}.
\]
For $V=\tau\Sigma_P$, the product rule gives
\[
\dot V=\Sigma_P+\tau\,\dot\Sigma_P.
\]
Also
\[
\dot\Sigma_P
=
\dot M^{PP}-\dot U\,U^\top-U\,\dot U^\top.
\]
Substituting the ODEs for $\dot M^{PP}$ and $\dot U$ and using
$M^{PP}=\Sigma_P+UU^\top$ yields
\begin{align*}
\tau\,\dot\Sigma_P
&=
PP^\top-(\Sigma_P+UU^\top)
-(P-U)U^\top-U(P-U)^\top \\
&=
(P-U)(P-U)^\top-\Sigma_P.
\end{align*}
Therefore
\[
\dot V
=
\Sigma_P+\big((P-U)(P-U)^\top-\Sigma_P\big)
=
(P-U)(P-U)^\top.
\]

\paragraph{Expressing $P(\tau)$ as a function of $(U,V)$.}
Fix $i\in[N]$ and write
\[
\bar U_{-i}:=\frac{1}{N-1}\sum_{j\ne i}U_j,
\qquad
\bar V_{i,-i}:=\frac{1}{N-1}\sum_{j\ne i}V_{ij}.
\]
The mean-field moment coordinates inherit the linear moment identities from the demand model:
\[
M_i^Q=a-b\,U_i+c\,\bar U_{-i},
\]
and
\[
M_i^{PQ}
=
a\,U_i-b\,M^{PP}_{ii}
+c\,\overline{M^{PP}}_{i,-i},
\qquad
\overline{M^{PP}}_{i,-i}
:=
\frac{1}{N-1}\sum_{j\ne i}M^{PP}_{ij}.
\]
Since
\[
M^{PP}=UU^\top+\frac{1}{\tau}V,
\]
we have
\[
M^{PP}_{ii}=U_i^2+\frac{1}{\tau}V_{ii},
\qquad
\overline{M^{PP}}_{i,-i}
=
U_i\bar U_{-i}+\frac{1}{\tau}\bar V_{i,-i}.
\]
Substituting into $M_i^{PQ}$ gives
\[
M_i^{PQ}
=
U_i\,M_i^Q
+
\frac{1}{\tau}\big(-bV_{ii}+c\,\bar V_{i,-i}\big),
\]
so
\[
M_i^{PQ}-M_i^P M_i^Q
=
\frac{-bV_{ii}+c\,\bar V_{i,-i}}{\tau}.
\]
Also
\[
M^{PP}_{ii}-(M_i^P)^2=\frac{V_{ii}}{\tau}.
\]
Hence the OLS slope estimate in moment form satisfies
\[
\hat\beta_i
=
\frac{M_i^{PQ}-M_i^P M_i^Q}{M^{PP}_{ii}-(M_i^P)^2}
=
\frac{-bV_{ii}+c\,\bar V_{i,-i}}{V_{ii}}.
\]

Thus:
\begin{itemize}
\item If $-bV_{ii}+c\,\bar V_{i,-i}\ge 0$, equivalently $\hat\beta_i\ge 0$, then predicted
profit $p_i(\hat\alpha_i+\hat\beta_i p_i)$ is nondecreasing at the relevant upper endpoint in the
sense established in Lemma~\ref{lem:ols_price_from_moments}, and the selected price is $P_{\max}$.

\item If $-bV_{ii}+c\,\bar V_{i,-i}<0$, equivalently $\hat\beta_i<0$, the unconstrained maximizer is
$-\hat\alpha_i/(2\hat\beta_i)$. Using
\[
\hat\alpha_i=M_i^Q-\hat\beta_i U_i,
\qquad
M_i^Q=a-bU_i+c\bar U_{-i},
\]
and the moment identities above, a direct substitution gives
\[
\widetilde P_i(U,V)
=
\frac{(a+c\,\bar U_{-i})V_{ii}-c\,U_i\,\bar V_{i,-i}}
{2\bigl(bV_{ii}-c\,\bar V_{i,-i}\bigr)}.
\]
Enforcing feasibility gives
\[
P_i=[\widetilde P_i(U,V)]_{[P_{\min},P_{\max}]}.
\]
\end{itemize}
This is exactly the pricing map stated in Lemma~\ref{lem:mean_field_to_price_moments}, so
$P(\tau)=\pi(M(\tau))$ is the posted-price coordinate of the price-moments ODE.

\paragraph{Initial conditions.}
Because $U=M^P$, we have
\[
U(1)=M^P(1)=\mu.
\]
Also,
\[
V(1)
=
1\cdot\bigl(M^{PP}(1)-U(1)U(1)^\top\bigr)
=
(\mu\mu^\top+\Sigma_{\exp})-\mu\mu^\top
=
\Sigma_{\exp}.
\]
Therefore $(U,V)$ solves
\[
    U(1)=\mu,
    \qquad
    V(1)=\Sigma_{\exp},
    \qquad
    \dot U=\frac{P-U}{\tau},
    \qquad
    \dot V=(P-U)(P-U)^\top,
\]
with the stated price map.

Finally, the recovery formulas follow from the same identities:
\[
    M_i^P(\tau)=U_i(\tau),
    \qquad
    M_i^Q(\tau)=a-bU_i(\tau)+c\bar U_{-i}(\tau),
\]
\[
    M^{PP}(\tau)=U(\tau)U(\tau)^\top+\frac{1}{\tau}V(\tau),
\]
and
\[
    M_i^{PQ}(\tau)
    =
    U_i(\tau)M_i^Q(\tau)
    +
    \frac{1}{\tau}\Bigl(-bV_{ii}(\tau)+c\bar V_{i,-i}(\tau)\Bigr).
\]
These dynamics and the posted-price coordinate $P(\tau)$ coincide with Definition~\ref{def:price_moments_ode}.
\end{myproof}
\section{Omitted Steps and Lemmas in Theorem~\ref{thm:supra_sufficient} (Sections~\ref{sec:supra_competitive_limiting_prices}, \ref{sec:supracomp_proof})}
\label{app:omitted_pf_supra_sufficient}

This appendix completes the proof of Theorem~\ref{thm:supra_sufficient} which was outlined in Section~\ref{sec:supracomp_proof}.
Throughout, $(U(\tau),V(\tau))$ denotes the solution to the price-moments ODE from Definition~\ref{def:price_moments_ode}, $P(\tau)$ denotes the corresponding posted-price vector, and
\[
e(\tau):=P(\tau)-U(\tau).
\]
We also write
\[
\bar U_{-i}:=\frac{1}{N-1}\sum_{j\ne i}U_j,
\qquad
\bar V_{i,-i}:=\frac{1}{N-1}\sum_{j\ne i}V_{ij}.
\]
The proof below strings together the main-text lemmas and states the remaining auxiliary lemmas at the points where they are used.
The proofs of all lemmas follow in the subsequent subsections: first the lemmas stated in the main text, then the auxiliary lemmas stated in this appendix.

\begin{myproof}
We first consider the lower cone, $\mu\in\mathcal{C}^-$.
At $\tau=1$, $V(1)=\Sigma_{\exp}$ is diagonal, so $\bar V_{i,-i}(1)=0$.
By Lemma~\ref{lem:corr_bias_jackie}, the initial OLS price is the correct best response:
\[
P_i(1)=\BR(\bar\mu_{-i}).
\]
The cap is inactive because $\mu_i<\BR(\bar\mu_{-i})$ implies $\BR(\bar\mu_{-i})>P_{\min}$, while
\[
\BR(\bar\mu_{-i})\leq \BR(P_{\max})<P_{\max}
\]
under the standing price-bound assumptions.
Since $\mu\in\mathcal{C}^-$, we therefore have
\[
e_i(1)=P_i(1)-U_i(1)=\BR(\bar\mu_{-i})-\mu_i>0
\qquad\text{for all }i.
\]
Lemma~\ref{lem:drift_orthant_invariant_jackie} then implies
\[
e(\tau)\succeq0
\qquad
\text{for all }\tau\ge1.
\]
Thus $U$ is componentwise nondecreasing, since $\dot U=e/\tau\succeq0$.

The same-sign gap generates the cross-firm covariance needed for the correlation-bias argument.
Because the covariance state evolves as $\dot V=e e^\top$, common-sign price movements imply nonnegative off-diagonal covariances; strict common-sign movement at the start makes these cross-covariances strictly positive immediately afterward.

\begin{lemma}[Common-sign gaps generate positive cross-covariances]
\label{lemma:nonnegative_gap_implies_correlation}
Assume $V_{ij}(1)=0$ for all $i\ne j$.
If $e(\tau)\succeq0$ for all $\tau\ge1$ or $e(\tau)\preceq0$ for all $\tau\ge1$, then $V_{ij}(\tau)\ge0$ for all $i\ne j$ and all $\tau\ge1$.
If, moreover, $e(\tau_0)\succ0$ or $e(\tau_0)\prec0$ for some $\tau_0\ge1$, then
\[
\bar V_{i,-i}(\tau)>0
\qquad
\text{for all }i\in[N]\text{ and all }\tau>\tau_0.
\]
\end{lemma}

Applying Lemma~\ref{lemma:nonnegative_gap_implies_correlation} with $\tau_0=1$ gives
\[
\bar V_{i,-i}(\tau)\ge0
\qquad
\text{for all }i\text{ and all }\tau\ge1,
\]
with strict positivity for every $\tau>1$.
The hypotheses of Lemma~\ref{lem:weak_supra_from_bias_jackie} are therefore satisfied: $U$ is componentwise monotone and the average cross-covariances are nonnegative.
Hence the limit
\[
U^\infty:=\lim_{\tau\to\infty}U(\tau)
\]
exists and satisfies the Nash lower bound
\[
U^\infty\succeq \pNE\mathbf{1}.
\]

It remains to show that the inequality is strict.
The first reduction shows that, once the Nash lower bound is known, any equality in a single coordinate forces the entire limiting vector to be exactly Nash.

\begin{lemma}[Reduction to the Nash boundary]
\label{lem:reduce_to_NE_boundary_jackie}
Suppose $U^\infty:=\lim_{\tau\to\infty}U(\tau)$ exists, $\bar V_{i,-i}(\tau)\ge0$ for all $i$ and all $\tau\ge1$, and
\[
U^\infty\succeq \pNE\mathbf{1}.
\]
If $U_j^\infty=\pNE$ for some $j$, then $U^\infty=\pNE\mathbf{1}$.
\end{lemma}

Suppose, toward a contradiction, that $U^\infty$ is not componentwise strictly above Nash.
Then some coordinate satisfies $U_j^\infty=\pNE$, and Lemma~\ref{lem:reduce_to_NE_boundary_jackie} implies
\[
U^\infty=\pNE\mathbf{1}.
\]
We now rule out this boundary case.

The boundary argument uses one additional monotonicity property.
The forward-invariance lemma gives $P-U\succeq0$; the next two lemmas show that, on the lower cone, the posted prices themselves are componentwise nondecreasing.
First, once $\dot P\succeq0$, this sign is forward invariant.

\begin{lemma}[Forward invariance of price monotonicity on $\mathcal{C}^-$]
\label{lem:Pdot_sign_invariant}
Assume $P(\tau)-U(\tau)\succeq0$ for all $\tau\geq \tau_0$.
If $\dot P(\tau_0)\succeq0$ at some $\tau_0\geq1$, then $P$ is componentwise nondecreasing on $[\tau_0,\infty)$.
\end{lemma}

The hypothesis of Lemma~\ref{lem:Pdot_sign_invariant} is initialized at the start of exploitation.

\begin{lemma}[Initialization of price monotonicity on $\mathcal{C}^-$]
\label{lem:Pdot_initial_lower_cone}
Suppose $\mu\in\mathcal{C}^-$ and $U(1)=\mu$, $V(1)=\Sigma_{\exp}$, where $\Sigma_{\exp}$ is positive diagonal.
Then $P$ is right-differentiable at $\tau=1$ and
\[
\dot P(1^+)\succ0.
\]
\end{lemma}

By Lemmas~\ref{lem:Pdot_initial_lower_cone} and~\ref{lem:Pdot_sign_invariant}, $P$ is componentwise nondecreasing.
If $U(\tau)\to\pNE\mathbf{1}$, this monotonicity forces $P(\tau)$ to converge to the same Nash vector and eventually lie below it.
The next lemma then shows that the covariance matrix stabilizes, while the accumulated cross-covariances remain strictly positive.

\begin{lemma}[Covariance convergence on the Nash boundary]
\label{lem:Vconv_on_NE_jackie}
Suppose that $U(\tau)\to\pNE\mathbf{1}$, $P(\tau)-U(\tau)\succeq0$ for all $\tau\ge1$, $P$ is componentwise nondecreasing, and $\bar V_{i,-i}(\tau)\ge0$ for all $i$ and all $\tau\ge1$.
Assume also that $e(\tau_0)\succ0$ for some $\tau_0\ge1$.
Then:
\begin{enumerate}[label=(\arabic*),leftmargin=*]
    \item $P(\tau)\to\pNE\mathbf{1}$ and $P(\tau)\preceq\pNE\mathbf{1}$ for all sufficiently large $\tau$;
    \item each $V_{ii}(\tau)$ converges to a finite limit $V_{ii}^\infty$, and each average cross term
    \[
    \bar V_{i,-i}(\tau)=\frac{1}{N-1}\sum_{j\ne i}V_{ij}(\tau)
    \]
    converges to a strictly positive limit $\bar V_{i,-i}^\infty>0$.
\end{enumerate}
\end{lemma}

Applying Lemma~\ref{lem:Vconv_on_NE_jackie} with $\tau_0=1$ gives
\[
P(\tau)\to\pNE\mathbf{1},
\qquad
V_{ii}(\tau)\to V_{ii}^\infty,
\qquad
\bar V_{i,-i}(\tau)\to \bar V_{i,-i}^\infty>0.
\]
But this is incompatible with the OLS price map.
A strictly positive limiting cross-covariance creates a persistent upward omitted-variable bias at the Nash vector, so the posted price cannot converge to Nash.

\begin{lemma}[Limiting bias excludes the Nash boundary]
\label{lem:exclude_NE_by_bias_jackie}
Suppose $\mu\in\mathcal{C}^-$.
There is no lower-cone trajectory such that
\[
U(\tau)\to\pNE\mathbf{1},\qquad
P(\tau)\to\pNE\mathbf{1},
\qquad
V_{ii}(\tau)\to V_{ii}^\infty,
\qquad
\bar V_{i,-i}(\tau)\to \bar V_{i,-i}^\infty>0
\quad\text{for every }i.
\]
\end{lemma}

This contradicts Lemma~\ref{lem:exclude_NE_by_bias_jackie}.
Therefore the limiting running mean must satisfy
\[
U^\infty\succ\pNE\mathbf{1}.
\]
Since $e(\tau)=P(\tau)-U(\tau)\succeq0$ on $\mathcal{C}^-$, there exists $\tau_0<\infty$ such that, for all $\tau>\tau_0$,
\[
P^{\mathrm{ODE}}(\tau;\mu,\Sigma_{\exp})
=
P(\tau)
\succeq U(\tau)
\succ\pNE\mathbf{1}.
\]
This proves Theorem~\ref{thm:supra_sufficient}(b).

It remains to handle the upper cone, $\mu\in\mathcal{C}^+$.
The proof is the same in spirit, but the invariant orthant is reversed.
At $\tau=1$, $V(1)$ is diagonal, so $P_i(1)=\BR(\bar\mu_{-i})$.
Since $\mu\in\mathcal{C}^+$, we have
\[
P_i(1)-U_i(1)=\BR(\bar\mu_{-i})-\mu_i<0
\qquad
\text{for all }i.
\]
Thus prices initially move downward relative to trailing means.
The required sign-reversed forward-invariance statement is the following.

\begin{lemma}[Negative-orthant forward invariance]
\label{lem:negative_orthant_invariant_jackie}
Fix $\tau_0\geq1$.
If $P(\tau_0)-U(\tau_0)\preceq0$, then $P(\tau)-U(\tau)\preceq0$ for all $\tau\geq \tau_0$.
\end{lemma}

By Lemma~\ref{lem:negative_orthant_invariant_jackie}, $e(\tau)\preceq0$ for all $\tau\ge1$ on $\mathcal{C}^+$.
Since covariance accumulation depends on products $e_i e_j$, not on the sign of $e_i$ alone, Lemma~\ref{lemma:nonnegative_gap_implies_correlation} again gives positive cross-covariances for all $\tau>1$.
The upper-cone case is therefore simpler than the lower-cone case: because $\mu\in\mathcal{C}^+$ already implies $\mu\succ\pNE\mathbf{1}$, and the induced correlation bias gives pointwise prices above Nash, no separate exclusion of the Nash boundary is needed.
We record this as the final auxiliary lemma.

\begin{lemma}[Upper-cone strictness]
\label{lem:upper_cone_strictness_jackie}
If $\mu\in\mathcal{C}^+$, then
\[
P^{\mathrm{ODE}}(\tau;\mu,\Sigma_{\exp})\succ\pNE\mathbf{1}
\qquad
\text{for every }\tau\in[1,\infty).
\]
\end{lemma}

Lemma~\ref{lem:upper_cone_strictness_jackie} proves Theorem~\ref{thm:supra_sufficient}(a).
Together with the lower-cone argument above, this completes the proof of Theorem~\ref{thm:supra_sufficient}.
\end{myproof}

\subsection{Proof of Lemma~\ref{lem:corr_bias_jackie}}
\label{app:corr_bias_proof}

\begin{myproof}
Fix $i\in[N]$ and hold $U$ and $V_{ii}$ fixed. Write $x:=\bar V_{i,-i}$ and $A:=a+c\,\bar U_{-i}$. Using the algebraic OLS price from Definition~\ref{def:price_moments_ode}, write
\[
\widetilde P_i(x):=\frac{A\,V_{ii}-c\,U_i\,x}{2(bV_{ii}-c x)},
\qquad
\text{on the domain }bV_{ii}-cx>0.
\]

\emph{Part (1).}
If $x=0$, then
\[
\widetilde P_i(0)=\frac{A\,V_{ii}}{2bV_{ii}}=\frac{A}{2b}=\BR(\bar U_{-i}).
\]

\emph{Part (2).}
Differentiate $\widetilde P_i$ with respect to $x$ using the quotient rule. Let
$N(x):=A V_{ii}-cU_i x$ and $D(x):=2(bV_{ii}-cx)$. Then $N'(x)=-cU_i$ and $D'(x)=-2c$, so
\[
\widetilde P_i'(x)=\frac{N'(x)D(x)-N(x)D'(x)}{D(x)^2}
=\frac{(-cU_i)\,2(bV_{ii}-cx)-\big(A V_{ii}-cU_i x\big)(-2c)}{4(bV_{ii}-cx)^2}.
\]
Expanding the numerator and canceling the $x$-terms gives
\[
(-2cU_i bV_{ii}+2cU_i c x) + (2cA V_{ii}-2c^2U_i x)=2cV_{ii}(A-bU_i).
\]
Therefore,
\[
\widetilde P_i'(x)=\frac{2cV_{ii}(A-bU_i)}{4(bV_{ii}-cx)^2}.
\]
On the domain $bV_{ii}-cx>0$, the denominator is strictly positive. Also $V_{ii}>0$ because $V_{ii}(1)=(\Sigma_{\exp})_{ii}>0$ and $\dot V_{ii}= (P_i-U_i)^2\ge 0$. Finally,
\[
A-bU_i=a+c\bar U_{-i}-bU_i \ge a-bP_{\max}+cP_{\min}>0,
\]
because $U_i,\bar U_{-i}\in[P_{\min},P_{\max}]$, $P_{\min}>0$, and $P_{\max}\le a/b$. Hence $\widetilde P_i'(x)>0$ and $\widetilde P_i$ is strictly increasing in $\bar V_{i,-i}$.
\end{myproof}

\subsection{Proof of the nonnegative-orthant part of Lemma~\ref{lem:drift_orthant_invariant_jackie}}
\label{app:drift_orthant_invariant_proof}

\begin{myproof}
Let $e(\tau):=P(\tau)-U(\tau)$. The claim is that if $e(\tau_0)\succeq 0$ for some $\tau_0\ge 1$, then $e(\tau)\succeq 0$ for all $\tau\ge \tau_0$.

Assume for contradiction that the trajectory leaves $\R_+^N$ after $\tau_0$, and define the first exit time
\[
\tau^\star:=\inf\{\tau\ge \tau_0:\exists i\text{ with }e_i(\tau)<0\}.
\]
By continuity of $e(\cdot)$, we have $e(\tau^\star)\succeq 0$ and there exists at least one index $i$ with $e_i(\tau^\star)=0$. Fix such an $i$.

We rule out each possible way in which the $i$th coordinate could be the first one to cross into the negative orthant.

\paragraph{Clipped and non-OLS branches.}
First consider the branches on which the posted price is locally fixed at a cap. If $P_i(\tau^\star)=P_{\min}$ and $e_i(\tau^\star)=0$, then $U_i(\tau^\star)=P_{\min}$. As long as the lower clip remains active, $U_i$ solves
\[
    \dot U_i(\tau)=\frac{P_{\min}-U_i(\tau)}{\tau},
\]
with initial value $U_i(\tau^\star)=P_{\min}$, and hence $U_i(\tau)=P_{\min}$ on that right-neighborhood. Thus the lower clip cannot by itself make $e_i$ negative. If the lower clip ceases to bind, the posted price moves weakly upward from $P_{\min}=U_i(\tau^\star)$, which is also not a first exit from $\R_+^N$.

If $P_i(\tau^\star)=P_{\max}$ and $e_i(\tau^\star)=0$, then $U_i(\tau^\star)=P_{\max}$. While the upper clip, or the non-OLS branch $P_i\equiv P_{\max}$, remains active, feasibility gives $U_i(\tau)\le P_{\max}$ and hence $e_i(\tau)=P_{\max}-U_i(\tau)\ge0$. Therefore this branch cannot itself generate a negative gap. The only remaining possibility is that the trajectory leaves the upper clipped branch and crosses through the algebraic OLS boundary $\widetilde P_i=U_i$. That crossing is covered by the energy calculation below.

\paragraph{OLS boundary.}
It remains to rule out a crossing through an OLS boundary point with
\[
    B_i:=bV_{ii}-c\bar V_{i,-i}>0,
    \qquad \widetilde P_i(\tau^\star)-U_i(\tau^\star)=0.
\]
This includes the interior OLS case and the instant at which a clip ceases to bind. Define
\[
\mathcal E_i:=2bV_{ii}\bigl(U_i-\BR(\bar U_{-i})\bigr)-cU_i\,\bar V_{i,-i}.
\]
A direct rearrangement of the formula for $\widetilde P_i$ gives the exact identity
\[
\widetilde P_i-U_i=\frac{-\mathcal E_i}{2B_i}.
\]
Since $B_i>0$, crossing from $\widetilde P_i-U_i\ge0$ to $\widetilde P_i-U_i<0$ is equivalent to crossing from $\mathcal E_i\le0$ to $\mathcal E_i>0$. At time $\tau^\star$ we have $\mathcal E_i(\tau^\star)=0$.

We now compute $\dot{\mathcal E}_i(\tau^\star)$ under the information $e(\tau^\star)\succeq0$ and $e_i(\tau^\star)=0$.
First, $e(\tau^\star)\succeq0$ implies $\dot U_j(\tau^\star)=e_j(\tau^\star)/\tau^\star\ge0$ for all $j$, so
$\dot{\bar U}_{-i}(\tau^\star)=\frac{1}{N-1}\sum_{j\ne i}\dot U_j(\tau^\star)\ge0$.
Also $e_i(\tau^\star)=0$ implies $\dot U_i(\tau^\star)=0$, and from $\dot V=(P-U)(P-U)^\top$ we get
$\dot V_{ii}(\tau^\star)=e_i(\tau^\star)^2=0$ and $\dot{\bar V}_{i,-i}(\tau^\star)=e_i(\tau^\star)\bar e_{-i}(\tau^\star)=0$.

Differentiating $\mathcal E_i$ along the ODE gives
\[
\dot{\mathcal E}_i
=2b\,\dot V_{ii}\big(U_i-\BR(\bar U_{-i})\big)
+2bV_{ii}\Big(\dot U_i-\BR'(\bar U_{-i})\,\dot{\bar U}_{-i}\Big)
-c\,\dot U_i\,\bar V_{i,-i}-cU_i\,\dot{\bar V}_{i,-i}.
\]
Evaluating at $\tau^\star$ and using $\dot V_{ii}(\tau^\star)=\dot U_i(\tau^\star)=\dot{\bar V}_{i,-i}(\tau^\star)=0$ gives
\[
\dot{\mathcal E}_i(\tau^\star)=-2bV_{ii}(\tau^\star)\,\BR'(\bar U_{-i}(\tau^\star))\,\dot{\bar U}_{-i}(\tau^\star)
=-c\,V_{ii}(\tau^\star)\,\dot{\bar U}_{-i}(\tau^\star)\le0.
\]
Therefore, at a boundary point where $\mathcal E_i(\tau^\star)=0$, the derivative points toward $\{\mathcal E_i\le0\}$ and cannot cross into $\{\mathcal E_i>0\}$. Equivalently, $\widetilde P_i-U_i$ cannot become negative immediately after $\tau^\star$, hence $e_i$ cannot become negative immediately after $\tau^\star$.

\paragraph{Conclusion.}
In all cases, $e_i$ cannot be the first coordinate to turn negative, contradicting the definition of $\tau^\star$. Hence $e(\tau)\succeq 0$ for all $\tau\ge \tau_0$.
\end{myproof}

\subsection{Proof of Lemma~\ref{lem:weak_supra_from_bias_jackie}}

\begin{myproof}
Assume (i) $\bar V_{i,-i}(\tau)\ge 0$ for all $i$ and $\tau\ge 1$, and (ii) each coordinate $U_i(\tau)$ is monotone in $\tau$.
Since $U_i(\tau)\in[P_{\min},P_{\max}]$ for all $\tau$ (because $U$ is a running average of prices in $[P_{\min},P_{\max}]$), every monotone $U_i$ has a finite limit; write $U_i^\infty:=\lim_{\tau\to\infty}U_i(\tau)$ and $U^\infty:=(U_i^\infty)_{i=1}^N$.

\paragraph{Step 1: show $P_i(\tau)\ge \BR(\bar U_{-i}(\tau))$ for all $i,\tau$.}
Fix $i$ and $\tau$. If the non-OLS branch is active, then $P_i(\tau)=P_{\max}$. Under the standing feasibility condition $\pNE\le P_{\max}$ and monotonicity of $\BR$, we have $\BR(\bar U_{-i}(\tau))\le \BR(P_{\max})\le P_{\max}=P_i(\tau)$.
If instead the OLS branch is active, then $bV_{ii}-c\bar V_{i,-i}>0$ and Lemma~\ref{lem:corr_bias_jackie} gives
$\widetilde P_i(U,V)\ge \BR(\bar U_{-i})$ because $\bar V_{i,-i}\ge 0$ and equality holds at $\bar V_{i,-i}=0$.
Clipping cannot reduce the value below $\BR(\bar U_{-i})$ since $\BR(\bar U_{-i})\le P_{\max}$ and $P_i=[\widetilde P_i]_{[P_{\min},P_{\max}]}$ (if $\BR(\bar U_{-i})<P_{\min}$ then $P_i\ge P_{\min}>\BR(\bar U_{-i})$). Hence $P_i(\tau)\ge \BR(\bar U_{-i}(\tau))$ in all cases.

\paragraph{Step 2: rule out $\min_i U_i^\infty<\pNE$.}
Let $m^\infty:=\min_i U_i^\infty$ and choose an index $k$ with $U_k^\infty=m^\infty$.
Suppose for contradiction that $m^\infty<\pNE$. Define the continuous function $\psi(u,v):=\BR(u)-v$.
We have $\psi(m^\infty,m^\infty)=\BR(m^\infty)-m^\infty>0$ because $\BR(u)-u$ is affine with unique zero at $u=\pNE$.
Therefore we can pick $\varepsilon>0$ small enough that
\[
\eta:=\BR(m^\infty-\varepsilon)-(m^\infty+\varepsilon)>0.
\]
Since $U_k(\tau)\to m^\infty$ and $\bar U_{-k}(\tau)\to \bar U_{-k}^\infty\ge m^\infty$, there exists $\tau_1$ such that for all $\tau\ge \tau_1$,
$U_k(\tau)\le m^\infty+\varepsilon$ and $\bar U_{-k}(\tau)\ge m^\infty-\varepsilon$. Using Step~1 and monotonicity of $\BR$,
\[
e_k(\tau):=P_k(\tau)-U_k(\tau)\ \ge\ \BR(\bar U_{-k}(\tau)) - U_k(\tau)\ \ge\ \BR(m^\infty-\varepsilon)-(m^\infty+\varepsilon)=\eta,\qquad \tau\ge \tau_1.
\]
Then $\dot U_k(\tau)=e_k(\tau)/\tau\ge \eta/\tau$ for all $\tau\ge \tau_1$, so integrating gives
$U_k(\tau)\ge U_k(\tau_1)+\eta\log(\tau/\tau_1)$, which diverges as $\tau\to\infty$, contradicting convergence of $U_k(\tau)$.
Hence $m^\infty\ge \pNE$, i.e.\ $U^\infty\succeq \pNE\mathbf{1}$.
\end{myproof}

\subsection{Proof of Lemma~\ref{lemma:nonnegative_gap_implies_correlation}}

\begin{myproof}
Write $e(\tau):=P(\tau)-U(\tau)$.
For $i\ne j$, the ODE gives
\[
\dot V_{ij}(\tau)=e_i(\tau)e_j(\tau).
\]
If $e(\tau)\succeq0$ for all $\tau\ge1$ or $e(\tau)\preceq0$ for all $\tau\ge1$, then $e_i(\tau)e_j(\tau)\ge0$ for all $i\ne j$ and all $\tau\ge1$.
Thus $V_{ij}$ is nondecreasing and $V_{ij}(\tau)\ge V_{ij}(1)=0$.

Now assume $e(\tau_0)\succ0$; the case $e(\tau_0)\prec0$ is identical after reversing signs.
Fix any $\tau>\tau_0$. The map $(U,V)\mapsto P(U,V)$ is continuous (it is a rational map on the OLS region, composed with coordinatewise clipping, and equals the constant $P_{\max}$ on the non-OLS region), and $(U,V)$ is continuous in time because it solves an ODE with locally bounded right-hand side. Hence $e(\cdot)$ is continuous. Since $e(\tau_0)\succ0$, there exist $\delta_\tau\in(0,\tau-\tau_0]$ and $\varepsilon_\tau>0$ such that
$e_i(s)\ge\varepsilon_\tau$ and $e_j(s)\ge\varepsilon_\tau$ for all $s\in[\tau_0,\tau_0+\delta_\tau]$. Therefore
\[
V_{ij}(\tau)=V_{ij}(1)+\int_1^\tau e_i(s)e_j(s)\,ds
\ \ge\ \int_{\tau_0}^{\tau_0+\delta_\tau}\varepsilon_\tau^2\,ds
\ =\ \varepsilon_\tau^2\delta_\tau\ >\ 0.
\]
Since $\tau>\tau_0$ was arbitrary, $V_{ij}(\tau)>0$ for every $i\ne j$ and every $\tau>\tau_0$. Averaging over $j\ne i$ gives
\[
\bar V_{i,-i}(\tau)=\frac{1}{N-1}\sum_{j\ne i}V_{ij}(\tau)>0
\]
for all $i$ and all $\tau>\tau_0$.
\end{myproof}

\subsection{Proof of Lemma~\ref{lem:reduce_to_NE_boundary_jackie}}

\begin{myproof}
Assume $U^\infty\succeq \pNE\mathbf{1}$ and $U_j^\infty=\pNE$ for some $j$.
Suppose for contradiction that $U^\infty\neq \pNE\mathbf{1}$.
Then there exists $k\ne j$ with $U_k^\infty>\pNE$, so $\bar U_{-j}^\infty>\pNE$.
Because $\BR$ is strictly increasing and $\BR(\pNE)=\pNE$, we have $\BR(\bar U_{-j}^\infty)>\pNE$.
Define
\[
\eta:=\frac12\bigl(\BR(\bar U_{-j}^\infty)-\pNE\bigr)>0.
\]
By convergence, for all sufficiently large $\tau$ we have
\[
\BR(\bar U_{-j}(\tau))\ge \pNE+\eta,
\qquad
U_j(\tau)\le \pNE+\eta/2.
\]
Also $\bar V_{j,-j}(\tau)\ge0$ implies $P_j(\tau)\ge \BR(\bar U_{-j}(\tau))$, by the same argument as in the proof of Lemma~\ref{lem:weak_supra_from_bias_jackie}.
Hence, for all sufficiently large $\tau$,
\[
e_j(\tau):=P_j(\tau)-U_j(\tau)
\ge
(\pNE+\eta)-(\pNE+\eta/2)
=
\eta/2.
\]
Therefore $\dot U_j(\tau)=e_j(\tau)/\tau\ge(\eta/2)/\tau$ for all large $\tau$, which implies $U_j(\tau)$ cannot converge to $\pNE$.
This contradiction shows that no such $k$ exists, hence $U^\infty=\pNE\mathbf{1}$.
\end{myproof}

\subsection{Proof of Lemma~\ref{lem:Pdot_sign_invariant}}

\begin{myproof}
Assume $e(\tau):=P(\tau)-U(\tau)\succeq 0$ for all $\tau\ge \tau_0$, as in the nonnegative-orthant part of Lemma~\ref{lem:drift_orthant_invariant_jackie}.
Fix $\tau_0\ge 1$ and assume $\dot P(\tau_0)\succeq 0$.

Because $P$ is defined by a piecewise-smooth map of $(U,V)$ (rational on the OLS region, constant on the non-OLS region, and coordinatewise clipped), it suffices to prove that on any open time interval on which \emph{(i)} every firm is on the OLS branch ($B_i>0$) and \emph{(ii)} clipping is inactive ($P_i=\widetilde P_i\in(P_{\min},P_{\max})$), the condition $\dot P\succeq 0$ is forward invariant. Once proved, crossing into the clipped or non-OLS regimes cannot create a negative $\dot P_i$ (in those regimes $P_i$ is locally constant), so the result extends to all $\tau\ge \tau_0$.

\paragraph{Step 1: compute $\dot P_i$ and isolate its sign.}
Work on an interval where $P_i=\widetilde P_i\in(P_{\min},P_{\max})$ and $B_i:=bV_{ii}-c\,\bar V_{i,-i}>0$ for every $i$.
Write $A_i:=a+c\,\bar U_{-i}$ and recall
\[
P_i=\widetilde P_i=\frac{A_iV_{ii}-cU_i\,\bar V_{i,-i}}{2B_i}.
\]
Define also the (strictly positive) conditional expected demand at $(U_i,\bar U_{-i})$:
\[
D_i:=A_i-bU_i=a+c\,\bar U_{-i}-bU_i>0,
\]
which holds because $U_i,\bar U_{-i}\in[P_{\min},P_{\max}]$ and $a-bP_{\max}+cP_{\min}>0$.

Differentiate $P_i$ along the ODE.
First, the reduced ODE gives $\dot U_i=e_i/\tau$ and $\dot{\bar U}_{-i}=\bar e_{-i}/\tau$, where $\bar e_{-i}:=\frac{1}{N-1}\sum_{j\ne i}e_j$.
Also $\dot V_{ii}=e_i^2$ and $\dot{\bar V}_{i,-i}=e_i\bar e_{-i}$ because $\dot V=e e^\top$.

Let $N_i:=A_iV_{ii}-cU_i\bar V_{i,-i}$, so $P_i=N_i/(2B_i)$.
Then $\dot P_i=\frac{\dot N_i B_i-N_i\dot B_i}{2B_i^2}$.
Compute
\[
\dot A_i=c\,\dot{\bar U}_{-i}=\frac{c}{\tau}\bar e_{-i},
\qquad
\dot B_i=b\,\dot V_{ii}-c\,\dot{\bar V}_{i,-i}=b e_i^2-c e_i\bar e_{-i}=e_i(b e_i-c\bar e_{-i}),
\]
and
\[
\dot N_i=\dot A_i V_{ii}+A_i\dot V_{ii}-c\,\dot U_i\,\bar V_{i,-i}-cU_i\,\dot{\bar V}_{i,-i}
=\frac{c}{\tau}V_{ii}\bar e_{-i}+A_i e_i^2-\frac{c}{\tau}\bar V_{i,-i}e_i-cU_i e_i\bar e_{-i}.
\]
Substituting into $\dot P_i$ and grouping terms yields the factorization
\[
\dot P_i=\frac{c}{2\tau\,B_i^2}\,\ell_i\,k_i,
\qquad
\ell_i:=B_i+\tau\,e_iD_i,
\qquad
k_i:=V_{ii}\bar e_{-i}-\bar V_{i,-i}e_i.
\]
(One can verify this by expanding $\dot N_i B_i-N_i\dot B_i$, then using $B_i=bV_{ii}-c\bar V_{i,-i}$ to rewrite the bracketed expression as $cD_i\,k_i$; the remaining $(c/\tau)B_i k_i$ term comes from the $\dot A_i$ contribution.)
Under $e_i\ge0$, $B_i>0$, and $D_i>0$, we have $\ell_i>0$.
Hence, on this interior OLS interval,
\[
\operatorname{sign}(\dot P_i)=\operatorname{sign}(k_i)
\qquad\text{for each }i.
\]

\paragraph{Step 2: derive the dynamics of $k$ and show $k\succeq0$ is forward invariant.}
Differentiate $k_i=V_{ii}\bar e_{-i}-\bar V_{i,-i}e_i$:
\[
\dot k_i=\dot V_{ii}\bar e_{-i}+V_{ii}\dot{\bar e}_{-i}-\dot{\bar V}_{i,-i}e_i-\bar V_{i,-i}\dot e_i.
\]
Using $\dot V_{ii}=e_i^2$ and $\dot{\bar V}_{i,-i}=e_i\bar e_{-i}$, the first and third terms cancel:
\[
\dot V_{ii}\bar e_{-i}-\dot{\bar V}_{i,-i}e_i
=
e_i^2\bar e_{-i}-e_i\bar e_{-i}e_i
=
0.
\]
Thus
\[
\dot k_i=V_{ii}\dot{\bar e}_{-i}-\bar V_{i,-i}\dot e_i.
\]
Now $\dot e_i=\dot P_i-\dot U_i=\dot P_i-e_i/\tau$ and
$\dot{\bar e}_{-i}=\dot{\bar P}_{-i}-\dot{\bar U}_{-i}=\dot{\bar P}_{-i}-\bar e_{-i}/\tau$, where
\[
\dot{\bar P}_{-i}:=\frac{1}{N-1}\sum_{j\ne i}\dot P_j.
\]
Therefore
\[
\dot k_i
=
V_{ii}\dot{\bar P}_{-i}-\bar V_{i,-i}\dot P_i
-\frac{1}{\tau}\bigl(V_{ii}\bar e_{-i}-\bar V_{i,-i}e_i\bigr)
=
V_{ii}\dot{\bar P}_{-i}-\bar V_{i,-i}\dot P_i-\frac{1}{\tau}k_i.
\]
Substituting
\[
\dot P_j=\frac{c}{2\tau B_j^2}\ell_j k_j
\]
gives a linear system $\dot k=A(\tau)k$ with Metzler off-diagonal structure: for $j\ne i$, the coefficient multiplying $k_j$ in $\dot k_i$ is
\[
\frac{V_{ii}}{N-1}\frac{c}{2\tau B_j^2}\ell_j\ge0.
\]
Hence the cone $\{k\succeq0\}$ is forward invariant on this interval: if $k_i$ hits $0$ while all $k_j\ge0$, then $\dot k_i\ge0$.

\paragraph{Step 3: conclude $\dot P\succeq0$ persists.}
At time $\tau_0$ we assume $\dot P(\tau_0)\succeq0$.
On the interior OLS interval, $\ell_i(\tau_0)>0$, so $\dot P_i(\tau_0)\ge0$ implies $k_i(\tau_0)\ge0$ for each $i$.
By forward invariance, $k(\tau)\succeq0$ for later times on the interval, and then Step~1 gives $\dot P(\tau)\succeq0$.

Finally, if at some later time the trajectory enters a regime where firm $i$ is clipped (so $P_i$ is locally constant) or enters the non-OLS branch ($P_i\equiv P_{\max}$), then $\dot P_i\ge0$ holds trivially there.
Thus $P$ is componentwise nondecreasing on $[\tau_0,\infty)$.
\end{myproof}

\subsection{Proof of Lemma~\ref{lem:Pdot_initial_lower_cone}}

\begin{myproof}
Let $e_i:=P_i-U_i$ and $\bar e_{-i}:=(N-1)^{-1}\sum_{j\ne i}e_j$.
Use the notation from the proof of Lemma~\ref{lem:Pdot_sign_invariant}:
\[
B_i:=bV_{ii}-c\bar V_{i,-i},
\qquad
D_i:=a+c\bar U_{-i}-bU_i,
\qquad
\ell_i:=B_i+\tau e_iD_i,
\qquad
k_i:=V_{ii}\bar e_{-i}-\bar V_{i,-i}e_i.
\]
At $\tau=1$, $U(1)=\mu$, $V(1)=\Sigma_{\exp}$, and $\Sigma_{\exp}$ is diagonal, so
\[
V_{ii}(1)=(\Sigma_{\exp})_{ii},
\qquad
\bar V_{i,-i}(1)=0.
\]
Hence $B_i(1)=b(\Sigma_{\exp})_{ii}>0$ and the OLS branch is active.
Also $\widetilde P_i(1)=\BR(\bar\mu_{-i})$.
The clipping is inactive: the lower clip is inactive because $\BR(\bar\mu_{-i})>\mu_i\ge P_{\min}$, while the upper clip is inactive because
\[
\BR(\bar\mu_{-i})\le \BR(P_{\max})<P_{\max}.
\]
Hence
\[
P_i(1)=\BR(\bar\mu_{-i}),
\qquad
e_i(1)=\BR(\bar\mu_{-i})-\mu_i>0
\]
for every $i$, since $\mu\in\mathcal{C}^-$.

Thus $\bar e_{-i}(1)>0$ and
\[
k_i(1)=(\Sigma_{\exp})_{ii}\bar e_{-i}(1)>0.
\]
The factorization in Lemma~\ref{lem:Pdot_sign_invariant} gives
\[
\dot P_i=\frac{c}{2\tau B_i^2}\ell_i k_i
\]
on the smooth OLS interior branch.
At $\tau=1$, we have $B_i(1)>0$, $k_i(1)>0$, and
\[
\ell_i(1)=B_i(1)+e_i(1)D_i(1)>0,
\]
since
\[
D_i(1)=a+c\bar\mu_{-i}-b\mu_i\ge a+cP_{\min}-bP_{\max}>0
\]
by the standing price-bound assumptions.
Therefore $\dot P_i(1^+)>0$ for every $i$, so $\dot P(1^+)\succ0$.
\end{myproof}

\subsection{Proof of Lemma~\ref{lem:Vconv_on_NE_jackie}}

\begin{myproof}
Assume $U(\tau)\to\pNE\mathbf{1}$, $e(\tau):=P(\tau)-U(\tau)\succeq0$ for all $\tau\ge1$, $P$ is componentwise nondecreasing, $\bar V_{i,-i}(\tau)\ge0$ for all $i,\tau$, and $e(\tau_0)\succ0$ for some $\tau_0$.

\paragraph{Part (1): $P\to\pNE\mathbf{1}$ and eventually $P\preceq\pNE\mathbf{1}$.}
Fix $i$.
Since $P_i$ is nondecreasing and $P_i(\tau)\in[P_{\min},P_{\max}]$, the limit
\[
P_i^\infty:=\lim_{\tau\to\infty}P_i(\tau)
\]
exists.
The ODE $\dot U_i=(P_i-U_i)/\tau$ implies the running-average identity
\[
U_i(\tau)=\frac{1}{\tau}\left(U_i(1)+\int_1^\tau P_i(s)\,ds\right).
\]
If $P_i(s)\to P_i^\infty$, then the Ces\`aro mean converges to the same limit:
\[
\frac{1}{\tau}\int_1^\tau P_i(s)\,ds\to P_i^\infty.
\]
Hence $U_i(\tau)\to P_i^\infty$ as well. Since we also assume $U_i(\tau)\to\pNE$, it follows that $P_i^\infty=\pNE$ for every $i$, i.e.
\[
P(\tau)\to\pNE\mathbf{1}.
\]
Because $P_i$ is nondecreasing and converges to $\pNE$, we must have $P_i(\tau)\le \pNE$ for all sufficiently large $\tau$.
Thus $P(\tau)\preceq\pNE\mathbf{1}$ eventually.

\paragraph{Part (2): convergence of $V$ and strict positivity of limiting cross terms.}
For large $\tau$, Part~(1) gives $P(\tau)\preceq\pNE\mathbf{1}$.
Since $e(\tau)\succeq0$, we also have $U(\tau)\preceq P(\tau)\preceq\pNE\mathbf{1}$ for large $\tau$.
Define
\[
g_i(\tau):=\pNE-U_i(\tau)\ge0,
\qquad
\delta_i(\tau):=\pNE-P_i(\tau)\ge0
\]
for all sufficiently large $\tau$.
Then
\[
e_i(\tau)=P_i(\tau)-U_i(\tau)=g_i(\tau)-\delta_i(\tau)\le g_i(\tau).
\]

Because $\bar V_{i,-i}(\tau)\ge0$, the same argument as in the proof of Lemma~\ref{lem:weak_supra_from_bias_jackie} gives
\[
P_i(\tau)\ge \BR(\bar U_{-i}(\tau))
\qquad
\text{for all }i,\tau.
\]
Therefore, for large $\tau$,
\[
\delta_i(\tau)
=
\pNE-P_i(\tau)
\le
\pNE-\BR(\bar U_{-i}(\tau))
=
\frac{c}{2b}\bigl(\pNE-\bar U_{-i}(\tau)\bigr)
=
\frac{c}{2b}\bar g_{-i}(\tau),
\]
where $\bar g_{-i}:=(N-1)^{-1}\sum_{j\ne i}g_j$.
Let $G(\tau):=\sum_{i=1}^N g_i(\tau)$.
Using $\dot U_i=e_i/\tau$ gives
\[
\dot g_i(\tau)=-\frac{g_i(\tau)-\delta_i(\tau)}{\tau},
\]
so
\[
\dot G(\tau)
=
-\frac{1}{\tau}\sum_i (g_i(\tau)-\delta_i(\tau))
\le
-\frac{1}{\tau}\left(G(\tau)-\frac{c}{2b}\sum_i \bar g_{-i}(\tau)\right).
\]
But $\sum_i\bar g_{-i}=G$, since each $g_j$ appears in exactly $N-1$ averages and the factor $1/(N-1)$ cancels.
Hence
\[
\dot G(\tau)\le -\frac{1}{\tau}\left(1-\frac{c}{2b}\right)G(\tau).
\]
Integrating yields
\[
G(\tau)\le C \tau^{-(1-c/(2b))}
\]
for some $C>0$.
Since $c<b$ implies $1-\frac{c}{2b}>\frac12$, we have $G\in L^2([1,\infty))$.
In particular, $e_i(\tau)\le g_i(\tau)\le G(\tau)$ implies $e_i\in L^2([1,\infty))$.

Now
\[
V_{ii}(\tau)=V_{ii}(1)+\int_1^\tau e_i(s)^2\,ds
\]
converges to a finite limit $V_{ii}^\infty$ because $e_i^2$ is integrable.
Similarly, for $i\ne j$,
\[
V_{ij}(\tau)=V_{ij}(1)+\int_1^\tau e_i(s)e_j(s)\,ds
\]
converges because $e_ie_j$ is integrable by Cauchy--Schwarz.
Finally, since $e(\tau_0)\succ0$, continuity implies that $e_i(s)e_j(s)>0$ on some interval of positive length for every $i\ne j$.
Thus
\[
\int_1^\infty e_i(s)e_j(s)\,ds>0,
\]
and, because $V_{ij}(1)=0$ for $i\ne j$, we get $V_{ij}^\infty>0$ for all $i\ne j$.
Averaging gives $\bar V_{i,-i}^\infty>0$ for all $i$.
\end{myproof}

\subsection{Proof of Lemma~\ref{lem:exclude_NE_by_bias_jackie}}

\begin{myproof}
Suppose, toward a contradiction, that along a lower-cone trajectory,
\[
U(\tau)\to\pNE\mathbf{1},
\qquad
P(\tau)\to\pNE\mathbf{1},
\qquad
V_{ii}(\tau)\to V_{ii}^\infty,
\qquad
\bar V_{i,-i}(\tau)\to \bar V_{i,-i}^\infty>0
\]
for every $i$.
Fix $i$.

For large $\tau$, the non-OLS branch is impossible: if $-bV_{ii}(\tau)+c\bar V_{i,-i}(\tau)\ge0$, then by definition $P_i(\tau)=P_{\max}$, contradicting $P_i(\tau)\to\pNE<P_{\max}$.
Thus for all sufficiently large $\tau$ we are on the OLS branch, so
\[
B_i(\tau):=bV_{ii}(\tau)-c\bar V_{i,-i}(\tau)>0.
\]

We first note that $B_i^\infty:=\lim_{\tau\to\infty}B_i(\tau)$ must be strictly positive.
If $B_i^\infty=0$, then since $V_{ii}^\infty\ge V_{ii}(1)>0$ and
$\bar V_{i,-i}^\infty>0$, the numerator of the algebraic OLS price satisfies
\[
\begin{aligned}
\lim_{\tau\to\infty}
\Big((a+c\bar U_{-i}(\tau))V_{ii}(\tau)-cU_i(\tau)\bar V_{i,-i}(\tau)\Big)
&=
(a+c\pNE)V_{ii}^\infty-c\pNE\bar V_{i,-i}^\infty\\
&=
2b\pNE V_{ii}^\infty-c\pNE\bar V_{i,-i}^\infty\\
&=
\pNE\bigl(2bV_{ii}^\infty-c\bar V_{i,-i}^\infty\bigr)\\
&=
\pNE bV_{ii}^\infty
>0,
\end{aligned}
\]
where the last equality uses $bV_{ii}^\infty=c\bar V_{i,-i}^\infty$.
Since the denominator $2B_i(\tau)$ would converge to $0$ from above, the algebraic OLS price would diverge to $+\infty$, and hence the clipped posted price would converge to $P_{\max}$, not to $\pNE$.
This contradiction shows $B_i^\infty>0$.

Therefore we may pass to the limit in the algebraic OLS price.
At $U=\pNE\mathbf{1}$ and $\bar V_{i,-i}^\infty>0$, Lemma~\ref{lem:corr_bias_jackie} gives
\[
\widetilde P_i(\pNE\mathbf{1},V^\infty)
>
\BR(\pNE)
=
\pNE.
\]
Thus there exists $\eta>0$ such that, for all sufficiently large $\tau$,
\[
\widetilde P_i(U(\tau),V(\tau))\ge \pNE+\eta.
\]
Clipping cannot turn this into a price converging to $\pNE$: if $\widetilde P_i(U(\tau),V(\tau))\le P_{\max}$, then
\[
P_i(\tau)=\widetilde P_i(U(\tau),V(\tau))\ge \pNE+\eta,
\]
while if $\widetilde P_i(U(\tau),V(\tau))>P_{\max}$, then $P_i(\tau)=P_{\max}>\pNE$.
In either case, $P_i(\tau)\not\to\pNE$, contradicting the assumption.
Therefore no such trajectory exists.
\end{myproof}

\subsection{Proof of Lemma~\ref{lem:negative_orthant_invariant_jackie}}
\label{app:supracomp_step1_adapt}
\label{app:negative_orthant_invariant_proof}

\begin{myproof}
Let $e(\tau):=P(\tau)-U(\tau)$.
Suppose $e(\tau_0)\preceq0$ and assume, toward a contradiction, that the trajectory leaves $\R_-^N$ after $\tau_0$. Define
\[
\tau^\star:=\inf\{\tau\ge \tau_0:\exists i\text{ with }e_i(\tau)>0\}.
\]
At $\tau=\tau^\star$, we have $e(\tau^\star)\preceq0$ and at least one coordinate satisfies $e_i(\tau^\star)=0$. Fix such an $i$.

The upper-cap and non-OLS cases are analogous to the lower-cone argument.
If $P_i(\tau^\star)=P_{\max}$ and $e_i(\tau^\star)=0$, then $U_i(\tau^\star)=P_{\max}$.
As long as the upper clip, or the non-OLS branch $P_i\equiv P_{\max}$, remains active, $U_i$ solves
\[
\dot U_i(\tau)=\frac{P_{\max}-U_i(\tau)}{\tau}
\]
with initial value $U_i(\tau^\star)=P_{\max}$, so $U_i(\tau)=P_{\max}$ and $e_i(\tau)=0$ on that right-neighborhood.
If the upper cap ceases to bind, the posted price moves weakly downward from $P_{\max}=U_i(\tau^\star)$, which points into $\R_-^N$, not out of it.

If $P_i(\tau^\star)=P_{\min}$ and $e_i(\tau^\star)=0$, then $U_i(\tau^\star)=P_{\min}$.
While the lower clip remains active, the same scalar argument gives $U_i(\tau)=P_{\min}$ and $e_i(\tau)=0$.
If the lower clip ceases to bind upward, any possible exit must pass through the algebraic OLS boundary $\widetilde P_i=U_i$.
Thus it remains only to rule out an OLS-boundary crossing with
\[
B_i:=bV_{ii}-c\bar V_{i,-i}>0,
\qquad
\widetilde P_i(\tau^\star)-U_i(\tau^\star)=0.
\]

Use the same energy as in the proof of the nonnegative-orthant part of Lemma~\ref{lem:drift_orthant_invariant_jackie},
\[
\mathcal E_i:=2bV_{ii}\bigl(U_i-\BR(\bar U_{-i})\bigr)-cU_i\,\bar V_{i,-i},
\]
for which
\[
\widetilde P_i-U_i=\frac{-\mathcal E_i}{2B_i}.
\]
Because $B_i>0$, crossing from $\widetilde P_i-U_i\le0$ to $\widetilde P_i-U_i>0$ is equivalent to crossing from $\mathcal E_i\ge0$ to $\mathcal E_i<0$. At the boundary time, $\mathcal E_i(\tau^\star)=0$.

Now $e(\tau^\star)\preceq0$ implies $\dot U_j(\tau^\star)=e_j(\tau^\star)/\tau^\star\le0$ for all $j$, so $\dot{\bar U}_{-i}(\tau^\star)\le0$.
Also $e_i(\tau^\star)=0$ implies $\dot U_i(\tau^\star)=0$, $\dot V_{ii}(\tau^\star)=0$, and $\dot{\bar V}_{i,-i}(\tau^\star)=0$.
The same derivative computation as in the proof of Lemma~\ref{lem:drift_orthant_invariant_jackie} gives
\[
\dot{\mathcal E}_i(\tau^\star)
=
-cV_{ii}(\tau^\star)\,\dot{\bar U}_{-i}(\tau^\star)
\ge0.
\]
Thus $\mathcal E_i$ cannot cross from $\mathcal E_i\ge0$ to $\mathcal E_i<0$, equivalently $\widetilde P_i-U_i$ cannot cross from $\widetilde P_i-U_i\le0$ to $\widetilde P_i-U_i>0$.
This contradicts the definition of $\tau^\star$.
Therefore $P(\tau)-U(\tau)\preceq0$ for all $\tau\ge \tau_0$.
\end{myproof}

\subsection{Proof of Lemma~\ref{lem:upper_cone_strictness_jackie}}
\label{app:supracomp_step2_adapt}
\label{app:upper_cone_strictness_proof}

\begin{myproof}
We prove the result in four short steps.

\paragraph{Step 1: $\mathcal{C}^+$ implies exploration means exceed Nash.}
If $\mu\in\mathcal{C}^+$, then $\mu\succ\pNE\mathbf{1}$.
To see this, let $m:=\min_i\mu_i$ and choose $k$ with $\mu_k=m$.
Since $\bar\mu_{-k}\ge m$ and $\BR$ is increasing,
\[
m=\mu_k>\BR(\bar\mu_{-k})\ge \BR(m).
\]
But $\BR(u)-u$ is affine with unique zero at $u=\pNE$, hence $m>\pNE$, and therefore $\mu_i\ge m>\pNE$ for all $i$.

At $\tau=1$, $V(1)=\Sigma_{\exp}$ is diagonal, so $\bar V_{i,-i}(1)=0$.
The zero-cross-covariance case of the pricing rule gives $P_i(1)=\BR(\bar\mu_{-i})$, with the cap inactive under the standing price bounds.
Since $\bar\mu_{-i}>\pNE$ and $\BR$ is increasing with $\BR(\pNE)=\pNE$, we have
\[
P_i(1)=\BR(\bar\mu_{-i})>\pNE.
\]
Also, since $\mu\in\mathcal{C}^+$,
\[
P_i(1)-U_i(1)=\BR(\bar\mu_{-i})-\mu_i<0
\qquad\text{for all }i.
\]

\paragraph{Step 2: the negative orthant is forward invariant.}
By Lemma~\ref{lem:negative_orthant_invariant_jackie},
\[
P(\tau)-U(\tau)\preceq0
\qquad
\text{for all }\tau\ge1.
\]
Thus $U$ is componentwise nonincreasing, since $\dot U=(P-U)/\tau\preceq0$.
Also, because $e(1)\prec0$, Lemma~\ref{lemma:nonnegative_gap_implies_correlation} gives
\[
\bar V_{i,-i}(\tau)>0
\qquad
\text{for all }i\text{ and all }\tau>1.
\]

\paragraph{Step 3: the running mean remains weakly above Nash.}
The hypotheses of Lemma~\ref{lem:weak_supra_from_bias_jackie} are satisfied on $\mathcal{C}^+$:
$U$ is componentwise monotone, and $\bar V_{i,-i}(\tau)\ge0$ for all $i,\tau$.
Hence
\[
U^\infty:=\lim_{\tau\to\infty}U(\tau)
\quad\text{exists and satisfies}\quad
U^\infty\succeq\pNE\mathbf{1}.
\]
Since $U$ is componentwise nonincreasing, this implies
\[
U(\tau)\succeq U^\infty\succeq\pNE\mathbf{1}
\qquad
\text{for all }\tau\ge1.
\]
Therefore $\bar U_{-i}(\tau)\ge\pNE$ for every $i$ and $\tau$.

\paragraph{Step 4: posted prices are strictly above Nash.}
Fix $\tau>1$ and $i\in[N]$.
If the non-OLS branch is active, then $P_i(\tau)=P_{\max}>\pNE$.
If the OLS branch is active, then $bV_{ii}-c\bar V_{i,-i}>0$ and, since $\bar V_{i,-i}(\tau)>0$, Lemma~\ref{lem:corr_bias_jackie} gives
\[
\widetilde P_i(\tau)>\BR(\bar U_{-i}(\tau)).
\]
By Step~3, $\bar U_{-i}(\tau)\ge\pNE$, so
\[
\BR(\bar U_{-i}(\tau))\ge \BR(\pNE)=\pNE.
\]
Thus $\widetilde P_i(\tau)>\pNE$.
Clipping cannot reduce the posted price below $\pNE$: if $\widetilde P_i(\tau)\le P_{\max}$ then $P_i(\tau)=\widetilde P_i(\tau)>\pNE$, while if $\widetilde P_i(\tau)>P_{\max}$ then $P_i(\tau)=P_{\max}>\pNE$.
Hence
\[
P_i(\tau)>\pNE
\qquad
\text{for all }i\text{ and all }\tau>1.
\]
Together with Step~1, which gives $P_i(1)>\pNE$, we conclude
\[
P^{\mathrm{ODE}}(\tau;\mu,\Sigma_{\exp})=P(\tau)\succ\pNE\mathbf{1}
\qquad
\text{for every }\tau\in[1,\infty).
\]
\end{myproof}
\section{Proofs for the Symmetric-History Results (Section~\ref{sec:supra_competitive_limiting_prices})}
\label{app:symmetric_history_proofs}

This appendix contains two symmetric-history calculations. First, we formalize the historical-correlation interpretation from \cref{sec:supra_competitive_limiting_prices}: in a symmetric history, the common pairwise price correlation $\rho$ generates the same symmetric price as an auxiliary conduct game in which each firm internalizes rivals' profits with weight $\rho$. Second, we prove the symmetric-exploration limit in \cref{thm:symmetric_explore_zero_noise}. Throughout, $\BR(x)=(a+cx)/(2b)$ and $\barpMNP:=\min\{\pMNP,P_{\max}\}$.

\subsection{Historical correlation as an implied conduct parameter}
\label{app:correlated_history_conduct}

Consider a symmetric price history in the price-moments state. The OLS demand regression is misspecified because each firm regresses demand on its own price while omitting rivals' prices, even though rivals' prices affect demand. The relevant moments are $(U,V)$: $U_i$ is firm $i$'s historical average price, $V_{ii}$ is its accumulated own-price variance, and $V_{ij}$ is the accumulated covariance between firms $i$ and $j$.

A symmetric equicorrelated history has $U_i=u$, $V_{ii}=v>0$, and $V_{ij}=\rho v$ for every pair $i\ne j$. Then
\[
    \frac{V_{ij}}{\sqrt{V_{ii}V_{jj}}}=\rho,
\]
so $\rho$ is the common pairwise correlation of firms' historical price paths. We show that this same $\rho$ also has a conduct interpretation: the self-consistent OLS price is the same as the symmetric price in an auxiliary game where each firm acts as if it internalizes a fraction $\rho$ of rivals' profits. Thus $\rho=0$ corresponds to Nash pricing, while $\rho=1$ corresponds to full joint-profit internalization.

For the calculation, recall the OLS pricing rule
\[
    P_i
    =
    \left[
    \frac{(a+c\bar U_{-i})V_{ii}-cU_i\bar V_{i,-i}}
    {2(bV_{ii}-c\bar V_{i,-i})}
    \right]_{[P_{\min},P_{\max}]},
    \qquad
    \bar U_{-i}:=\frac{1}{N-1}\sum_{j\ne i}U_j,\quad
    \bar V_{i,-i}:=\frac{1}{N-1}\sum_{j\ne i}V_{ij},
\]
on the branch where $bV_{ii}-c\bar V_{i,-i}>0$. In the symmetric equicorrelated state below, this condition holds because $bV_{ii}-c\bar V_{i,-i}=(b-c\rho)v>0$.

Now let
\[
    \pi_i(p)=p_i(a-bp_i+c\bar p_{-i}),
    \qquad
    \bar p_{-i}:=\frac{1}{N-1}\sum_{j\ne i}p_j .
\]
The auxiliary $\rho$-conduct game is the static game in which firm $i$ chooses $p_i\in[P_{\min},P_{\max}]$ to maximize
\[
    \pi_i(p)+\rho\sum_{j\ne i}\pi_j(p).
\]
The equivalence below is an outcome equivalence: the OLS rule and this auxiliary conduct game have the same symmetric fixed-point price, although they need not have the same off-equilibrium best-response map.

\begin{lemma}[Historical correlation as an implied conduct parameter]
\label{lem:correlated_history_conduct}
Fix $N\ge2$ and $\rho\in[0,1]$. At any symmetric equicorrelated price-moments state with $U_i=u$, $V_{ii}=v>0$, and $V_{ij}=\rho v$ for all $i\ne j$, the misspecified OLS rule reduces to
\[
    B_\rho(u)
    =
    \left[
    \frac{a+c(1-\rho)u}{2(b-c\rho)}
    \right]_{[P_{\min},P_{\max}]} .
\]
Its symmetric self-consistent price is
\[
    p_\rho
    =
    \min\left\{
    P_{\max},
    \frac{a}{2b-c(1+\rho)}
    \right\}.
\]
This same $p_\rho$ is the symmetric Nash price of the auxiliary $\rho$-conduct game.
\end{lemma}

\begin{myproof}
At the stated symmetric equicorrelated state, $\bar U_{-i}=u$, $\bar V_{i,-i}=\rho v$, and $bV_{ii}-c\bar V_{i,-i}=(b-c\rho)v>0$. Substituting into the OLS pricing rule gives
\[
P_i
=
\left[
\frac{(a+cu)v-cu\rho v}{2(bv-c\rho v)}
\right]_{[P_{\min},P_{\max}]}
=
\left[
\frac{a+c(1-\rho)u}{2(b-c\rho)}
\right]_{[P_{\min},P_{\max}]}
=:B_\rho(u).
\]

A symmetric self-consistent OLS price solves $p=B_\rho(p)$. Ignoring the price caps,
\[
    2(b-c\rho)p=a+c(1-\rho)p,
    \qquad\text{so}\qquad
    p^*_\rho=\frac{a}{2b-c(1+\rho)} .
\]
As $\rho$ increases from $0$ to $1$, this price moves from the Nash price $a/(2b-c)$ to the monopoly price $a/[2(b-c)]$. Thus, under the maintained assumption that the lower cap does not bind at Nash, only the upper cap can bind, giving
\[
    p_\rho
    =
    \min\left\{
    P_{\max},
    p^*_\rho
    \right\}
    =
    \min\left\{
    P_{\max},
    \frac{a}{2b-c(1+\rho)}
    \right\}.
\]

It remains to check that the auxiliary conduct game has the same symmetric price. Differentiating firm $i$'s auxiliary objective with respect to $p_i$ gives
\[
\frac{\partial}{\partial p_i}
\left(
\pi_i(p)+\rho\sum_{j\ne i}\pi_j(p)
\right)
=
a-2bp_i+c\bar p_{-i}
+
\rho c\bar p_{-i}
=
a-2bp_i+c(1+\rho)\bar p_{-i}.
\]
Therefore the unconstrained best response in the auxiliary game is
\[
    p_i=\frac{a+c(1+\rho)\bar p_{-i}}{2b}.
\]
At a symmetric Nash equilibrium, $p_i=\bar p_{-i}=p$, so
\[
    (2b-c(1+\rho))p=a,
\]
which gives the same un-capped price $p^*_\rho$. Applying the same upper price cap yields $p_\rho$.
\end{myproof}

Thus $\rho$ gives a simple calibration of the implied degree of collusion: $\rho=0$ rationalizes the Nash price, $\rho=1$ rationalizes the capped monopoly benchmark, and intermediate values rationalize intermediate symmetric prices.

\subsection{Proof of Theorem~\ref{thm:symmetric_explore_zero_noise}: symmetric-exploration price limit}
\label{app:symmetric_explore_zero_noise_proof}

\begin{myproof}
The proof reduces the symmetric price-moments ODE to one scalar trajectory. We then use the fixed sign of the price-minus-mean gap to get convergence, relate terminal displacement to accumulated co-movement, and pass to the vanishing-exploration-noise limit.

Fix $\sigma_{\exp}>0$ and set $\Sigma_{\exp}=\sigma_{\exp}^2 I_N$. By permutation equivariance of the price-moments ODE in \cref{def:price_moments_ode} and of the posted-price map \eqref{eq:posted_price_formula}, the symmetric initial condition $U(1)=s\mathbf{1}$ and $V(1)=\sigma_{\exp}^2 I_N$ remains symmetric. Thus there are scalar functions\footnote{The lowercase notation is local to the symmetric reduction. The full ODE variables remain the vector/matrix objects $U(\tau)$, $V(\tau)$, and $P(\tau)$.} $u,v,C,p$ such that $U_i(\tau)=u(\tau)$, $V_{ii}(\tau)=v(\tau)$, $V_{ij}(\tau)=C(\tau)$ for $i\ne j$, and $P_i(\tau)=p(\tau)$. They satisfy
\[
\dot u=\frac{p-u}{\tau},\qquad
\dot v=\dot C=(p-u)^2,\qquad
u(1)=s,\quad v(1)=\sigma_{\exp}^2,\quad C(1)=0,
\]
so $v-C\equiv\sigma_{\exp}^2$. Since $-bv+cC=-(b-c)C-b\sigma_{\exp}^2<0$, the non-OLS branch never occurs. The common unclipped price is
\[
\widetilde p_{\sigma_{\exp}}(u,C)=
\frac{a(C+\sigma_{\exp}^2)+c\sigma_{\exp}^2u}
{2((b-c)C+b\sigma_{\exp}^2)},
\qquad
p(\tau)=
\bigl[\widetilde p_{\sigma_{\exp}}(u(\tau),C(\tau))\bigr]_{[P_{\min},P_{\max}]}.
\]

Let $e(\tau):=p(\tau)-u(\tau)$. At $\tau=1$, the unclipped price is $\BR(s)$, and clipping preserves the sign of $\BR(s)-s$. Since $\BR(s)-s=(a-(2b-c)s)/(2b)$, this sign is $-\operatorname{sign}(s-\pNE)$. By the sign-invariance lemma, $e(\tau)\ge0$ if $s<\pNE$, $e(\tau)\le0$ if $s>\pNE$, and $e(\tau)\equiv0$ if $s=\pNE$. Hence $u(\tau)$ is monotone and bounded, so $u_\infty(\sigma_{\exp}):=\lim_{\tau\to\infty}u(\tau)$ exists; also $C(\tau)$ increases to some $C_\infty(\sigma_{\exp})\in[0,\infty]$.

We first record the terminal relation between $u_\infty$ and $C_\infty$. If $C_\infty=\infty$, then $\widetilde p_{\sigma_{\exp}}(u(\tau),C(\tau))\to\pMNP$, so $p(\tau)\to\barpMNP$; since $u$ converges and $du/d\log \tau=e$ has fixed sign, $e(\tau)\to0$ and therefore $u_\infty=\barpMNP$. If instead $u_\infty<\barpMNP$, then $C_\infty<\infty$, the upper clip is eventually inactive, and $u_\infty=\widetilde p_{\sigma_{\exp}}(u_\infty,C_\infty)$. Solving this identity and using $\dot C=e^2=\tau^2\dot u^2$ gives
\[
C_\infty
=
\sigma_{\exp}^2
\frac{2b-c}{2(b-c)}
\frac{u_\infty-\pNE}{\pMNP-u_\infty},
\qquad
|u_\infty-s|^2
=
\left|\int_1^\infty r^{-1}(r\dot u(r))\,dr\right|^2
\le C_\infty .
\]
Consequently, whenever $u_\infty<\barpMNP$,
\[
|u_\infty-s|
\le
\sigma_{\exp}
\left(
\frac{2b-c}{2(b-c)}
\frac{u_\infty-\pNE}{\pMNP-u_\infty}
\right)^{1/2}.
\]
The same relation also shows $u_\infty\ge\pNE$; otherwise the displayed formula for $C_\infty$ would be negative.

We also need an upper bound. If $\pMNP\ge P_{\max}$, then $u_\infty\le P_{\max}=\barpMNP$. If $\pMNP<P_{\max}$, then
\[
\widetilde p_{\sigma_{\exp}}(u,C)-u
=
\frac{
2(b-c)(\pMNP-u)C+(2b-c)(\pNE-u)\sigma_{\exp}^2
}{
2((b-c)C+b\sigma_{\exp}^2)
}.
\]
Thus $u(\tau)\ge\pMNP+\delta$ eventually would imply $p(\tau)-u(\tau)\le-\eta$ eventually, uniformly over $C(\tau)\ge0$, contradicting convergence of $u(\tau)$. Hence $\pNE\le u_\infty(\sigma_{\exp})\le\barpMNP$ for every $\sigma_{\exp}>0$.

Now let $\sigma_{\exp}\downarrow0$. If $s=\pNE$, then $e\equiv0$ and $u_\infty(\sigma_{\exp})=s$. If $s\in(\pNE,\barpMNP)$, monotonicity gives $u_\infty(\sigma_{\exp})\le s$, and the factor in the width bound is uniformly bounded, so $u_\infty(\sigma_{\exp})\to s$. If $s=\barpMNP$, the same conclusion follows by contradiction: any subsequence with $u_\infty(\sigma_k)\le s-\varepsilon$ has a uniformly bounded width factor and is forced to converge to $s$. Therefore $u_\infty(\sigma_{\exp})\to s$ for all $s\in[\pNE,\barpMNP]$.

If $s<\pNE$, then $u_\infty(\sigma_{\exp})\ge\pNE>s$. A subsequence satisfying $u_\infty(\sigma_k)\le\barpMNP-\varepsilon$ would again have a uniformly bounded width factor and would be forced to converge to $s$, a contradiction. Hence $u_\infty(\sigma_{\exp})\to\barpMNP$. Finally, if $s>\barpMNP$, then necessarily $\barpMNP=\pMNP<P_{\max}$; the upper bound gives $u_\infty(\sigma_{\exp})\le\pMNP$, and any subsequence bounded below $\pMNP$ by a fixed $\varepsilon$ is ruled out by the same width-bound argument. Thus $u_\infty(\sigma_{\exp})\to\pMNP=\barpMNP$.

For each fixed $\sigma_{\exp}>0$, the preceding terminal argument gives $p(\tau)-u(\tau)\to0$. Since the symmetric trajectory satisfies $P^{\mathrm{ODE}}(\tau;s\mathbf{1},\sigma_{\exp}^2 I_N)=p(\tau)\mathbf{1}$, we have
\[
\lim_{\tau\to\infty}
P^{\mathrm{ODE}}(\tau;s\mathbf{1},\sigma_{\exp}^2 I_N)
=
u_\infty(\sigma_{\exp})\mathbf{1}.
\]
Combining this identity with the scalar limits above gives exactly the two cases in \cref{thm:symmetric_explore_zero_noise}.
\end{myproof}
\section{Proofs for the Best-Response Cone Characterizations (Section~\ref{sec:cone_discussion})}
\label{app:cone_discussion_proofs}

This appendix proves the two auxiliary propositions stated in
Section~\ref{sec:cone_discussion}. Throughout, $\BR(x)=(a+cx)/(2b)$ and
$C^+,C^-$ are the best-response cones defined in
\eqref{eq:upper_cone}--\eqref{eq:lower_cone}. We write $P(\tau)$ for the posted-price coordinate of
the price-moments ODE and $U(\tau)$ for the running price mean.

\subsection{Proof of \cref{prop:duopoly_limit_points_cones}}
\label{app:duopoly_limit_points_cone_proof}

\begin{myproof}
Consider the duopoly case \(N=2\). Let \(\Pi(U,V)\) denote the posted-price map in the
price-moments ODE. Write \((U(\tau),V(\tau))\) for the ODE solution, \(P(\tau):=\Pi(U(\tau),V(\tau))\), and
\(e(\tau):=P(\tau)-U(\tau)\). We first show that \(P(\tau)\) converges.

For log time \(s\ge0\), define
\[
    U^\ell(s):=U(e^s),\qquad
    Z^\ell(s):=e^{-s}V(e^s),\qquad
    P^\ell(s):=\Pi(U^\ell(s),Z^\ell(s)),
\]
and set \(e^\ell(s):=P^\ell(s)-U^\ell(s)\). Since \(\Pi\) is homogeneous of degree zero in its
second argument, \(P^\ell(s)=P(e^s)\). The log-time dynamics are
\[
    \dot U^\ell=e^\ell,\qquad
    \dot Z^\ell=e^\ell(e^\ell)^\top-Z^\ell,
\]
where dots denote derivatives in \(s\). The trajectory \((U^\ell,Z^\ell)\) is bounded: \(U^\ell(s)\in
[P_{\min},P_{\max}]^2\), and
\[
    Z^\ell(s)=e^{-s}\Sigma_{\exp}
    +\int_0^s e^{-(s-r)}e^\ell(r)e^\ell(r)^\top\,dr .
\]

We use the following elementary consequence of this autonomous log-time system. If \(U^\ell(s)\)
has a componentwise limit, then \(e^\ell(s)\to0\). Indeed, every point in the \(\omega\)-limit set
has the same \(U^\ell\)-coordinate, equal to that limit. Since the \(\omega\)-limit set is invariant,
the \(U^\ell\)-coordinate must remain fixed along the vector field on this set. But
\(\dot U^\ell=e^\ell\), so every \(\omega\)-limit point has \(e^\ell=0\). Hence, by continuity along
the trajectory, \(e^\ell(s)\to0\).

We now consider the sign pattern of \(e^\ell\). By
\cref{lem:drift_orthant_invariant_jackie}, if \(e^\ell(s_0)\succeq0\) for some \(s_0\), then
\(e^\ell(s)\succeq0\) for all \(s\ge s_0\). By the sign-reversed argument in
Appendix~\ref{app:supracomp_step1_adapt}, if \(e^\ell(s_0)\preceq0\), then \(e^\ell(s)\preceq0\) for all
\(s\ge s_0\). Thus, if the trajectory ever enters one of the two closed same-sign orthants, then
\(U^\ell\) is componentwise monotone and bounded on the tail. Hence \(U^\ell(s)\) has a
componentwise limit, so \(e^\ell(s)\to0\), and \(P^\ell(s)=U^\ell(s)+e^\ell(s)\) converges.

It remains to consider the case where the trajectory never enters either closed same-sign orthant.
Then \(e^\ell_1(s)e^\ell_2(s)<0\) for every \(s\ge0\). Moreover, the mixed-sign orientation cannot
change without first entering one of the two closed same-sign orthants. Hence either
\(e^\ell_1(s)>0>e^\ell_2(s)\) for all \(s\ge0\), or \(e^\ell_2(s)>0>e^\ell_1(s)\) for all \(s\ge0\).
In the first case \(U^\ell_1\) is increasing and bounded while \(U^\ell_2\) is decreasing and bounded;
in the second case the roles of the two coordinates are reversed. Thus \(U^\ell(s)\) again has a
componentwise limit, so \(e^\ell(s)\to0\), and \(P^\ell(s)=U^\ell(s)+e^\ell(s)\) converges.

Therefore \(P(\tau)\) converges as \(\tau\to\infty\), equivalently
\[
    P^\infty(\mu,\Sigma_{\exp})
    :=\lim_{\tau\to\infty}P^{\mathrm{ODE}}(\tau;\mu,\Sigma_{\exp})
\]
exists. The preceding argument also gives \(U(\tau)\to P^\infty(\mu,\Sigma_{\exp})\).

It remains to identify the limit. Define
\[
    \Delta_1(\tau):=P_1(\tau)-\BR(U_2(\tau)),\qquad
    \Delta_2(\tau):=P_2(\tau)-\BR(U_1(\tau)).
\]
We show that \(\Delta_1(\tau)\) and \(\Delta_2(\tau)\) always have the same weak sign. On the OLS
branch, a direct rearrangement of the price-moments pricing formula gives
\[
    \widetilde P_1-\BR(U_2)
    =
    \frac{cV_{12}(a+cU_2-bU_1)}{2b(bV_{11}-cV_{12})},
    \qquad
    \widetilde P_2-\BR(U_1)
    =
    \frac{cV_{12}(a+cU_1-bU_2)}{2b(bV_{22}-cV_{12})}.
\]
The terms \(a+cU_2-bU_1\) and \(a+cU_1-bU_2\) are strictly positive because
\(U_i\in[P_{\min},P_{\max}]\) and \(a-bP_{\max}+cP_{\min}>0\). Hence, on the OLS branch, both
unclipped deviations have the sign of \(V_{12}\).

Clipping preserves this weak sign. Since \(\BR\) is increasing with fixed point
\(\pNE\in[P_{\min},P_{\max}]\), we have \(\BR(x)\in[P_{\min},P_{\max}]\) for every
\(x\in[P_{\min},P_{\max}]\). Therefore projecting \(\widetilde P_i\) onto
\([P_{\min},P_{\max}]\) cannot change the weak sign of
\(\widetilde P_i-\BR(U_{-i})\). Finally, on the non-OLS branch, the pricing rule sets
\(P_i=P_{\max}\); this branch can occur only when the corresponding cross-covariance term is
nonnegative, and then \(P_i-\BR(U_{-i})\ge0\). Therefore, for every \(\tau\),
\[
    \Delta_1(\tau)\Delta_2(\tau)\ge0 .
\]

Passing to the limit and using \(U(\tau)\to P^\infty\) and continuity of \(\BR\), we obtain
\[
    \bigl(P^\infty_1-\BR(P^\infty_2)\bigr)
    \bigl(P^\infty_2-\BR(P^\infty_1)\bigr)\ge0 .
\]
Thus either both deviations are weakly positive, in which case
\(P^\infty(\mu,\Sigma_{\exp})\in\overline{\mathcal{C}^+}\), or both deviations are weakly negative,
in which case \(P^\infty(\mu,\Sigma_{\exp})\in\overline{\mathcal{C}^-}\). Hence
\[
    P^\infty(\mu,\Sigma_{\exp})
    \in \overline{\mathcal{C}^+}\cup\overline{\mathcal{C}^-}.
\]

If both best-response equalities bind, then
\[
    P^\infty_1=\BR(P^\infty_2),\qquad
    P^\infty_2=\BR(P^\infty_1).
\]
Since \(\BR(x)=(a+cx)/(2b)\) is affine with slope \(c/(2b)<1\), this fixed-point system has the
unique solution \(P^\infty_1=P^\infty_2=a/(2b-c)=\pNE\). Thus the only point at which both
best-response equalities bind is \(\pNE\mathbf 1\). Consequently, if the limiting point is not on a
one-sided best-response boundary, then it lies in \(\mathcal{C}^+\) or \(\mathcal{C}^-\).
\end{myproof}

\subsection{Proof of Proposition~\ref{prop:random_interval_cone_probability}}
\label{app:random_interval_cone_probability_proof}

\begin{myproof}
It is convenient to recenter prices around Nash. Let
\[
z_i:=\mu_i-p^{\NE},
\qquad
\lambda:=\frac{c}{2b}=\frac{r}{2}.
\]
Since $\BR(p^{\NE}+z)-p^{\NE}=\lambda z$, the cone conditions become
\[
C^+
=
\left\{
z_i>\lambda\frac{1}{N-1}\sum_{j\ne i}z_j\ \text{ for all }i
\right\},
\qquad
C^-
=
\left\{
z_i<\lambda\frac{1}{N-1}\sum_{j\ne i}z_j\ \text{ for all }i
\right\}.
\]

Write the outer band in the random interval prior as
$[\underline P,\overline P]=[p^{\NE}-L,p^{\NE}+H]$, with $L,H>0$. Let $x,y$ be the two outer-band draws, and let
$\ell:=\min\{x,y\}$ and $u:=\max\{x,y\}$ as in the main-body definition of the prior.
Conditional on both recentered anchors being positive, equivalently $\ell>p^{\NE}$, write
\[
z_\ell:=\ell-p^{\NE},
\qquad
z_u:=u-p^{\NE},
\]
so $0<z_\ell<z_u$. The remaining $N-2$ firms draw independently from $[\ell,u]$, so after recentering we may write
\[
z_k=z_\ell+(z_u-z_\ell)Y_k,\qquad k=3,\dots,N,
\]
where the $Y_k$ are i.i.d. $\Unif[0,1]$. Let
\[
q:=\frac{z_\ell}{z_u},
\qquad
Y:=\sum_{k=3}^N Y_k.
\]
Conditional on both recentered anchors being positive, the ratio $q$ is uniform on $[0,1]$ and independent
of $Y$.

When all $z_i$ are positive, membership in $C^+$ is determined by the smallest coordinate
$z_\ell$: since the inequality
\[
z_i>\lambda\frac{\sum_{j\ne i}z_j}{N-1}
\]
is most stringent for the smallest $z_i$, it is necessary and sufficient to check it at $z_i=z_\ell$. Thus
$\mu\in C^+$ whenever
\[
z_\ell>\lambda\frac{z_u+\sum_{k=3}^N z_k}{N-1}.
\]
Dividing by $z_u$ and substituting $z_k/z_u=q+(1-q)Y_k$, this condition is equivalent to
\[
q>\theta(Y),
\qquad
\theta(Y):=
\frac{\lambda(1+Y)}
{(N-1)-\lambda(N-2-Y)}.
\]
Since $q\sim\Unif[0,1]$ independently of $Y$,
\[
\Pr(\mu\in C^+\mid \text{both recentered anchors positive})
=
\E[1-\theta(Y)].
\]
The function $y\mapsto\theta(y)$ is concave on $[0,N-2]$, because
\[
\theta''(y)
=
-\frac{2\lambda^2\bigl((N-1)-\lambda(N-1)\bigr)}
{\bigl((N-1)-\lambda(N-2-y)\bigr)^3}
<0.
\]
Therefore Jensen's inequality gives
\[
\E[\theta(Y)]
\le
\theta(\E Y)
=
\theta\!\left(\frac{N-2}{2}\right)
=
\frac{N\lambda}{2(N-1)-\lambda(N-2)}.
\]
Hence
\[
\Pr(\mu\in C^+\mid \text{both recentered anchors positive})
\ge
1-
\frac{N\lambda}{2(N-1)-\lambda(N-2)}
=
\frac{2(N-1)(1-\lambda)}
{2(N-1)-\lambda(N-2)}.
\]

The same argument applies conditional on both recentered anchors being negative, equivalently $u<p^{\NE}$, after replacing $z_i$ by
$-z_i$, and gives the identical lower bound for $C^-$. If
\[
\pi_+:=\frac{H}{L+H},
\qquad
\pi_-:=\frac{L}{L+H},
\]
then the probability that both recentered anchors are positive is $\pi_+^2$ and the probability that both
recentered anchors are negative is $\pi_-^2$. Therefore
\[
\Pr(\mu\in C^+\cup C^-)
\ge
(\pi_+^2+\pi_-^2)
\frac{2(N-1)(1-\lambda)}
{2(N-1)-\lambda(N-2)}.
\]
Since $\pi_++\pi_-=1$, we have $\pi_+^2+\pi_-^2\ge1/2$. Thus
\[
\Pr(\mu\in C^+\cup C^-)
\ge
\frac{(N-1)(1-\lambda)}
{2(N-1)-\lambda(N-2)}.
\]
Substituting $\lambda=r/2$ yields
\[
\Pr(\mu\in C^+\cup C^-)
\ge
\frac{(N-1)(2-r)}
{4(N-1)-r(N-2)}.
\]

It remains to show that this lower bound is at least $1/4$. Using $\lambda=r/2<1/2$,
\[
\frac{(N-1)(1-\lambda)}
{2(N-1)-\lambda(N-2)}
\ge
\frac14
\]
is equivalent to
\[
4(N-1)(1-\lambda)\ge 2(N-1)-\lambda(N-2),
\]
or
\[
2(N-1)\ge \lambda(3N-2).
\]
This holds for every $N\ge2$ because $\lambda<1/2$ implies
\[
\lambda(3N-2)<\frac{3N-2}{2}\le 2(N-1).
\]
Therefore
\[
\Pr\bigl(\mu\in C^+\cup C^-\bigr)
\ge
\frac{(N-1)(2-r)}
{4(N-1)-r(N-2)}
\ge
\frac14,
\]
as claimed.
\end{myproof}

\end{document}